\documentclass[11pt]{article}

\usepackage[body={16cm,23cm}]{geometry}
\usepackage{epsfig,amsfonts,amssymb}
\usepackage{enumerate}
\usepackage{subeqn}
\usepackage{cite}
\usepackage[hang,small,bf
]{caption}
\usepackage{epic}
\usepackage{eepic}
\usepackage{color}
\usepackage[all,cmtip,line]{xy}
\newcommand{\arr}[4]{\begin{picture}(20,20)\drawline(0,0)(#1,#2)
\drawline(0,0)(#3,#4)\drawline(#1,#2)(#3,#4)\end{picture}}

\definecolor{pink}{rgb}{1,0.5,0.5} 

\def\Cbbd{\mathbb{C}}

\def\Zbbd{\mathbb{Z}}
\def\Tb{\mathbf{T}}
\def\Qb{\mathbf{Q}}

\def\Pcal{\mathcal{P}}

\def\Rcal{\mathcal{R}}
\def\Scal{\mathcal{S}}

\def\Dcal{\mathcal{D}}

\def\Pcal{\mathcal{P}}

\def\Rcal{\mathcal{R}}
\def\Scal{\mathcal{S}}

\def\a{\alpha}
\def\b{\beta}
\def\g{\gamma}

\def\l{\lambda}
\def\t{\theta}
\def\vt{\vartheta}
\def\s{\sigma}
\def\o{\omega}
\def\beq{\begin{equation}}
\def\eeq{\end{equation}}
\def\be{\begin{displaymath}}
\def\ee{\end{displaymath}}
\def\bea{\begin{eqnarray}}
\def\eea{\end{eqnarray}}
\def\ov{\overline}
\def\bmat{\left(\begin{array}}

\def\ds{\displaystyle}

\def\Wp{ \raise.4ex\hbox{\textrm{\Large $\wp$}}}

\def\TJ{{\mathsf \Theta}}
\def\HJ{{\mathsf H}}

\def\K{{\mathsf K}}
\def\La{{\mathsf \Lambda}}
\def\Q{{\bf Q}}
\def\T{{\bf T}}

\def\R{{\mathbf R}}

\def\eqref#1{(\ref{#1})}
\def\?{(?)\marginpar{|?}}

\renewcommand{\author}[1]{\large\rm #1\\ \bigskip}
\newcommand{\address}[1]{{\normalsize\it #1\\}\bigskip}
\renewcommand{\title}[1]{\bigskip\bigskip\Large\bf #1\bigskip\bigskip\\}
\renewcommand\theequation{\thesection.\arabic{equation}}
\def\nsection#1{\setcounter{equation}{0}\section{#1}}
\def\As{{\mathsf A}}

\def\Cs{{\mathsf C}}
\def\Fs{{\mathsf F}}
\def\Gs{{\mathsf G}}

\def\Ts{{\mathsf T}}

\def\Qs{{\mathsf Q}}
\def\Ws{{\mathsf W}}
\def\Ms{{\mathsf M}}
\def\Ls{{\mathsf \Lambda}}
\def\Es{{\mathsf E}}
\def\hs{{\mathsf h}}
\def\fs{{\mathsf f}}
\def\Qn{{\mathsf Q}_0}
\def\Qo{\overline{\mathsf Q}}
\def\Qbar{\overline{{\mathsf Q}}_0}
\def\hf{{\textstyle {\frac{1}{2}}}}
\def\q{{\mathsf q}}

\newcounter{app}
\newcounter{sapp}[app]
\def\theapp{\Alph{app}}
\newcommand{\app}[1]{
\refstepcounter{app}{\vspace{7mm}
\noindent\Large\bf Appendix
\theapp.
 \ #1 \par \vspace{5mm}}
\setcounter{equation}{0}
\def\theequation{\Alph{app}.\arabic{equation}}}

\begin{document}
\vglue .3 cm

\begin{center}
\title{Analytic theory of the eight-vertex model}

\author{        Vladimir V. Bazhanov\footnote[1]{email:
                {\tt Vladimir.Bazhanov@anu.edu.au}} and
                Vladimir V. Mangazeev\footnote[2]{email:
                {\tt vladimir@maths.anu.edu.au}}}

\address{Department of Theoretical Physics,\\
         Research School of Physical Sciences and Engineering,\\
    Australian National University, Canberra, ACT 0200, Australia.}

\end{center}

\setcounter{footnote}{0}
\vspace{5mm}

\begin{abstract}
We observe that the exactly solved
{\em eight-vertex solid-on-solid model\/}
contains an hitherto unnoticed arbitrary field parameter, similar
to the horizontal field in the six-vertex model. The parameter is required
to describe a continuous spectrum of the {\em unrestricted}\/  solid-on-solid
model, which has an infinite-dimensional space of states even for a
finite lattice. The introduction of the continuous field parameter allows us
to completely review the theory of functional relations in
the eight-vertex/SOS-model from a uniform analytic point of view.
We also present a number of analytic and numerical techniques for the
analysis of the Bethe Ansatz equations.
It turns out that different solutions of
these equations can be obtained
from each other by  analytic continuation.
In particular, for
small lattices we explicitly demonstrate that the largest and smallest
eigenvalues of the transfer matrix of the eight-vertex model
are just different branches of the
same multivalued function of the field parameter.
\end{abstract}

\vspace{0.5cm}
\vspace{0.5cm}


\newpage


\nsection{Introduction}
The powerful analytic and algebraic techniques discovered by Baxter
in his pioneering papers \cite{Bax72,Bax73a,Bax73b,Bax73c}
on the exact solution of the eight-vertex lattice
model laid the foundation for many important applications in the theory
of integrable systems of statistical mechanics and quantum field theory.

This paper concerns one of these techniques --- the method of functional
relations. Over the last three decades, since Baxter's original works
\cite{Bax72,Bax73a,Bax73b,Bax73c}, this method has been substantially
developed and applied to a large number of various solvable models.
However, the status of this method in the eight-vertex model itself
with an account of all subsequent developments has not been recently
reviewed. This paper is intended to (partially) fill this gap.
Here we will adopt an analytic approach exploiting the existence
of an (hitherto unnoticed) continuous field parameter in the solvable
eight-vertex solid-on-solid model of ref.\cite{Bax73b}.

For the purpose of the following discussion it will be useful to
first summarize the key results of \cite{Bax72,Bax73a,Bax73b,Bax73c}.
Here we will use essentially the same notations
as those in \cite{Bax72}.
Consider the homogeneous eight-vertex (8V) model
on a square lattice of $N$ columns, with periodic boundary conditions.
The model contains three arbitrary parameters $u$, $\eta$ and
$\q=e^{i\pi\tau}$,
${\rm Im}\,\tau>0$, which enter the parametrization of the Boltzmann weights
(the parameter $\q$ enters as the nome for the elliptic theta-functions).
The parameters $\eta$ and $\q$ are considered as constants
and the spectral parameter $u$ as a complex variable.
We assume that the parameter $\eta$ is real and positive,
$0<\eta<\pi/2$,\ which corresponds to the disordered regime
\cite{Bax82}
of the model.

The row-to-row transfer
matrix of the model, $\Tb(u)$, possesses
remarkable analytic properties. Any of its
eigenvalues, $\Ts(u)$, is both
(i)  an entire function of the variable $u$, and
(ii) satisfies Baxter's famous functional equation,
\beq
\Ts(u)\,\Qs(u)=\fs(u-\eta)\,\Qs(u+2\eta)+\fs(u+\eta)\,\Qs(u-2\eta),
\label{TQei}
\eeq
where
\footnote{\label{foot1}Here
we use the standard theta-functions \cite{WW},
$\vt_i(u\,|\,\q)$, \ $i=1,\ldots,4$, \ $\q=e^{{\rm i} \pi \tau}$,\
$\rm{Im}\,\tau>0$, with the periods $\pi$ and $\pi \tau$.
Our spectral parameter $u$ is  shifted with respect to that in
\cite{Bax72} by a half of the imaginary period, see Sect~\ref{e8-model}
 for further details.}
\beq
\fs(u)=\big(\vartheta_4(u\,|\,\q)\big)^N\ ,\label{phi-def}
\eeq
and $\Qs(u)$ is an entire quasi-periodic function of $u$, such that
\beq
\Qs(u+\pi)=\pm (-1)^{N/2}\,\Qs(u),\qquad
\Qs(u+2\pi\tau)=\q^{-2N}\,e^{-2iu N}\,\Qs(u).\label{qper-1}
\eeq
These analytic properties completely determine all eigenvalues
of the transfer matrix $\T(u)$. Indeed, Eq.\eqref{TQei} implies that
the zeroes $u_1,u_2,\ldots,u_n$, of $\Qs(u)$
satisfy the Bethe Ansatz equations,
\beq
\frac{\fs(u_k+\eta)}{\fs(u_k-\eta)}=-\frac{\Qs(u_k+2\eta)}{\Qs(u_k-2\eta)},
\qquad    \Qs(u_k)=0,\qquad k=1,\ldots,n \ .
\label{BAei}\eeq
These equations, together with the periodicity relations \eqref{qper-1},
define the entire function $\Qs(u)$ (there will be many solutions
corresponding to different eigenvectors). Once
$\Qs(u)$ is known the eigenvalue $\Ts(u)$ is
evaluated from \eqref{TQei}.

The entire functions $\Qs(u)$ appearing in \eqref{TQei} are, in fact,
eigenvalues of another matrix, $\Qb(u)$, called the $\Qb$-matrix.
Originally it was constructed \cite{Bax72} in terms of some special
transfer matrices.  A different, but related, construction of the
$\Qb$-matrix was given in \cite{Bax73a} and later on used
in the book \cite{Bax82}. An alternative approach to the 8V-model
was developed in
\cite{Bax73b,Bax73c} where Baxter invented the {\em ``eight-vertex''
  solid-on-solid}\  (SOS) model and solved it exactly by means of the
co-ordinate Bethe Ansatz.  This approach provided
another derivation of the same result (\ref{TQei})-(\ref{BAei}), since the
8V-model is embedded within the SOS-model.

Baxter's $\Qb$-matrix (or the $\Qb$-operator) possesses various exceptional
properties and plays an important role in many aspects of the
theory of integrable systems.
A complete theory of the $\Qb$-operator in the 8V-model is not yet developed.
However for the simpler models related with the quantum affine
algebra $U_q(\widehat{sl_2})$ (where the fundamental ${\bf
  L}$-operators \cite{Faddeev:1979}
are intertwined by the ${\bf R}$-matrix of the six-vertex
model) the properties of the $\Qb$-operator are very well understood
\cite{BLZ97a}.
In this case the $\Qb$-operators (actually, there are two different
$\Qb$-operators, $\Qb_+$ and $\Qb_-$) are defined as traces of
certain monodromy matrices associated with infinite-dimensional
representations of the so-called $q$-oscillator algebra. The main algebraic
properties of the $\Qb$-operators can be concisely expressed by a single
factorization relation
\beq
\Tb_j^+(u)=\Qb_+(u+(2j+1)\eta)\Qb_-(u-(2j+1)\eta)\label{tjplus}
\eeq
where  $\Tb_j^+(u)$ is the transfer matrix associated with the
infinite-dimensional highest weight representation of $U_q({sl_2})$
with an arbitrary (complex) weight $2j$.
Remarkably, this relation
alone leads to a simple derivation of {\em all}
functional relations involving various ``fusion'' transfer matrices
and $\Q$-operators \cite{BLZ97a,BLZ99a}.
For this reason Eq.\eqref{tjplus} can be regarded as a {\em
fundamental fusion relation\/}: once it is derived, no further algebraic
work is required.

An important part of the theory of the $\Qb$-operators
belongs to their analytic properties
  with respect to a certain
parameter, which we call here the ``field parameter".
In the context of conformal field theory (considered
in \cite{BLZ97a,BLZ99a})
this is the ``vacuum parameter'', which determines
the Virasoro highest weight $\Delta$; in the six-vertex model it corresponds
to the horizontal field. In fact, the very existence of two
different solutions \cite{BLZ97a,KLWZ97} of the TQ-equation \eqref{TQei}
can be simply illustrated by the fact that the spectrum of the transfer matrix
does not depend on the sign of the field, whereas the spectrum of the
$\Qb$-operator does.

It is well known that it is impossible to introduce an arbitrary field
parameter into the ``zero-field'' or
``symmetric'' eight vertex model of \cite{Bax72}
without destroying its integrability.
However, such parameter is intrinsically present in the solvable
SOS-model. It does
not enter the Boltzmann weights, but arises from a proper definition of
the space of states of the model. To realize this recall that
the SOS-model \cite{Bax73c} is an interaction-round-a-face
model where the face variables $\ell_i$ (called the heights)
take arbitrary integer values
$-\infty<\ell_i<+\infty$. Its transfer matrix acts in an
infinite-dimensional space of states even for a finite
lattice.
It has a continuous spectrum, parameterized by the eigenvalue of the
operator which simultaneously increments  all
height variables, $\ell_i\to\ell_i+1$, on the lattice.
Indeed, taking into account the results of \cite{TF79,FV96},
it is not difficult to conclude
 that the calculations of \cite{Bax73c}
require only a very simple modification to deduce
that the eigenvalues of the SOS transfer matrix enjoy
the same TQ-equation \eqref{TQei}, but require different periodicity
properties
\beq
\Qs_\pm(u+\pi)=e^{\pm i\varphi}\, \Qs_\pm(u),\qquad
\Qs_\pm(u+2\pi\tau)=\q^{-2N}\,e^{\pm \psi}\,e^{-2iu N}\,\Qs_\pm(u),
\label{qper2}
\eeq
where the exponent $\varphi$ is arbitrary. It is determined by the
eigenvalue  $\omega=e^{2i\eta\varphi/\pi}$ of the height translation
operator\footnote{%
In \cite{Bax73c} Baxter restricted the parameter $\eta$ to the
``rational'' values $L\,\eta=m_1\pi+m_2\pi\tau$, \
$L,m_1,m_2\in\Zbbd$ and considered
a finite-dimensional subspace of the whole space of states,
regarding the values of heights {\it modulo} $L$.
In this case the phase factors $\omega=e^{2i\eta\varphi/\pi}$
take quantized values $\omega^L=1$ (see \cite{TF79, Bax02} for further
discussion of this point).
Apart from providing the conceptual advantage of a finite-dimensional
space of states, the above restriction on $\eta$ and $\omega$
was not used anywhere
else in \cite{Bax73c} and, therefore, can be removed.
The transfer matrix of the
  8V-vertex model (reformulated as the SOS-model) acts only in the
  finite-dimensional subspace of the SOS space of states,
  corresponding to a discrete set of exponents $\varphi=k\pi$
and $\psi=0$ (the value of
  $N$ is assumed to be even).}
  (the second exponent $\psi$ is dependent on $\varphi$).
It is natural to assume, that
the functions $\Qs_\pm(u)$
solving these equations, are eigenvalues of the $\Q$-operators for the
SOS-model. Of course, it would be very desirable to obtain their
explicit definition (and generalize the
algebraic result \eqref{tjplus} to the SOS-model), however,
many properties of these operators can already be deduced from the information
about their eigenvalues.

In this paper we will develop the analytic theory
of the functional relations for the SOS-model starting from the eigenvalue
equations \eqref{TQei} and \eqref{qper2}. Bearing in mind that
the TQ-equation \eqref{TQei} arises from very non-trivial
algebraic ``fusion'' relations \cite{Bax72},
it is not surprising that it implies
all other functional relations. The required calculations are
essentially the same as those in \cite{BLZ97a,BLZ99a}, apart from
trivial modifications arising in the context of lattice models.

The eigenvalues $\Qs_\pm(u)$ are two linear independent
``Bloch wave'' solutions \cite{BLZ97a,KLWZ97}
of the finite difference equation \eqref{TQei}
for the same $\Ts(u)$.  Their
quantum Wronskian $\Ws(\varphi)$, defined as,
\beq
2i\,\Ws(\varphi)\,{\fs(u)}={\Qs_+(u+\eta)\,\Qs_-(u-\eta)
-\Qs_+(u-\eta)\,\Qs_-(u+\eta)}\ ,\label{wrons}
\eeq
is a complicated function of $\varphi$, $\eta$
and $\q$, depending on the eigenvalue  $\Ts(u)$.
The Bloch solutions $\Qs_\pm(u)$ are well defined
provided the exponent $\varphi$ does not take some
``singular values'' (see Eq.\eqref{dang} below), where
$\Ws(\varphi)$ vanishes.
Otherwise Eq.\eqref{TQei} has only one quasi-periodic
solution, while the second linear independent solution does not
possess  any simple periodicity properties.

All singular cases (in fact, they split into different classes)
can be effectively studied with a limiting
procedure starting from a non-singular value of $\varphi$.
In the simplest case, when $\eta$ is generic and $\varphi$
approaches the points $\varphi=k\pi$, $k\in\Zbbd$, the
solutions $\Qs_+(u)$ and $\Qs_-(u)$ smoothly approach the same value
(which for even $N$ coincides with the eigenvalue $\Qs(u)$ of the
8V-model).

A more complicated situation occurs when the field $\varphi$ tends to a
singular value, say $\varphi=0$,
while the parameter $\eta$ simultaneously approaches some
rational fraction of $\pi$, where
the transfer matrix of
the 8V-model has degenerate eigenvalues. The limiting value of
$\Ts(u)$ is always uniquely defined. However, if $\Ts(u)$ is a
degenerate eigenvalue, the limiting values of
$\Qs_\pm(u)$ are not uniquely defined.
They have ``complete exact strings''
of zeroes whose position can be made arbitrary
by changing the direction of the
two-parameter $(\eta,\varphi)$-limit.
Obviously, this
reflects a non-uniqueness of eigenvectors for degenerate states
\cite{Bax02}. An immediate consequence of this phenomenon is that, for
rational $\eta$, there is no
unique algebraic definition of the $\Q$-operator
in the symmetric 8V-model.
This explains an important observation of \cite{McCoy1},
that Baxter's two $\Q$-operators, constructed in \cite{Bax72} and
\cite{Bax73a},  are actually different operators, with different
eigenvalues for degenerate eigenstates.

Further, the eigenvalues $\Qs_\pm(u),$ considered as functions of
$\varphi$, have rather complicated analytic properties. Besides having
the (relatively simple) singular points discussed above, they
are multivalued functions with algebraic branching points in the
complex $\varphi$-plane. The corresponding multi-sheeted Riemann
surface appears to be extremely complicated; we were only able to
numerically explore it for some particular eigenvalues.

It is easy to see that the roots of the Bethe
Ansatz equations \eqref{BAei}, considered as functions of $\varphi$
satisfy a system of the first order ordinary differential equations,
$du_k/d\varphi=U_k(u_1,u_2,\ldots,u_n)$, \  where  $U_k$ are
meromorphic functions of their arguments.
Using these equations one can analytically continue any particular
solution of \eqref{BAei} along a continuous path between
two points, corresponding to the same value of $\varphi$
on different sheets of
the Riemann surface.
In general, the resulting set of roots $u_1,u_2,\ldots,u_n$
differs from the initial one, but, of course, satisfies
exactly the same equation \eqref{BAei}.
In other words,
different solutions of the Bethe Ansatz equations and, therefore,
different eigenvalues of the transfer matrix can be obtained from each
other by the analytic continuation in the parameter $\varphi$.

Guided by the above observation one might be tempted to suggest that
the Bethe Ansatz equations have {\em only one solution, considered as a
function of $\varphi$}.  Undoubtedly, that could be an elegant
resolution to the problem
of completeness, which traditionally attracts a lot of attention in the
literature (see, e.g., the recent papers \cite{FM01,Bax02}
and references therein). At the moment we cannot prove or disprove
the above assertion. The analytic structure of the eigenvalues of the
eight-vertex SOS model is quite complicated and
certainly deserves further detailed studies.

This material was planned as an introductory part for an
   extended version of our previous work \cite{BM05,BM06a} devoted to
   the connection of the 8V-model with the Painlev\'e transcendents.
However, in the
   course of writing, we realized that a review of the theory of
   the functional relations in the 8V-model
   could be of interest to a much wider audience than originally
   intended and deserves a separate publication.
   A detailed account of the results presented in \cite{BM05,BM06a} will be
   given in the second paper of this series \cite{BM06b}, which is
   totally devoted to the special $\eta=\pi/3$
   case of the eight vertex model with an odd number of sites. There we will
   consider remarkable connections of this special model with
   various differential equations, including the celebrated Lam\'e, Mathieu,
Painlev\'e III and Painlev\'e VI equations.

The organization of the paper is as follows. In Sect.~\ref{general-sect} we
present the analytic theory of the functional relations in the 8V/SOS-model.
In Sect.~\ref{wron-sect} we discuss some applications of the quantum
Wronskian relation. In particular, we show how it can be used for the
analysis of the degenerate states.
In Sect.~\ref{cont-sect} we completely classify
eigenvalues of the transfer matrix of the symmetric 8V-model for
small lattices of the size $N\le4$. We then study the analytic
properties of the eigenvalues with respect to the field variable
$\varphi$ with a combination of analytic and numerical techniques.
In Conclusion we briefly summarize obtained results.
Basic properties of the 8V-model are reviewed in the Appendix.
\nsection{Functional relations in the eight-vertex SOS model.}
\label{general-sect}
\subsection{Overview}
In this section we will outline the analytic theory of the functional
relations in the SOS-model (which also covers the symmetric 8V-model).
Actually most of the functional relations discussed below are quite
universal and apply to a wider class of related model.
They include the six-vertex model in a field \cite{Baxter:1971sam},
the restricted solid-on-solid (RSOS) model \cite{ABF84} and
some integrable models of quantum field theory: the $c<1$ conformal
field theory \cite{BLZ97a}
and the massive sine-Gordon model in a finite volume \cite{BLZ97b}.

Let $\Ts(u)$ and $\Qs(u)$ denote the eigenvalues of the transfer
matrix and the $\Q$-operator respectively and $\eta$ is an arbitrary real
parameter in the range
\beq
0<\eta<\pi/2\ .\label{eta-range}
\eeq
In all subsequent derivations we will use only one
general assumption about the properties of the eigenvalues:
\begin{quote}
\em

\noindent
We assume that $\Ts(u)$ and $\fs(u)$ are {\em entire periodic} function of the
variable $u$,
\beq
\Ts(u+\pi)=\Ts(u), \qquad \fs(u+\pi)=\fs(u)\ ,\label{per2}
\eeq
and that the function $\Qs(u)$ solving the TQ-equation,
\beq
\Ts(u)\,\Qs(u)=\fs(u-\eta)\,\Qs(u+2\eta)+\fs(u+\eta)\,\Qs(u-2\eta)\ ,
\label{TQ2}
\eeq is a also an {\em entire} (but not necessarily periodic)
function of $u$.
\end{quote}

For every particular model the above requirements are
supplemented by additional, model-specific analyticity properties of $\Qs(u)$
(such as, for example, the imaginary period relation \eqref{qper-1} for the
8V-model). These properties will only be used in
Sect.~\ref{wron-sect} and \ref{cont-sect};
they are discussed at the end of this section.

As explained in the Introduction, once the additional analyticity
properties are fixed, the functional equation \eqref{TQ2}
completely determines all eigenvalues $\Ts(u)$ and $\Qs(u)$.
For certain applications, however, it is more convenient to use other
functional equations in addition to (or instead of) \eqref{TQ2}.
We will show that all such additional functional relations in the
SOS-model (and in the related models mentioned above)
follow elementary from two ingredients:
\begin{enumerate}[(i)]
\item
the TQ-equation itself (Eqs. \eqref{per2} and \eqref{TQ2} above),
and
\item
the fact that for the same eigenvalue $\Ts(u)$
this equation has two  different \cite{BLZ97a,KLWZ97}
linearly independent solutions  for $\Qs(u)$
which are entire functions of $u$.
\end{enumerate}

The only property of the function $\fs(u)$ essentially
used in this Section is its periodicity \eqref{per2}.
For technical reasons we will also assume
that $\fs(u-\eta)$ and $\fs(u+\eta)$ do not have common
zeroes.
This is a very mild assumption, excluding rather exotic
row-inhomogeneous models, which are beyond the scope of this paper.

\subsection{General functional relations.}\label{general-rel}
Since $\Ts(u)$ is an entire function, Eq.\eqref{TQ2}  implies that
the zeroes
$u_1,u_2,\ldots,u_n$ of any eigenvalue $\Qs(u)$
satisfy the same set of the Bethe Ansatz equations
\beq
\frac{\fs(u_k+\eta)}{\fs(u_k-\eta)}=-\frac{\Qs(u_k+2\eta)}{\Qs(u_k-2\eta)},
\qquad    \Qs(u_k)=0,\qquad k=1,\ldots,n \ ,
\label{BA2}\eeq
where the number of zeroes, $n$, \  is determined by the model-specific
analyticity properties.

For any given eigenvalue $\Ts(u)$ introduce an infinite set of
functions $\Ts_k(u)$, $k=3,4,\ldots\infty$, defined by the recurrence
relation
\beq
\Ts_k(u+\eta)\,\Ts_k(u-\eta)
=\fs({u+k\,\eta})\,\fs({u-k\,\eta})
+\Ts_{k-1}(u)\,\Ts_{k+1}(u),\qquad k\ge2\ , \label{fus1}
\eeq
where
\beq
\Ts_0(u)\equiv 0,\qquad \Ts_1(u)\equiv\fs(u), \qquad
\Ts_2(u)\equiv\Ts(u)\ .\label{init}
\eeq
This relation can be equivalently rewritten as
\begin{subequations}\label{fus23}
\beq
\Ts(u)\,\Ts_k(u+k\eta)=\fs(u-\eta)\,\Ts_{k-1}(u+(k+1)\eta)+
\fs(u+\eta)\,\Ts_{k+1}(u+(k-1)\eta)\ ,\label{fus2}
\eeq
or as
\beq
\Ts(u)\,\Ts_k(u-k\eta)=\fs(u+\eta)\,\Ts_{k-1}(u-(k+1)\eta)+
\fs(u-\eta)\,\Ts_{k+1}(u-(k-1)\eta)\ .\label{fus3}
\eeq
\end{subequations}
Using the definition \eqref{fus1} one can easily express $\Ts_k(u)$ in terms of
$\Ts(u)$ as a determinant
\beq
\Ts_k(u)=\big(\fs^{(k)}(u)\big)^{-1}\
\det\Big\|\,\Ms_{ab}(u+k\eta)\,\Big\|_{1\le a,b \le k-1},
\qquad k\ge2, \label{deter}
\eeq
where the $(k-1)$ by $(k-1)$ matrix $\Ms(u)_{ab}$, \ ${1\le a,b \le
  k-1}$, is given by
\beq
\Ms_{ab}(u)=\delta_{a,b}\,
\Ts(u-2a\eta)-\delta_{a,b+1}\,\fs(u-(2a+1)\eta)
-\delta_{a+1,b}\,\fs(u-(2a-1)\eta)\ ,\label{M-def}
\eeq
while the normalization factor reads
\beq
\fs^{(k)}(u)=\prod_{\ell=0}^{k-3}\ \fs(u-(k-3-2\ell)\eta)\ .\label{fk}
\eeq
Finally, expressing $\Ts(u)$ from \eqref{TQ2} through the
corresponding eigenvalue $\Qs(u)$ one arrives to the formula
\beq
\Ts_k(u)=\Qs(u-k\eta)\,\Qs(u+k\eta)\,\sum_{\ell=0}^{k-1}
\frac{\fs({u+(2\ell-k+1)\eta})}
{\Qs({u+(2\ell-k)\eta})\,\Qs({u+(2\ell-k+2)\eta})}\ ,\label{tkq}
\eeq
valid for $k=1,2,\ldots,\infty$.
Note, that the Bethe Ansatz equations \eqref{BA2} guarantee that all
the higher $\Ts_k(u)$ with $k\ge3$ are entire functions of $u$ as well
as $\Ts(u)$.
It is worth noting that these functions are actually
eigenvalues of the ``higher'' transfer matrices, obtained through the
algebraic fusion procedure \cite{KRS81}. In our analytic approach
this information is, of course, lost. Nevertheless it will be useful to
have in mind that the index $k$ in the notation $\Ts_k(u)$
refers to the dimension of the
``auxiliary'' space in the definition of the corresponding transfer matrix.
Another convenient scheme of notation for higher transfer matrices
(used, e.g., in \cite{BLZ99a}) is based on
(half-)integer spin labels $j$, such that $k=2j+1$.

In a generic case Eq.\eqref{TQ2} has
two linear independent ``Bloch wave'' solutions $\Qs_\pm(u)$,
defined by their quasi-periodicity properties,
\beq
\Qs_\pm(u+\pi)=e^{\pm i\varphi}\Qs_\pm(u),\label{Bloch}
\eeq
where the exponent $\varphi$ depends on the eigenvalue $\Ts(u)$. These
solutions satisfy the quantum Wronskian relation
\beq
2i\,\Ws(\varphi)\,{\fs(u)}={\Qs_+(u+\eta)\,\Qs_-(u-\eta)
-\Qs_+(u-\eta)\,\Qs_-(u+\eta)}\ ,\label{wrons1}
\eeq
where $\Ws(\varphi)$ does not depend on $u$.
Indeed, equating the two
alternative expressions for $\Ts(u)$,
\beq
\Ts(u)\,\Qs_+(u)=\fs(u-\eta)\,\Qs_+(u+2\eta)+\fs(u+\eta)\,\Qs_+(u-2\eta)\ ,
\label{TQp}
\eeq
and
\beq
\Ts(u)\,\Qs_-(u)=\fs(u-\eta)\,\Qs_-(u+2\eta)+\fs(u+\eta)\,\Qs_-(u-2\eta)\ ,
\label{TQm}
\eeq
and writing $\Ws(\varphi)$  as $\Ws(\varphi|u)$  (to assume its
possible $u$-dependence,
which cannot be ruled out just from the definition \eqref{wrons1}) one gets
$\Ws(\varphi|u+\eta)=\Ws(\varphi|u-\eta)$.
On the other hand,
Eqs.\eqref{wrons1}, \eqref{per2} and \eqref{Bloch}
imply a different periodicity relation
$\Ws(\varphi|v+\pi)=\Ws(\varphi|v)$
. For generic real $\eta$, these two periodicity
relations can only be compatible
if $\Ws(\varphi|u)$ is independent of $u$,
\beq
\Ws(\varphi|u )\equiv\Ws(\varphi)\ .\label{wcons}
\eeq
When $\eta$ and $\varphi$ are in general position,
the eigenvalues $\Qs_\pm(u)$ are locally analytic functions of $\eta$,
therefore, by continuity, Eq.\eqref{wcons} at generic $\varphi$ holds also
when $\eta/\pi$ is a rational number. However, when $\varphi$ takes special
values (for example, in the symmetric 8-vertex model) Eq.\eqref{wcons}
for rational $\eta/\pi$ cannot be established by the analytic arguments only.

Obviously, the condition \eqref{Bloch} defines $\Qs_\pm(u)$ up to arbitrary
$u$-independent normalization factors. Using this freedom, it is
convenient to assume the normalization\footnote{%
In the context of the 8V/SOS-model this is the most natural normalization.
The eigenvalues $\Qs_\pm(u)$ are factorized in products of theta functions
(see \eqref{wei} below)
and the variation of $\varphi$ only affects
positions of zeroes. Obviously, the transfer matrix
eigenvalues, $\Ts(u)$, do not
have any singularities in $\varphi$.}
 such that neither of $\Qs_\pm(u)$
vanishes identically (as a function of $u$) or diverges at any
value of $\varphi$.
Then the quantum Wronskian $\Ws(\varphi)$
will take finite values, but  still can
vanish at certain isolated values of the exponent $\varphi$.
These values are called {\em singular} in the sense that
there is only one quasi-periodic solution \eqref{Bloch},
while the second linear
independent solution of \eqref{TQ2}
does not possess the simple periodicity
properties \eqref{Bloch}.
As argued in \cite{BLZ97a}, the singular exponents take
values in the ``dangerous'' set
\beq
{\varphi}_{dang}
=k{\pi}+\frac{\pi^2}{2\eta}\,\ell\ ,\qquad
k,\ell \in{\mathbb Z}\ .\label{dang}
\eeq
However, each eigenvalue has its own set of singular exponents, being
a subset of \eqref{dang}.

Evidently, $\Qs(u)$ in \eqref{tkq} can be substituted by any of the
two Bloch solutions $\Qs_\pm(u)$, so there are two
alternative expression for each $\Ts_k(u)$.
Further, multiplying \eqref{TQp} and \eqref{TQm} by $\Qs_-(u-2\eta)$ and
$\Qs_+(u-2\eta)$ respectively, subtracting resulting equations
and using \eqref{wrons1} one obtains
\beq
2i\,\Ws(\varphi)\,\Ts(u)=
\Qs_+(u+2\eta)\,\Qs_-(u-2\eta)
-\Qs_+(u-2\eta)\,\Qs_-(u+2\eta)\ .\label{tqpm}
\eeq
The last result, combined with the determinant formula
\eqref{deter}, gives
\beq
2i\,\Ws(\varphi)\,\Ts_k(u)=
\Qs_+(u+k\eta)\,\Qs_-(u-k\eta)
-\Qs_+(u-k\eta)\,\Qs_-(u+k\eta), \label{tkqpm}
\eeq
where $k=0,1,2,\dots,\infty$.

All the functional relations presented above
are general corollaries of the
TQ-equation \eqref{per2}, \eqref{TQ2}.

\subsection{Rational values of $\eta$}
\label{rational-sect}
Let us now assume that
\beq
2L\eta =m\pi, \qquad 1\le m\le L-1, \qquad L\ge 2,\label{rat-eta}
\eeq
where $m$ and $L$ are mutually prime integers. Evidently,
\beq
2 k\eta\not=0\pmod{\pi},\qquad 1\le k \le L-1\ .
\eeq
Combining the expression \eqref{tkq} with \eqref{per2} and
\eqref{Bloch}, one immediately  obtains the following functional
relation,
\beq
\Ts_{L+k}(u)=2\cos(m\,\varphi)\,\Ts_k(u+\hf m\pi)+\Ts_{L-k}(u), \qquad
k=1,2,\ldots\ .\label{simk}
\eeq
which shows that, for the rational $\eta$ of the form \eqref{rat-eta},
all higher $\Ts_k(u)$ with $k\ge L$
are the linear combinations of a finite number of the lower
$\Ts_k(u)$ with  $k\le L$.
This relation is a simple corollary of the
  TQ-equation. It always holds for the rational values
of $\eta$ and does not require the existence of the second Bloch
  solution in \eqref{Bloch} (indeed, Eq.\eqref{simk}
is independent of the sign of  $\varphi$).
Setting $k=1$ in \eqref{simk} one obtains
\beq
\Ts_{L+1}(u)=2\cos(m\,\varphi)\,\fs(u+\hf m\pi)+\Ts_{L-1}(u)\ .\label{sim1}
\eeq
This allows one to bring Eq.\eqref{fus1} with $k=L$ to the form
\beq
\Ts_L(u+\eta)\,\Ts_L(u-\eta)=
\Big(\fs(u+\hf m\pi)+e^{im\varphi}\,\Ts_{L-1}(u)\Big)
\Big(\fs(u+\hf m\pi)+e^{-im\varphi}\,\Ts_{L-1}(u)\Big) \label{fusL}
\eeq
where the periodicity \eqref{per2} of the function $\fs(u)$ was taken
into account. Thus, for the rational $\eta$, the
equations \eqref{fus1} with $k=2,3,\ldots,L-1$ together with
Eq.\eqref{fusL}
form a closed system of functional equations for a set of $L-1$ eigenvalues
$\{\Ts(u),\Ts_3(u),\ldots,\Ts_L(u)\}$. Given that all $\Ts_k(u)$ with
$k\ge3$ are
recursively defined through $\Ts(u)$, this system of equation leads
to a single equation involving $\Ts(u)$ only. Indeed, substituting the
determinant formulae \eqref{deter} into \eqref{sim1} one obtains
\beq
\det\Big\|\,\overline{\Ms}_{ab}(u)\,\Big\|_{1\le a,b \le L}=0\ ,\label{det2}
\eeq
where the $L$ by $L$ matrix reads
\beq
\overline{\Ms}_{ab}(u)=\Ms_{ab}(u)
-\omega\,\delta_{a,1}\,\delta_{b,L}\,\fs(u-3\eta)
-\omega^{-1}\,\delta_{a,L}\,\delta_{b,1}\,\fs(u+\eta)
\eeq
with $\Ms_{ab}(u)$ given by \eqref{M-def} and $\omega=e^{\pm im\varphi}$.
\subsubsection{Non-zero quantum Wronskian}
Continuing the consideration of the rational case \eqref{rat-eta}, let
us additionally assume that both quasi-periodic
solutions \eqref{Bloch} exist and that
their quantum Wronskian \eqref{wrons1} is non-zero.
It is worth noting that the functions $\Qs_\pm(u)$ in this case
cannot contain {\em complete exact strings}.
A complete exact string (or, simply, a {\em complete string})
  is a ring of $L$ zeroes
$u_1,\ldots,u_L$, where each consecutive zero differs from the previous one
by $2\eta$, closing over the period $\pi$,
\beq
u_{k+1}=u_k+2\eta,\qquad k=1,\ldots,L, \qquad u_{L+1}= u_1\pmod{\pi}\
.\label{compstr}
\eeq
It is easy to see that any such string manifests itself as a factor
in the RHS of \eqref{wrons1}, but not in its LHS (unless, of course,
$\Ws(\varphi)=0$).

It follows from \eqref{Bloch} that
\beq
\Qs_\pm(u+m\pi)=e^{\pm im\varphi}\Qs_\pm(u)\ .\label{Bloch2}
\eeq
Using \eqref{tkqpm} and \eqref{rat-eta} one easily obtains
the two equivalent relations,
\begin{subequations}\label{rel12}
\beq
e^{+im\varphi}\,
\Ts_k(u)+\Ts_{L-k}(u+\hf m\pi)=\Cs(\varphi)\
\Qs_+(u+k\eta)\,\Qs_-(u-k\eta)\ ,\label{rel1}
\eeq
and
\beq
e^{-im\varphi}\,
\Ts_k(u)+\Ts_{L-k}(u+\hf m\pi)=\Cs(\varphi)\
\Qs_+(u-k\eta)\,\Qs_-(u+k\eta)\ ,\label{rel2}
\eeq
\end{subequations}
where $k=0,1,\ldots,L$ and
\beq
\Cs(\varphi)=\frac{\sin(m\varphi)}{\Ws(\varphi)}\ .\label{A-def}
\eeq
In particular, for $k=0$ one gets,
\beq
\Ts_{L}(u+\hf m\pi)=\Cs(\varphi)\
\Qs_+(u)\,\Qs_-(u)\ .\label{tl}
\eeq
Quote also one simple but useful\footnote{%
Namely this relation with $k=1$ was used in \cite{BLZunpub}
to show that for rational values of $\eta$
the expression for the non-linear mobility for the
quantum Brownian particle in a periodic potential obtained in \cite{BLZ97a}
exactly coincide with that of \cite{FLS95b}
found from the thermodynamic Bethe Ansatz.}
consequence of \eqref{rel12},
\beq
\log\frac{\Qs_+(u+k\eta)}{\Qs_-(u+k\eta)}-
\log\frac{\Qs_+(u-k\eta)}{\Qs_-(u-k\eta)}=
\log\left(\frac{e^{+im\varphi}\,
\Ts_k(u)+\Ts_{L-k}(u+\hf m\pi)}{e^{-im\varphi}\,
\Ts_k(u)+\Ts_{L-k}(u+\hf m\pi)}\right)
\eeq
This first-order
finite difference equation relates the ratio $\Qs_+/\Qs_-$
with the eigenvalues of the (higher) transfer matrices.

Introduce the meromorphic functions
\beq
{\mathsf \Psi}_\pm(u)=e^{\pm im\varphi}\,\sum_{\ell=0}^{L-1}
\frac{\fs({u+(2\ell+1)\eta})}
{\Qs_\pm({u+2\ell\eta})\,\Qs_\pm({u+(2\ell+2)\eta})},\label{Phi}
\eeq
such that
\beq
\Ts_L(u+\hf m\pi)=\Big(\Qs_+(u)\Big)^2 {\mathsf \Psi}_+(u)=\Big(\Qs_-(u)\Big)^2
{\mathsf \Psi}_-(u)\ .\label{tl2}
\eeq
With this definition all the relations \eqref{rel12}
reduce to a single
relation which again can be written in two equivalent forms
\beq
{\mathsf \Psi}_+(u)=\Cs(\varphi)\
\frac{\Qs_-(u)}{\Qs_+(u)},\qquad
{\mathsf \Psi}_-(u)=\Cs(\varphi)\
\frac{\Qs_+(u)}{\Qs_-(u)}\ .\label{phirel}
\eeq
Obviously,
\beq
{\mathsf \Psi}_+(u){\mathsf \Psi}_-(u)
=\Big(\Cs(\varphi)\Big)^2,\qquad
\frac{{\mathsf \Psi}_+(u)}{{\mathsf \Psi}_-(u)}
=\Big(\frac{\Qs_-(u)}{\Qs_+(u)}\Big)^2\ .
\label{obv}
\eeq

\subsubsection{The RSOS regime and its vicinity}\label{rsos-sect}

Further reduction of the functional relation in the rational case
\eqref{rat-eta} occurs for certain special values of the field from
the set
\beq
m\,\varphi=(r+1)\,\pi,\qquad r=0,1,2,\ldots \ .\label{fields}
\eeq
Consider the effect of varying $\varphi$ in the relation
\eqref{tl}. The eigenvalue $\Ts_L(u)$ in the LHS will remain finite,
so as the eigenvalues $\Qs_\pm(u)$ in the RHS. The latter also
do not vanish identically (as functions of $u$) at any value of
$\varphi$ (see the discussion of our normalization assumptions before
\eqref{dang} above). Therefore the coefficient $\Cs(\varphi)$, defined
in \eqref{A-def}, is always finite. This means that in the rational
case \eqref{rat-eta},
the quantum Wronskian, $\Ws(\varphi)$, can only vanish at zeroes of
the numerator in \eqref{A-def}.
However, the converse is not true: $\Ws(\varphi)$ does not necessarily vanish
when $\Cs(\varphi)=0$. Here we are interested in this latter case
where
\beq
\Cs(\varphi)=0,\qquad \Ws(\varphi)\not=0
\eeq
with $\varphi$ from the set \eqref{fields}. By definition we call
it the {\em RSOS regime}.
The relations \eqref{rel12} and \eqref{tl}
reduce to
\begin{subequations}\label{rsos}
\beq
\Ts_k(u)=(-1)^r\,\Ts_{L-k}(u+\hf m\pi),\qquad k=1,\ldots,L-1,\label{rsos1}
\eeq
and
\beq
\Ts_L(u)=0.\label{rsos2}
\eeq
\end{subequations}
All these relations can be written as a single relation (in two
equivalent forms involving only $\Qs_+(u)$ or $\Qs_-(u)$
respectively),
\beq
{\mathsf \Psi}_+(u)={\mathsf \Psi}_-(u)=0,\label{rsos3}
\eeq
with ${\mathsf \Psi}_\pm(u)$ defined by \eqref{Phi}.

The special
``truncation'' relations \eqref{rsos},
exactly coincide with those appearing in  the RSOS-model \cite{ABF84}.
These were obtained \cite{BP82,BR89}
by the algebraic fusion procedure \cite{DJMO86}
and hold for all eigenvalues of the RSOS model.
The above analysis shows that {\em all eigenvalues of the RSOS model
  are non-singular}.  The quantum Wronskian of the Bloch solutions
\eqref{Bloch} is always non-zero (otherwise the coefficient
$\Cs(\varphi)$ in \eqref{rel12} would not have vanished).
For this reason the solutions of the
Bethe Ansatz equations for the RSOS model  cannot contain complete
strings. Since, as argued in \cite{Bax02} the complete strings are
necessary attributes of degenerate states, one arrives to a rather
non-trivial statement: {\em the
spectrum of the transfer matrix in the RSOS model is non-degenerate}.

Consider now the vicinity of the RSOS regime, when $\eta$ and
$\varphi$ are approaching their limiting values given by
\eqref{rat-eta} and \eqref{fields} respectively.
Interestingly, one can express some $\eta$- and $\varphi$-derivatives
\beq
\partial_{\eta}\Ts_k(u)=\frac{\partial}{\partial{\eta}}\,
\Ts_k(u\,|\eta,\varphi),\qquad
\partial_{\varphi}\Ts_k(u)=\frac{\partial}{\partial{\varphi}}\,
\Ts_k(u\,|\eta,\varphi)\ ,
\eeq
calculated at the ``RSOS point'',
\beq
(\eta,\,\varphi)=(m\pi/2L,\,\pi(r+1)/m)\ ,\label{point}
\eeq
in terms of the corresponding values of $\Qs_\pm(u)$
and their first order $u$-derivatives
\beq
\Qs'_\pm(u)=\frac{\partial}{\partial{u}}\,
\Qs_\pm(u|\eta,\varphi)\ .
\eeq
Using \eqref{tkqpm} one obtains,
\begin{equation}
\begin{array}{ll}
&\partial_\eta
\Big[\Ts_k(u)-(-1)^r\,\Ts_{L-k}(u+m\pi/2)\Big]+L\,\partial_v
\Ts_k(u)=\\
&\qquad\qquad\qquad\qquad=\ds\frac{L}{i\Ws(\varphi)}\,
\Big[\Qs'_+(u+k\eta)\,\Qs_-(u-k\eta)
-\Qs_+(u-k\eta)\,\Qs'_-(u+k\eta)\Big]\ ,\\&\\
&\partial_\varphi \Big[\Ts_k(u)-(-1)^r\,\Ts_{L-k}(u+m\pi/2)\Big]=\\
&\qquad\qquad\qquad\qquad\ds
=\frac{m}{2\Ws(\varphi)}\,\Big[\Qs_+(u+k\eta)\,\Qs_-(u-k\eta)+
\Qs_+(u-k\eta)\,\Qs_-(u+k\eta)\Big]\ ,
\end{array}\label{derivat}
\eeq
where the expressions in the RHS are calculated directly at the point
\eqref{point}.
According to the definitions
\eqref{init}, $\Ts_0(u)$ and $\Ts_1(u)$ do not depend on $\eta$ and
$\varphi$ at
all, therefore, one can express $\eta$- and $\varphi$-derivatives
of $\Ts_{L-1}(u)$ and $\Ts_L(u)$ at the RSOS point \eqref{point} in terms of
the of values $\Qs_\pm(u)$ and $\Qs'_\pm(u)$.

\subsection{Zero field case}\label{sect-zero-field}
Consider now the zero field limit $\varphi=0$.
Let us return to the case of
an irrational $\eta/\pi$ where the spectrum of
the transfer matrix is non-degenerate. The eigenvalues
$\Qs_\pm(u)$, corresponding to the same eigenstate smoothly
approach the same value at $\varphi=0$.
Moreover, adjusting a $\varphi$-dependent
normalization of $\Qs_\pm(u)$ one can bring their small $\varphi$ expansion
to the form
\beq
\Qs_\pm(u)=\Qn(u)\mp\varphi\/\,\Qbar(u)/2+O(\varphi^2),
\qquad \varphi\to 0\ ,\label{phi-exp}
\eeq
where
\beq
\Qn(u)=\Qs_\pm(u)\vert_{\varphi=0},\qquad
\Qbar(u)=-2\,\frac{d\,\Qs_+(u)}{d\varphi}\Big{\vert}_{\varphi=0}=
2\,\frac{d\,\Qs_-(u)}{d\varphi}\Big{\vert}_{\varphi=0}\ .\label{q0-def}
\eeq
{}From \eqref{Bloch} it follows that
\beq
\Qn(u+\pi)=\Qn(u),\qquad \Qbar(u+\pi)=\Qbar(u)+2i\Qn(u)\ .\label{zfper}
\eeq
It it easy to see that the quasi-periodic part of $\Qbar(u)$ is totally
determined by $\Qn(u)$,
\beq
\Qbar(u)=\frac{2 i u}{\pi}\, \Qn(u) + \Qbar^{(per)}(u)\ .\label{qbdecom}
\eeq
However, the periodic part
\beq
\Qbar^{(per)}(u+\pi)=\Qbar^{(per)}(u),
\eeq
can only be determined up to an additive term proportional to $\Qn(u)$.
Indeed, consider the effect of
an inessential normalization transformation
\beq
\Qs_\pm(u)\to e^{\pm\alpha\varphi}\Qs_\pm(u)\ ,
\eeq
where $\alpha$ is a constant. The value of $\Qs_0(u)$ remains unchanged
while the periodic part of $\Qbar(u)$  transforms
as
\beq
\Qbar^{(per)}(u)\to\Qbar^{(per)}(u)-2 \alpha\, \Qn(u)\ .\label{alpha-trans}
\eeq

The quantum Wronskian  relation \eqref{wrons1} reduces to
\beq
\Qn(u+\eta)\,\Qbar(u-\eta)-\Qn(u-\eta)\,\Qbar(u+\eta)=2i\,{\dot{\Ws}(0)}
\,\fs(u)\ ,
\label{zfwron}
\eeq
where
\beq
{\Ws}'(0)=\frac{d\,\Ws(\varphi)}{d\varphi}\Big|_{\varphi=0}\ .\label{wprime}
\eeq
The expression \eqref{tkqpm} now becomes
\beq
2i{\Ws}'(0) \Ts_k(u)=\Qn(u+k\eta)\,\Qbar(u-k\eta)-\Qn(u-k\eta)
\,\Qbar(u+k\eta)\ .
\eeq
It is easy to see that  at $\varphi=0$ the TQ-equation \eqref{TQ2} is satisfied
if $\Qs(u)$ there is replaced by either of $\Qn(u)$ or $\Qbar(u)$.
The same remark
applies to the more general equation \eqref{tkq}.

The Bethe Ansatz equations \eqref{BA2} for the zeroes of
$\Qn(u)$ are the standard
equations \cite{Bax72} arising in the analysis of the symmetric 8V-model.
Exactly the same equations also hold
for the zeroes of $\Qbar(u)$, but their
usefulness is very limited. Even though $\Qbar(u)$ is an entire function of
$u$, it lacks the simple periodicity (cf. \eqref{zfper}) and, therefore,
does not possess any convenient product representation. Moreover, the
transformation \eqref{alpha-trans} affects the position of zeros of
$\Qbar(u)$, making them ambiguous. All this renders the Bethe Ansatz equations
for $\Qbar(u)$ useless. Fortunately, these equations are not really required
for determination of $\Qbar(u)$. Once the zeros of $\Qn(u)$ are known
the function $\Qbar(u)$ is explicitly  calculated
from \eqref{zfwron} (see \eqref{qbar} below).

Additional functional relations arise in the
rational case \eqref{rat-eta}.
These relations are straightforward corollaries of \eqref{rel12},
\eqref{tl} and \eqref{phirel}. For instance, Eq.\eqref{rel12} gives
\beq
\Ts_k(u)+\Ts_{L-k}(u+m\pi/2)=\Cs(0)\
\Qn(u+k\eta)\,\Qn(u-k\eta),\qquad \varphi=0\ ,\label{zfrel1}
\eeq
where
\beq
\Cs(0)=m/\Ws'(0)\ .
\eeq
All these relations (with different $k$)
can  be equivalently
re-written as a single relation
\beq
\sum_{\ell=0}^{L-1}
\frac{\fs({u+(2\ell+1)\eta})}
{\Qn({u+2\ell\eta})\,\Qn({u+(2\ell+2)\eta})}=\Cs(0),\qquad \varphi=0\ ,
\label{Phi0}
\eeq
which is the $\varphi=0$ version of \eqref{phirel}.
Setting $k=0$ in \eqref{zfrel1} one gets
\beq
\Ts_L(u+m\pi/2)=\Cs(0)\,(\Qn(u))^2,\qquad \varphi=0\ .\label{zftl}
\eeq
Thus, at $\varphi=0$ the eigenvalue $\Ts_L(u+m\pi/2)$
becomes a perfect square. It
only has {\em double} zeroes, whose positions
coincide with the zeroes of $\Qn(u)$.

As is well known, in the rational case \eqref{rat-eta} the
transfer matrix of the 8V-model has a degenerate
spectrum (for sufficiently large values of $N\ge 2L$).
We would like to stress here that the above relations
\eqref{zfrel1}--\eqref{zftl} hold only for {\em non-degenerate} states.
Actually, the assumption made in the beginning of this subsection,
that $\Qs_+(u)$ coincides with $\Qs_-(u)$ when $\varphi=0$, is true only
for non-degenerate states. Removing this assumption and taking $\varphi\to0$
limit in \eqref{tl}, while keeping $\eta$ fixed by \eqref{rat-eta}, one obtains
\beq
\Ts_L(u+m\pi/2)=\Cs(0)\, \Qo_+(u) \Qo_-(u),\qquad \Qo_\pm=\lim_{\varphi\to0}
\Qs_\pm(u)\ .\label{zftl1}
\eeq
For a degenerate state the eigenvalues
$\Qo_+(u)$ and $\Qo_-(u)$ can only differ by positions of
{\em complete exact strings}.
This ambiguity does not affect any transfer matrix eigenvalues
$\Ts_k(u)$, since the complete strings trivially
cancel out from \eqref{tkq}.
In principle, the complete strings can take arbitrary positions,
however, for $\Qo_\pm(u)$ they take rather distinguished positions. Indeed,
due to \eqref{zftl1}, the zeroes of $\Qo_\pm(u)$ manifest themselves as
zeroes of $\Ts_L(u+m\pi/2)$ which are uniquely defined even for the
degenerate states.
{}From the above discussion it is clear  that
$\Ts_L(u)$ has either double zeroes or complete strings of zeroes.
Further analysis of the degenerate case
is contained in Sect.\ref{sing-sect}.

\subsection{Particular models}

So far our considerations were rather general and covered several
related models at the same time.
For each particular model, one needs to specify
additional properties, namely,
(i) the explicit form of the function $\fs(u)$ and
(ii) detailed analytic properties of the eigenvalues $\Qs_\pm(u)$.
In this Section we will do this for three different models:
the 8V/SOS-model, the 6V-model and the $c<1$ conformal field theory.

\subsubsection{\bf The symmetric eight-vertex model}\label{e8-model}
The basic properties of the 8V-model are briefly reviewed
in the Appendix~\ref{appA}. Readers who are not well familiar with the subject
will benefit from reading this Appendix prior to the rest of the paper.
Our notations are slightly
different from those in Baxter's original papers
\cite{Bax72,Bax73a,Bax73b,Bax73c}.
The variables $q$, $\eta$, $v$ and
$\rho$ used therein (hereafter denoted as
$\q_B$, $\eta_B$, $v_B$ and $\rho_B$) are related to our variables
$\q$, $\eta$, $v$ and $\rho$ as
\beq
\q^2=\q_B=e^{-{\pi\K_B'}/{\K_B}},\qquad \eta=\frac{\pi \eta_B}{2\K_B},
\qquad  v=\frac{\pi v_B}{2\K_B},\qquad \rho=\rho_B\ ,\label{baxvar0}
\eeq
where $\K_B$ and $\K'_B$ are the complete elliptic integrals associated to the
nome $\q_B$.
Here we the fix the normalization of the Boltzmann
weights as
\beq
\rho=2\,\ \vt_2(0\,|\,\q)^{-1}\ \vt_4(0\,|\,\q^2)^{-1}\ ,
\eeq
where
\beq
\vt_i(u\,|\,\q),\>\> i=1,\ldots,4,\quad  \q=e^{{\rm i} \pi \tau},
\qquad \rm{Im}\,\tau>0,
\eeq
are the standard theta functions \cite{WW}
with the periods $\pi$ and $\pi \tau$.

We denote the transfer matrix $\T$ and the $\Q$-matrix
from \cite{Bax72,Bax73a} as $\T^B(v)$ and $\Q^B(v)$,
remembering that our variable $v$ is related to $v_B$ by \eqref{baxvar0}.
Below we often use a shifted spectral parameter
\beq
u=v-\pi\tau/2\ ,\label{uv-rel0}
\eeq
simply connected to the variable $v$ in \eqref{baxvar0}.
We also consider the re-defined matrices
\beq
\T(u)=(-i\,\q^{-1/4})^N\, e^{i v N} \, \T^B(v),\qquad
\Q(u)= e^{i v N/2} \Q^B(v)\label{tq-redef}
\eeq
where $N$ is the number of columns of the lattice.
The eigenvalues $\Ts(u)$ and $\Qs(u)$ of these new matrices
enjoy the following
periodicity properties
\beq
\Ts(u+\pi)=\Ts(u),\qquad \Ts(u+\pi\tau)=(-\q)^{-N}\,e^{-2iuN}\,
\Ts(u)\ ,\label{tper3}
\eeq
and
\beq
\qquad \mbox{\sl 8V-model:}\qquad\Qs(u+\pi)=s\, e^{i\pi N/2} \,\Qs(u),\qquad
\Qs(u+2\pi\tau)=\q^{-2N}\, e^{-2iNu} \,\Qs(u)
\ . \label{qper3}
\eeq
Here the ``quantum number''\  $s=\pm1$,\
is the eigenvalue of the operator $\Scal$, defined
in \eqref{rs-def}.  This operator always commutes with $\T(u)$ and $\Q(u)$.

Baxter's  TQ-equation (Eq.(4.2) of \cite{Bax72} and Eq.(87)
of \cite{Bax73a}) now
takes the form \eqref{TQ2} with
\beq
\fs(u)=\big(\vartheta_4(u\,|\,\q)\big)^N\ .\label{phi-def1}
\eeq
The main reason for the above redefinitions is to bring the
TQ-equation to the universal form \eqref{TQ2},
where $\Ts(u)$ and $\fs(u)$ are
periodic functions of $u$ (see Eq.\eqref{per2}) for an arbitrary, odd or
even, number of sites, $N$. This also helps to facilitate the considerations
of the scaling limit in our next paper \cite{BM06b}.

Comparing the first equation in \eqref{qper3} with the
periodicity of the Bloch solutions
\eqref{Bloch} one concludes that the exponents $\varphi$ read
\beq
\varphi^{(8V)}=
\left\{\begin{array}{ll}
0\pmod{\pi}, &N={\rm even}\strut
\\\\
\ds\frac{\pi}{2}\pmod{\pi},& N={\rm odd}
\end{array}\right.\label{fi8v}
\eeq
Thus, for an even $N$ the exponents of the symmetric 8V-model,
with the cyclic boundary conditions, always belong to the
``dangerous'' set \eqref{dang}.
For an odd $N$ the exponents \eqref{fi8v} fall into this set
only for certain rational values of $\eta/\pi$.
A notable example is the case $\eta=\pi/3$,
considered in \cite{BM05,BM06a,BM06b}.

The imaginary period relations in \eqref{tper3} and \eqref{qper3}
certainly deserve a detailed consideration.
First, note that in \eqref{qper3}
  we only stated the periodicity with respect to the {\em
  double} imaginary period $2\pi\tau$, which always holds in all cases
  when the 8V-model has been exactly solved%
\footnote{Ref.\cite{Bax72}
  applies to rational $\eta$ and arbitrary values of $N$,
  while ref.\cite{Bax73a} applies to arbitrary $\eta$ and even values
  $N$. It is reasonable to assume that \eqref{qper3} holds in
  general, however, the case of an arbitrary $\eta$ and an odd $N$ has
  never been considered.
}. Actually, this is a rather overcautious statement which can be
easily specialized further.
For the following discussion assume a generic (i.e., irrational)
value of $\eta/\pi$.
Then for even $N$ the Bloch solutions \eqref{Bloch}
always coincide
(just as in the zero-field case of Sect.\ref{sect-zero-field}).
For odd $N$ there are always two linearly independent Bloch
solutions for each eigenvalue $\Ts(u)$, one with $s=+1$ and one
with $s=-1$ (remind that in this case each eigenvalue of the transfer
matrix is double-degenerate \cite{FM05}).
The existence of the ``imaginary''
period imposes rather non-trivial restrictions
on the properties of the eigenvalues. Indeed,
the second relation in \eqref{tper3} immediately
implies that the function
\beq
\widetilde{\Qs}(u)=r\,\q^{N/2}\,e^{i u N} \Qs(u+\pi\tau)\label{newsol}
\eeq
where $r$ is a constant,
satisfies the TQ-equation \eqref{TQ2} as well as  $\Qs(u)$.
Further, if  $\Qs(u)$ is a Bloch solution
\beq
 \Qs(u+\pi)=e^{i\varphi}\,  \Qs(u)
\eeq
with some $\varphi$ then  $\widetilde{\Qs}(u)$
is also such a solution with the exponent
\beq
\widetilde{\varphi}={\varphi}+N\pi \pmod{2\pi}.
\eeq
Obviously, there are two options, either
$\widetilde{\Qs}(u)$ is proportional to ${\Qs}(u)$
or it is proportional to the other linearly independent Bloch
solution with the negated exponent ``$-\varphi$''.
The first option is realized for even $N$,
\beq
\qquad \mbox{\sl 8V-model, $N$ even:}\qquad\qquad
\Qs(u+\pi\tau)=r\,\q^{-N/2}\,e^{-i u N} \Qs(u),
\phantom{\qquad \mbox{\sl 8V-model, $N$ even:}}
\label{Neven}
\eeq
The constant $r=\pm1$ is then the
eigenvalue of the spin-reversal operator $\Rcal$
(cf. \eqref{qpera}).
The second option requires the exponent $\varphi$ to be
a half-an-odd integer fraction of $\pi$, it  is realized for odd $N$,
\beq
\qquad \mbox{\sl 8V-model, $N$ odd:}\qquad
\qquad\Qs_\pm(u+\pi\tau)=\q^{-N/2}\,e^{-i u N}
\Qs_\mp(u).\phantom{\qquad \mbox{\sl 8V-model, $N$ even:}}
\label{Nodd}
\eeq

The above relations \eqref{Neven} and \eqref{Nodd} were derived
for irrational values of $\eta/\pi$, however they also hold in the rational
case \eqref{rat-eta}, if no additional degeneracy of the eigenvalues
of the transfer matrix occurs (apart from the one related with the
spin-reversal symmetry for odd $N$).
The functional relation \eqref{tl} can be then written in the form
\beq
\Ts_L(u+\hf m\pi)=\As\, e^{iN u}\, \Qs_+(u)\, \Qs_+(u+\pi\tau)\label{FM}
\eeq
where $\As$ is a constant.  This relation
is identical to the one conjectured in \cite{McCoy1}\footnote{%
The conjecture of \cite{McCoy1}
also covers a special case of degenerate states for rational values of
$\eta$, where the relation \eqref{Nodd} holds for the eigenvalues
of the $\Q$-matrix of \cite{Bax72} for even $N$.
}.

\subsubsection{The solid-on-solid model}
The main idea of this paper is to study deformations of the
eigenvalues $\Ts(u)$ and $\Qs(u)$ under continuous variations
of the exponents
$\varphi$ from their discrete values \eqref{fi8v}.
As explained in the
Introduction the resulting eigenvalues correspond to the unrestricted
SOS-model.
We will therefore assume the more general periodicity
relations \eqref{qper2} for the Bloch wave solutions $\Qs_\pm(u)$,
which hold for both odd and even $N$,
\beq\quad\mbox{\sl SOS-model:}\qquad
\Qs_\pm(u+\pi)=e^{\pm i\varphi}\, \Qs_\pm(u),\qquad
\Qs_\pm(u+2\pi\tau)=\q^{-2N}\,e^{\pm \psi}\,e^{-2iu N}\,
\Qs_\pm(u),
\label{qper4}
\eeq
where the exponent $\varphi$ is arbitrary. The second exponent $\psi$
is not an independent parameter, it is determined by $\varphi$
(see the discussion in Section~\ref{wron-8v-sect} below).

The second relation in
\eqref{qper4} can be further refined for even $N$
\beq\quad\mbox{\sl SOS-model, N even:}\qquad\qquad
\Qs_\pm(u+\pi\tau)=\q^{-N/2}\,e^{\pm \psi/2}\,e^{-iu N}\,
\Qs_\pm(u),\phantom{
\Qs_\pm(u)=e^{\pm i\varphi}\,}
\label{qper5}
\eeq
whereas the periodicity of $\Ts(u)$ remains the same \eqref{tper3} as in
the 8V-model.
However, there is no a general SOS-model analog of \eqref{Nodd},
as it is specific to half-odd exponents only. As a result Eq.\eqref{tper3}
is replaced with
\beq
\quad\mbox{\sl SOS-model, N odd:}\qquad
\Ts(u+\pi)=\Ts(u),\qquad \Ts(u+2\pi\tau)=\q^{-4N}\,e^{-4iuN}\,
\Ts(u)\ .\label{tper5}
\eeq
Strictly speaking the use of the term ``SOS-model'' here
is justified for even $N$ only \cite{Bax73b}.
Nonetheless, we will use this term to indicate arbitrary
values of the field parameter $\varphi$ in general.

\subsubsection{Six-vertex model in a horizontal field}\label{6v-model}
The allowed vertex configurations of the six-vertex model form a subset
of those shown in Fig.\ref{8-vertices}. Namely, the Boltzmann weights
$\o_7$ and $\o_8$ are equal to zero. The remaining six weights will be
parameterized as
\beq
\renewcommand\arraystretch{1.5}
\begin{array}{rclrclrcl}
\o_1&=&e^{+H-i\eta}\,a,&
\o_2&=&e^{-H-i\eta}\,a,& \o_3&=&e^{+H+i\eta}\,b,\\
\o_4&=&e^{-H+i\eta}\,b,&
\o_5&=&e^{iu-2i\eta}\,c,&\o_6&=&e^{iu-2i\eta}\,c,
\end{array}
\renewcommand\arraystretch{1.0}
\eeq
where $H$ stands for the horizontal field
\beq
a=\hs(u+\eta),\qquad b=\hs(u-\eta),\qquad
c=\hs(2\eta),\qquad\hs(u)=1-e^{2iu}\ .\label{h6-def}
\eeq
The above parametrization is simply related to that given in
Eq.(12) of \cite{Bax02}  (where the vertical field $V$ is set to zero).
The TQ-equation (eq.(11) of \cite{Bax02}) takes the form
\eqref{per2}, \eqref{TQ2}, where
\beq
\fs(u)=\left(\hs(u)\right)^N\ .
\eeq
The Bloch solutions \eqref{Bloch},
corresponding to the eigenvectors of the transfer matrix with
$n$ ``up-spins'',  can be written as
\beq
\Qs_\pm(u)=e^{\pm i \varphi u /\pi}\,\As_\pm(e^{2iu})\ ,\label{A6-def}
\eeq
where $\As_+(x)$ and $\As_-(x)$ are polynomials in $x$ of the degrees $n$
and $(N-n)$, respectively, and
\beq
\varphi=\frac{i \pi H N}{2\eta} +\frac{\pi}{2}\,(N-2 n)\ .
\eeq
Introduce new variables\footnote{The parameter $q$ should not be
  confused with the nome $\q$ in the 8V-model}
\beq
x=e^{2iu},\qquad q=e^{2i\eta},\qquad  z=e^{2i\eta\varphi/\pi}\ .\label{6v-var}
\eeq
Regarding $x$ as a new spectral parameter instead of $u$ and writing
$\Ts(u)$ and  $\fs(u)$ as $\Ts(x)$ and $\fs(x)$,
respectively,  one can rewrite \eqref{TQ2} in the form
\beq
\Ts(x)\,\As_\pm(x)=z^{\pm 1}\,\fs(q^{-1}\,x)\,\As_\pm(q^2\,x)
+z^{\mp 1}\,\fs(q\,x)\,\As_\pm(q^{-2}\,x)\ ,
\eeq
where the polynomials $\As_\pm(x)$ are defined in \eqref{A6-def}.
This form is particularly convenient for the 6V-model.
\subsubsection{Conformal field theory}
The continuous quantum field theory version of Baxter's commuting transfer
matrices of the lattice theory was developed in \cite{BLZ96,BLZ97a,BLZ99a}.
These papers were devoted to the
$c<1$ conformal field theory (CFT).
The parameters $\beta$ and $p$ used there define the central charge $c$ and
the Virasoro highest weight $\Delta$,
\beq
c=1-6(\beta-\beta^{-1})^2, \qquad \Delta =\left(\frac{p}{\beta}\right)^2
+\frac{c-1}{24}\ .
\eeq
They are related to our $\eta$ and $\varphi$ as
\beq
2\eta=\beta^2\pi,\qquad \varphi={2\pi p}/{\beta^2}\ .
\eeq
The multiplicative spectral parameter $\lambda$ used in those papers
is related to our variable $u$ as
\beq
\lambda^2=-e^{-2iu}\ .
\eeq
The eigenvalues of the CFT $\Q$-operators
$\Qs_\pm(u)$ are entire functions of the variable $u$,
satisfying the periodicity relation
\eqref{Bloch}. Their leading asymptotics at large positive imaginary $u$ read
\beq
\log \Qs_\pm(u)=\frac{A}{\cos(\frac{\pi\eta}{\pi-2\eta})}\,
 e^{i\pi u/(\pi-2\eta)} +O(1), \qquad u\to+i\infty,\quad
|{\rm Re} u|<\pi/2\ ,\label{asymp}
\eeq
where $A$ is a known constant \cite{BLZ97a}.
Here we assumed that $\eta$ does not belong to the set
\beq
\eta=\frac{\pi}{2}\Big(1-\frac{1}{2k}\Big),
 \qquad k=1,2,\ldots,\infty\ .\label{log-case}
\eeq
At these special values of $\eta$ the theory
 contains logarithmic divergences and
the asymptotics \eqref{asymp} should be replaced with
\beq
\log \Qs_\pm(u)=2i(-1)^k \,A\, u\,
e^{2iuk/\pi}+\Cbbd\,e^{2iuk} +O(1), \qquad u\to+i\infty,\quad
|{\rm Re} u|<\pi/2\ ,\label{asymp1}
\eeq
where $\Cbbd$ is a regularization-dependent constant.
The factorization formulae read\footnote{%
Here we assumed that $0<\eta<\pi/4$. When $\pi/4<\eta<\pi/2$
the product in \eqref{wei2} should contain the standard Weierstrass
regularization factors \cite{BLZ97a}.}
\beq
\Qs_\pm(u)=e^{\pm iu\varphi/\pi}\, \As_\pm(u),\qquad
\As_\pm(u)=\prod_{k=1}^\infty
\big(\,1-e^{-2i(u-u_k^\pm)}\,\big)\ ,\label{wei2}
\eeq
where the zeroes $u_1^\pm,u_2^\pm,\ldots$ accumulate at infinity along
the straight line
\beq
u=\pi/2+iy,\qquad y\to+\infty\ .
\eeq
Finally, the function $\fs(u)$ in the case of CFT should be set to
one\footnote{%
Again, we have assumed that $\eta$ does not fall into the set \eqref{log-case},
otherwise $\fs(u)=\exp(4A\eta e^{2iuk}/\pi)$.},
\beq
\fs(u)\equiv1\ .
\eeq
With these specializations the functional relations given above
become identical to those previously obtained in \cite{BLZ96,BLZ97a,BLZ99a}.

\subsection{Related developments and bibliography}

The literature on the functional relation in solvable models is
huge; therefore it would not be practical to mention all
papers in the area. Our brief review is restricted only to a subset of
publications directly related to the eight-vertex/six-vertex models
and associated models of quantum field theory.

\subsubsection{Transfer matrix relations}

In the above presentation the entire functions $\Ts_k(u)$ with $k\ge3$
were defined by the recurrence relation \eqref{fus1}, which allows one
to express them solely in terms of $\Ts(u)$, as in \eqref{deter}.
No other additional properties of $\Ts_k(u)$ were used. However, as is
well known, these functions are eigenvalues of the higher transfer
matrices, usually associated with the so-called fusion procedure.
This algebraic procedure
provides a derivation of the functional relations for the higher
transfer matrices based on decomposition properties of
products of representations of the affine quantum groups.
Originally,  all these ``transfer matrix relations'' were obtained
essentially in this way. We would like to stress that the logic of
these developments was exactly opposite to that employed in our
review. The goal was to find new techniques, independent
of the TQ-relation, rather than to deduce everything from the latter.
The first important contribution was made by Stroganov \cite{Str79}.
He gave an algebraic derivation of the first nontrivial
relation in \eqref{fus1} (with $k=2$),
\beq
\Ts(u+\eta)\,\Ts(u-\eta)-\fs(u+2\eta)\,\fs(u-2\eta) =
O\Big((u-u_0)^N\Big)\label{strog1}
\eeq
in the vicinity of the point $u=u_0$ where the transfer matrices
$\Ts(u_0+\eta)$ and $\Ts(u_0-\eta)$
become shift operators. Remarkably, this single relation alone
contains almost all information about the eigenvalues $\Ts(u)$. To
illustrate this point consider, for instance, the 6V-model. For a
chain of the length $N$ each
eigenvalue $\Ts(u)$ is a trigonometric polynomial of the degree $N$,
determined by $N+1$ unknown coefficients. The mere fact that the LHS of
\eqref{strog1} has an $N$-th order zero immediately
gives $N$ algebraic equations
for these unknowns. Similar arguments, obviously, apply to the
8V-model. One additional equation is usually easy to find from
some elementary considerations (e.g., from the large $u$ asymptotics
in the 6V-model). Further, in the thermodynamic limit, $N\to\infty$
with $u$ kept fixed,
Eq.\eqref{strog1} becomes a closed
functional relation for the eigenvalues (its RHS vanishes).
This is the famous ``inversion relation'' \cite{Str79,Zam79,Bax82inv}.
With additional analyticity assumptions it can be effectively
used to calculate the eigenvalues of the transfer matrix at $N=\infty$.
Recently, Eq.\eqref{strog1} was used to derive a new non-linear
integral equation \cite{Tak00}, especially suited for the analysis of
high-temperature properties of lattice models.

Soon after \cite{Str79} Stroganov derived \cite{Str79a}
a particular case of
\eqref{sim1} for the 6-vertex model with $\eta=\pi/6$ (i.e., for
$L=3$ and $m=1$). He also found that
for the case of an odd number of sites\footnote{%
In our notations this corresponds to $\varphi=\pi/2\pmod{\pi}$.}
Eq.\eqref{det2} takes the form
\beq
\Ts(u-2\eta)\,\Ts(u)\,\Ts(u+2\eta)=
\fs(u)\fs(u+2\eta)\, \Ts(u-2\eta)+
\fs(u-2\eta)\fs(u+2\eta)\, \Ts(u)+
\fs(u-2\eta)\fs(u)\, \Ts(u+2\eta)\ .
\eeq
He then used this equation to obtain Bethe Ansatz type equations
for the zeroes of $\Ts(u)$ and to reproduce Lieb's celebrated result
\cite{Lieb67} for the residual entropy of the two-dimensional ice.
Unfortunately, these results were left unpublished.

The ideas of \cite{Str79,Str79a} were further developed in
the {\em analytic Bethe Ansatz} \cite{Res83} where
the TQ-equation (or an analogous equation) is
used essentially as a formal substitution to solve
the transfer matrix functional equations.
The notion of ``higher'' or ``fused'' R-matrices was developed
in \cite{KRS81} from the point of view of representation theory.
These R-matrices
were calculated in \cite{KR87a} for the 6V-model,
in \cite{Skl82,Skl83,KZ99,FM04} for the 8V-model and in \cite{DJMO86}
for the SOS-model.
The functional relations \eqref{fus23} were given in \cite{KR87a}
for the 6V-model and in \cite{BR89} for 8V/SOS-model.
The determinant identity \eqref{det2} was discussed in \cite{BR89,Bax89}.
An algebraic derivation of the truncation relations \eqref{rsos} for the
RSOS model \cite{ABF84} was given in \cite{BR89}.
A particular case of this truncation for the hard hexagon model
\cite{Bax80} was previously discovered in \cite{BP82}. An algebraic derivation
of \eqref{sim1} in the zero-field
six-vertex model is given in \cite{Nep03}.
The idea of calculation of $\varphi$- and $\eta$-derivatives \eqref{derivat}
at the RSOS point given in Sect.\ref{rsos-sect}
is borrowed from \cite{Fen99} and \cite{Str01b}.

Remarkably, the same functional equations \eqref{fus1} (along with all
their specializations in the rational case) arise in a
related, but different context of the thermodynamic Bethe Ansatz
\cite{YY69}; see \cite{TS72} for its application
to the 8V-model.  Usually this approach in lattice models is associated
with non-linear integral equations.
Here we refer to the functional form of these equations discovered in
\cite{ZamTBA}.
Further discussion of the correspondence of the functional relation
method with the thermodynamic Bethe Ansatz and its
generalizations for excited states can be found in
\cite{KP92,KNS94,FLS95c,BLZ96,BLZ97b,DT96}.

\subsubsection{Q-matrix and TQ-relations}\label{Qsect}

As noted before, a full algebraic theory of the $\Q$-matrix in the 8V-model
is not yet developed.
The idea of the construction of the $\Q$-matrix in terms of some
special transfer
matrices belongs to Baxter.
It is a key element of his original solution of the 8V-model.
Readers interested in details should familiarize
themselves with the Appendix~C of
\cite{Bax72} (along with other four appendices and, of course, the main
text of that paper, which contain a wealth of important information on
the subject). The results of \cite{Bax72} only apply for certain rational values
of $\eta$. The construction of \cite{Bax72} and the set of allowed values of $\eta$
were recently revised in \cite{McCoy1}.
A different construction for the $\Q$-matrix, which
works for an arbitrary $\eta$,  was given \cite{Bax73b}.
Some remarks on a comparison between the two $\Q$-matrices
are given in Appendix~\ref{appA}.

There are many related
solvable models connected with the R-matrix of the 8V-model but
having different ${\bf L}$-operators and different
{\em quantum spaces}.
The general structure of the functional
relations in all such models remains the same.
In particular, they all possess a TQ-relation (though it may
contain different scalar factors and
require different analytic properties of the eigenvalues).
In \cite{Bax73b} Baxter also presented an extremely simple explicit formula
for the matrix elements of the $\Q$-matrix for the zero-field
6V-model in the sector with $N/2$ ``up-spins''
(the half filling). However, no such expression  is known for the
8V-model, or the other sectors of the 6V-model.
The quantum space of the 6V-model is build from the two-dimensional
highest weight representation of $U_q(sl_2))$ at every site of the lattice.
Curiously enough, if this representation
is replaced with the general cyclic representation
(arising at roots of unity, $q^L=1$) then all matrix elements of the
$\Q$-matrix can be explicitly calculated \cite{BS90}
as a simple product involving
only a two-spin interaction\footnote{%
The factorization of the
matrix elements of the $\Q$-matrix is typical for quantum space
representations without highest and lowest weights.
}. Remarkably,
the resulting $\Q$-matrix exactly coincides with the transfer matrix of
the chiral Potts model \cite{VG85,AuY87,BPA88};
this allows one to view the latter
as a ``descendant of the six-vertex model'' \cite{BS90}.
The generalization of this construction to the eight-vertex and
the Kashiwara-Miwa model \cite{Kas86} is considered in \cite{HY90}.
Further developments of the theory of the ${\bf
Q}$-matrix and related topics (along with many important
applications to various
solvable models) can be found in \cite{
Gau91,PG92,Der99,Skl00,DFM01,Kas01,FM02Proc,Smidual,Zab00,FM06}.

Baxter's original idea of the construction of $\Q$-operators which
utilizes traces of certain monodromy matrices was extended
in \cite{BLZ97a,BLZ99a} for trigonometric models related with
the quantum affine algebra $U_q(\widehat{sl}(2))$.
It turned out that in the trigonometric case the situation is
considerably simpler than for the 8-vertex model and the ${\bf
  Q}$-operators coincide with some special transfer matrices.
The corresponding ${\bf L}$-operators are obtained as
specializations of the universal $\R$-matrix \cite{KST95} to
infinite-dimensional representations of the $q$-oscillator algebra
in the ``auxiliary space''.
Although the calculations
of \cite{BLZ97a,BLZ99a} were specific to the continuous
quantum field theory, the same procedure can readily be
applied to lattice models (see, e.g., \cite{BOU02,Kor05a,Kor05b,BJMS06}
for the corresponding results for the 6V-model).
In the case of the 6V-model with non-zero
horizontal field this construction leads to two $\Q$-matrices\footnote{%
As noted in \cite{BLZ99a}, for the ``half-filled'' sector of the
zero-field 6V-model these $\Q$-matrices reduce to Baxter's
expression \cite{Bax73b} mentioned above,
as they, of course, should.},
whose eigenvalues $\Qs_\pm$ are precisely the ``Bloch wave'' solutions of
the TQ-equation.

Note that functional relations which involves bi-linear
combinations of $\Qs_\pm$, namely \eqref{tkqpm}, \eqref{rel12} and
\eqref{tl}
are universal in the sense that they do not involve the model-specific
function $\fs(u)$. These relations were derived in
\cite{BLZ97a,BLZ99a} in the context of the conformal field theory.
Similar relations previously appeared in the chiral Potts model
\cite{AMP89,BS90,BBP90}, though the correspondence is not exact
because there is no an additive spectral parameter in that model.
Eq.\eqref{tkqpm} in the eight-vertex and
the XXX-models was found in \cite{KLWZ97} and \cite{PS99a}.
A special case \eqref{FM} of the relation \eqref{tl} involving
Baxter's original $\Q$-matrix \cite{Bax72} for the 8V-model was
conjectured in \cite{McCoy1}.
Another special (zero-field) case \eqref{zfrel1} of the same relation
\eqref{tl} in conformal field theory was conjectured in \cite{FLS95c}.

\nsection{Quantum Wronskian relation and singular eigenvalues}
\label{wron-sect}

The quantum Wronskian relation \eqref{wrons1}, naturally arising
in the above analysis of the TQ equation, is a very non-trivial
functional relation.  In this Section we show how this relation can used
for the analysis of the eigenvalues.
In particular, we consider a practical question of
the calculation of $\Qs_-(u)$ from a known $\Qs_+(u)$. Next, we
study certain singular eigenvalues in the zero-field limit.

\subsection{Solving the quantum Wronskian relation}

\subsubsection{Six-vertex model (multiplicative spectral parameter)}
Let us first consider the example of the 6V-model.
It is convenient to present the polynomials $\As_\pm(x)$, defined in
\eqref{A6-def} in a factorized form,
\beq
\As_+(x)=\rho_+\,\prod_{k=1}^n\ (1-x/x^{+}_k)\ ,\qquad
\As_-(x)=\rho_-\,\prod_{k=1}^{N-n}\ (1-x/x^{-}_k)\ ,\label{Apm}
\eeq
and rewrite the quantum Wronskian relation \eqref{wrons1} as
\beq
z\,\As_+(xq)\As_-(x/q)-z^{-1}\,\As_+(x/q)\As_-(xq)=2i\,\Ws(z)\,\fs(x)\ .
\label{wr6}
\eeq
This single
functional equation determines both unknown polynomials $\As_\pm(x)$,
as well as the function $\Ws(z)$,
except for an obvious freedom to choose arbitrary $x$-independent
factors $\rho_\pm$ without changing the normalization-independent
combination
\beq
2i\,\Ws(z)/(\rho_+\rho_-)=z -z^{-1} \ ,\label{wr-inv}
\eeq
fixed by Eq.\eqref{wr6} at $x=0$.
There will be many solution to \eqref{wr6} corresponding to different
eigenvectors of the
transfer matrix. Indeed, setting there $x=q\,x^+_k$ and
$x=q^{-1}\,x^+_k$\ ,
one obtains two relations
\beq
z\, \As_-(x_k^+)\,\As_+(x^+_k\,q^2)=2i\,\Ws(z)\,\fs(x^+_k\,q),\qquad
z^{-1}\, \As_-(x_k^+)\,\As_+(x^+_k\,q^{-2})
=-2i\,\Ws(z)\,\fs(x^+_k\,q^{-1}) \ .\label{two}
\eeq
Taking their ratio, one comes back to the Bethe
Ansatz equations for the zeroes $x^+_1,x^+_2,\ldots,x^+_n$\  of $\As_+(x)$,
\beq
\frac{\fs(x_k^+\,q\
  )}{\fs(x_k^+\,q^{-1})}=-z^2\,\frac{\As_+(x_k^+\,q^2\ )}
{\As_+(x_k^+\,q^{-2})},
\qquad k=1,2,\ldots,n\label{ba6}
\eeq
Similar equations (with $z$ replaced by $z^{-1}$) hold
for the zeroes $x^-_1,x^-_2,\ldots,x^-_{N-n}$ of $\As_-(x)$.

Given a solution of \eqref{ba6} (which defines $\As_+(x)$)
the corresponding polynomial $\As_-(x)$ is uniquely determined
by \eqref{wr6}. For example, consider the case $N=2\,n$. Comparing the
leading power of $x$ in \eqref{wr6} one gets
\beq
s\mathop{=}^{{\rm def}}(x^+_1\,x^+_2\cdots x_n^+)^{-1}=
x^-_1\,x^-_2\cdots x_n^-\ .\label{s-def}
\eeq
Let us rewrite \eqref{wr6} as a functional difference equation
\beq
z\, \Fs(x/q)-z^{-1}\,\Fs(xq)
=\frac{\fs(x)}{\As_+(xq)\As_+(x/q)}\label{difeq}
\eeq
for the rational function
\beq
\Fs(x)=\frac{1}{2i\,\Ws(z)}\ \frac{\As_-(x)}{\As_+(x)}\ .\label{F-def}
\eeq
Performing the partial fraction decomposition of the RHS of \eqref{difeq}
and sharing the poles between $\Fs(xq)$ and $\Fs(x/q)$ one obtains
\bea
\Fs(x)&=&\frac{1}{z\,(1-s^2)}\
\sum_{k=1}^n\ \frac{\fs(x^+_k\,q)}{
x^+_k\,\As_+(x^+_k\,q^2)\,\As'_+(x^+_k)}\
\left(\frac{x-s^2\,x^+_k}{x-x^+_k}\right)\nonumber \\
&=&-\frac{z}{(1-s^2)}\ \sum_{k=1}^n\ \frac{\fs(x^+_k\,q^{-1})}
{x^+_k\,\As_+(x^+_k\,q^{-2})\,
\As'_+(x^+_k)}\ \left(\frac{x-s^2\,x^+_k}{x-x^+_k}\right)\label{F-sol}
\eea
where $\As_+'(x)$ denotes the derivative of $\As_+(x)$ with respect to $x$.
The two alternative expressions given above are equivalent in virtue of the
Bethe Ansatz equations \eqref{ba6}. The formula \eqref{F-sol} can be easily
generalized to arbitrary values of $n$. We leave this as an exercise for the
reader.

The above derivation has several limitations. Evidently the field
parameter $z$ should not take the values $z=\pm 1$, where the
denominator in \eqref{F-def} vanishes. Further, the above sharing of
the poles in \eqref{difeq} is only possible if all the roots $x_k^+$
are distinct, finite and non-zero. Moreover no pair of the roots could
satisfy the condition $x_j^+/x^+_k=q^{\pm2}$. In all these cases
the expression \eqref{F-sol} becomes meaningless (typically the
summand therein diverges).

The quantum Wronskian relation can also be solved directly for the
coefficients of the polynomials \eqref{A6-def}
\beq
\As_+(x)=\sum_{k=0}^n a^+_k\, x^k,\qquad \As_-(x)=\sum_{k=0}^{N-n} a^-_k \,x^k,
\eeq
This approach lead to a system of algebraic equations which sometimes
easier to analyze since it involves only symmetric functions of the
Bethe roots, but not the roots themselves.
As an example, consider the trivial eigenvalue
\beq
\Ts(x)=z\fs(xq^{-1})+z^{-1}\fs(x q)
\eeq
corresponding to the ferromagnetic ground state (all spins down) of the
six-vertex. One solution of
\eqref{wr6} is obvious
\beq
\As_+(x)=\rho_+=1\ .\label{rhoplus}
\eeq
The finite difference equation \eqref{difeq} can be readily solved
\beq
\frac{\As_-(x)}{2i\,\Ws(z)}
=-z\sum_{k=0}^N {N\choose k}\ \frac{(-x\,q)^k}{z^2- q^{2k}}\ .
\eeq
If $z$ takes one the values $z=\pm q^k$, $k=0,\ldots,N$, \ the
polynomial $\As_-(x)$ reduces to a single power, $x^k$, and the corresponding
Bloch solutions \eqref{A6-def} become linearly dependent.
These are the singular cases
discussed in Sect.\ref{general-rel}. Indeed, choosing the normalization
\beq
\rho_-=\prod_{k=1}^N (z^2-q^{2k})
\eeq
so that $\As_-(x)$ remains finite and does not vanish identically for
all $z$ and taking into account the definition of $z$ in
\eqref{6v-var}, one can
easily see from \eqref{wr-inv} and \eqref{rhoplus}
that the singular exponents form a subset
of \eqref{dang}, as expected.

\subsubsection{Eight-vertex/SOS model}\label{wron-8v-sect}
The formula similar to \eqref{F-sol} holds for the 8-vertex/SOS-model as well.
The periodicity relations \eqref{qper4}
imply that the entire functions $\Qs_\pm(u)$
can always be factorized as
\beq
\Qs_\pm(u)=e^{i\varphi_\pm u/\pi}\,\As_\pm(u),\qquad \As_\pm(u)=
\prod_{k=1}^N \hs(u-u_k^\pm), \label{wei}
\eeq
where $u^\pm_1, u^\pm_2, \ldots,u^\pm_N$ are zeroes of $\Qs_\pm(u)$ and the
elementary factors
\beq
\hs(u)=\vartheta_1(u\,|\,\q^2)\ \label{elem}
\eeq
have the {\em double} imaginary period\footnote{%
This choice
allows a uniform treatment of even and odd $N$, though for even $N$ the
zeroes $\Qs_\pm$
split into $N/2$ equidistant pairs with the separation $\pi\tau$.
Actually, we found this redundancy quite useful in controlling the consistency
of numerical calculations. Such pairing occurs automatically,  without placing
any constraints on the positions of zeroes.}. We will assume
that the roots $u^\pm_k$ lie in the {\em fundamental domain}
\beq
0<{\rm Re}\,u^\pm_k<\pi, \qquad -\pi|\tau|<{\rm Im}\,u^\pm_k<+\pi|\tau|,\qquad
k=1,\ldots,N\ .
\eeq

There are two independent sets of the Bethe Ansatz equations \eqref{BA2},
\beq
\frac{\fs(u_k^\pm+\eta)}{\fs(u_k^\pm-\eta)}=
-e^{4i\eta\varphi_\pm/\pi}\,
\frac{\As_\pm(u_k^\pm+2\eta)}{\As_\pm(u_k^\pm-2\eta)},
\qquad k=1,2,\ldots,N\label{BApm}
\eeq
with all upper and with
all lower sings for the zeroes of $\Qs_+(u)$ and $\Qs_-(u)$,
respectively.
The relations \eqref{qper4} impose the following constrains on the
parameters entering \eqref{wei},
\beq
\varphi_\pm=\pm\varphi+N\pi\pmod{2\pi}\ ,\label{1rel}
\eeq
\beq
2\tau\varphi_\pm=\mp i \psi-2\sum_{k=1}^N u^\pm_k +N\pi\pmod{2\pi}\ .
\label{2rel}
\eeq
Combining these relations one obtains
\beq
\varphi_++ \varphi_-=-2\pi m_2,\qquad
\overline{u}_++\overline{u}_-=\pi m_1+2\pi\tau m_2,\qquad m_1,m_2\in\Zbbd\ ,
\eeq
where
\beq
\overline{u}_+=\sum_{k=1}^N\, u^+_k\ ,
\qquad
\overline{u}_-=\sum_{k=1}^N\, u^-_k\ .
\eeq
Note that
for any pair of eigenvalues $\Qs_\pm(u)$
the parameter $\psi$ in \eqref{qper4}
can be regarded as
a function of $\varphi$. Indeed, the Bethe Ansatz equations \eqref{BApm}
only contain the exponents $\varphi_\pm$, which up to multiples of $\pi$
coincide with $\pm\varphi$. Once these equations are solved for $u^\pm_k$
the exponent $\psi$ is determined from \eqref{2rel}.

With the above  notations the 8V/SOS-model analog of the formula \eqref{F-sol}
can be written as
\bea
\Gs(u)
&=&+\frac{{\ds e^{-2i\varphi_+u/\pi}}\,\hs'(0)}{\hs(2\overline{u}_+)}\
\sum_{k=1}^N\ \frac{{\ds e^{2i\varphi_+\, u_k/\pi}}\ \fs(u^+_k+\eta)}{
\Qs_+(u^+_k+2\eta)\,\Qs'_+(u^+_k)}\
\frac{\hs(u+2\overline{u}_+-u^+_k)}{\hs(u-u^+_k)}\nonumber \\
&=&-\frac{{\ds e^{-2i\varphi_+u/\pi}}\,\hs'(0)}{\hs(2\overline{u}_+)}\
\sum_{k=1}^N\ \frac{{\ds e^{2i\varphi_+\, u_k/\pi}}\ \fs(u^+_k-\eta)}{
\Qs_+(u^+_k-2\eta)\,\Qs'_+(u^+_k)}\
\frac{\hs(u+2\overline{u}_+-u^+_k)}{\hs(u-u^+_k)}\label{qpm-ell}
\eea
where
\beq
\Gs(u)=\frac{1}
{2i\,\Ws(\varphi)}\ \frac{\Qs_-(u)}{\Qs_+(u)}
\eeq
and the prime now denotes the derivative with respect to $u$.
This formula is a simple consequence of Bethe Ansatz equations
\eqref{BApm}
and the following elegant identity for the elliptic theta functions%
\footnote{%
In the trigonometric limit $\q\to0$ this identity reduces to the
modified variant of Ex.2 on page 140 of \cite{WW}, obtained if the constant
term therein is ``smartly'' distributed among the terms in the sum.}
\beq
\prod_{k=1}^n\frac{\vartheta_1(x-y_k)}{\vartheta_1(x-z_k)}=
\frac{1}{\vartheta_1(\overline{z}-\overline{y})}
\sum_{k=1}^n\left(
\prod_{j\neq k}\frac{\vartheta_1(z_k-y_j)}{\vartheta_1(z_k-z_j)}\right)
\frac{\ds\vartheta_1(z_k-y_k)\vartheta_1(x-z_k+\overline{z}-\overline{y})}
{\vartheta_1(x-z_k)}\label{ident}
\eeq
Here $x$ is an independent variable,
$y_1,\ldots,y_n$ and $z_1,\ldots,z_n$ are arbitrary constants,
\beq
\overline{y}=\sum_{k=1}^n y_k,\qquad \overline{z}=\sum_{k=1}^n z_k \ ,
\eeq
and for brevity the theta-function $\vartheta_1(x\,|\,\q)$ is written
as $\vartheta_1(x)$. To obtain \eqref{qpm-ell} one needs to apply the
identity \eqref{ident} to the RHS of the quantum Wronskian relation, written
in the form
\beq
\Gs(u-\eta)-\Gs(u+\eta)=\frac{\fs(u)}{\Qs_+(u+\eta)\,\Qs_+(u-\eta)}\ .
\eeq

The zero-field variant of \eqref{qpm-ell} relates $\Qn(u)$ and $\Qbar(u)$
defined in \eqref{q0-def}. Let $u_k$ denote the zeroes of $\Qn(u)$,
\beq
\Qn(u)=
\prod_{k=1}^N \hs(u-u_k^\pm),\qquad \zeta(u)=\frac{d \log\hs(u)}{du}.
 \label{wei0}
\eeq
>From \eqref{zfwron} one can show that
\beq
\Qbar(u)=2\,\Qn(u)\,\Big\{\frac{i u}{\pi}
+\sum_{k=1}^N\,\zeta(u-u_k)\ {u}'_k(0)\Big\}+\Cbbd\,\Qn(u)\label{qbar}
\eeq
where $\Cbbd$ is an arbitrary constant and
\beq
{u}'_k(0)=\frac{d u_k}{d \varphi}\Big\vert_{\varphi=0}=
\frac{i\, \fs(u_k+\eta)\,{\Ws}'(0)}{\Qn(u_k+2\eta)\,\Qn'(u_k)}\ .
\eeq
The unknown constant ${\Ws}'(0)$, defined in \eqref{wprime},
can be expressed through the zeros $u_k$ from the differential
equations \eqref{odeal}.

\subsection{Singular eigenvalues}\label{sing-sect}
When $\eta/\pi$ is a rational number the spectrum of the transfer
matrix of the symmetric 8V-model is degenerate (for sufficiently
large $N$).
Solutions of
the Bethe Ansatz equations \eqref{BA2} for degenerate states are not
unique.
They contains arbitrary continuous parameters,
which determine positions of the complete strings \eqref{compstr}.
There is nothing wrong with this. In particular, this is not an
indication that the Bethe Ansatz is ``incomplete'' \cite{FM01}.
Quite to the contrary, as explained in \cite{Bax02}, the appearing continuous
parameter are precisely those that are needed to describe the embedding of
the corresponding eigenvectors into the eigenspace
of the degenerate eigenvalue.

It is not immediately clear, however, how the eigenvalues
$\Qs_\pm(u)$, which have no ambiguities at generic values
of the parameters $\eta$ and $\varphi$, could acquire continuous degrees of
freedom for the degenerate states at special values of $\eta$ and
$\varphi$. The explanation is simple: the limiting values of
$\Qs_\pm(u)$ are not uniquely defined; they depend on the details of
the limiting procedure.
Here we will not study this phenomenon in its full generality, but
give a particular example.
For simplicity we consider the 6V-model in a field, where most
calculations can be done analytically.
Assume the same notations as in
Sect.\ref{6v-model}. In particular, recall the definitions
of the multiplicative spectral parameter $x$, the field variable $z$
and the parameter $q$ given in \eqref{6v-var}.
For generic $z$ and $q$
the transfer matrix of the 6V-model is completely non-generate
(the usual degeneracy associated with the reversal of all spins
is broken in the presence of the field).
Consider the case $N=4$ and $n=2$.
This sector contains six different
eigenvectors.
The corresponding polynomials $\As_\pm(x)$, defined in \eqref{A6-def},
are all of the second degree.
Let us parameterize them as
\beq
\As_+(x)=1+a_1 x + a_2 x^2,\qquad
\As_-(x)=1+b_1 x + b_2 x^2\ .\label{a6pm}
\eeq
Substituting these expressions into the quantum Wronskian relation
\eqref{wr6} with $\rho_+=\rho_-=1$, one gets four
equations for the unknown coefficients $a_1,a_2,b_1,b_2$,
\bea
1&=&a_2 b_2,\nonumber\\
z^2&=&\frac{q^2 a_1 b_2 +4 q + a_2 b_1}{q^2 a_2 b_1 +4 q + a_1 b_2},\nonumber\\
z^2&=&\frac{q^4b_2 +q^2(a_1b_1-6)+a_2}{q^4a_2 +q^2(a_1b_1-6)+b_2},
\label{syst}\\
z^2&=&\frac{q^2 b_1 +4 q + a_1}{q^2 a_1 +4 q + b_1}\ .\nonumber
\eea
Excluding
$a_2,b_1,b_2$ one obtains a six-degree polynomial equation for
$a_1$, which factors into two linear and two quadratic equations.
One of the latter reads
\beq
q(1-q^2 z^2)\, a_1^2 +
[{4q^2(1-z^2)+z(1-q^4)}]\,a_1+4q(1-z)(q^2+z)=0,\label{quad}
\eeq
while the other is obtained by the substitution $z\to-z$.
Altogether one gets six different solutions for the coefficient
$a_1$. Once it is found, the remaining coefficients
are unambiguously determined by \eqref{syst}.

Consider the limit
\beq
\eta\to\pi/4,\qquad \varphi\to0\ .\label{eflim}
\eeq
In terms of $q$ and
$z$ it corresponds to $q\to i$, $z\to 1$.
In this limit the two eigenvalues of the transfer matrix,
$\Ts^{(1)}(x)$ and $\Ts^{(2)}(x)$,
corresponding to two different solutions of \eqref{quad}, smoothly
approach the same value
\beq
\Ts(x)=2x^4-12x^2+2 = 2\,(x^2-b_0^2)\,(x^2-b_0^{-2}),\qquad b_0=1+\sqrt{2}\ .
\label{b0-def}\eeq
To calculate $\Qs_\pm(u)$ one also needs to describe the first order
deviations of $\eta$ and $\varphi$ from their limiting values \eqref{eflim}.
This requires two arbitrary small parameters $\epsilon_1$ and
$\epsilon_2$, or just one such parameter and the ratio
$\epsilon_1/\epsilon_2$.
We found it convenient to use the following parametrization
\beq
\eta=\frac{\pi}{4}+(b^4-6 b^2+1)\epsilon
+O(\epsilon^2),\qquad
\varphi=8 (b^4-1)\epsilon
+O(\epsilon^2)\label{efexp}
\eeq
where $\epsilon\to0$ and $b$ is an arbitrary complex parameter, which
is kept finite.
The coefficients of the linear terms in $\epsilon$
have been specifically chosen to
simplify subsequent expressions.
The corresponding expansions for $q$ and $z$ simply follow from
\eqref{6v-var}. Substituting them into Eq.\eqref{quad}, solving it for $a_1$
and then determining the remaining coefficients in \eqref{a6pm} from
\eqref{syst}, one obtains
\bea
\As_+^{(1)}(x)=1-\frac{x^2}{b^2}
+O(\epsilon),
&&\qquad
\As_-^{(1)}(x)=1-b^2\,x^2
+O(\epsilon),\nonumber\\
&&\label{qpmlimit}\\
\As_+^{(2)}(x)=1-\frac{x^2}{\overline{b}^2}
+O(\epsilon),
&&\qquad
\As_-^{(2)}(x)=1-\overline{b}^2\,x^2
+O(\epsilon),\nonumber
\eea
where $\overline{b}$ is related to $b$ by a self-reciprocal transformation
\beq
\overline{b}^2=\frac{3-b^2}{1-3b^2},\label{b-bar}
\eeq
which exchanges the two solutions.
The corresponding eigenvalues and eigenvectors of the transfer
matrix read
\bea
&&\Ts^{(1)}(x)=\Ts(x)-16(1+b^2)^2x(1+x^2)\epsilon+O(\epsilon^2)\ ,\nonumber\\
&&\\
&&\Ts^{(2)}(x)=\Ts(x)+32(1-b^2)^2x(1+x^2)\epsilon+O(\epsilon^2)\ ,\nonumber
\eea
where $\Ts(x)$ is given by \eqref{b0-def} and
\beq
|\Psi^{(a)}\rangle=
|\uparrow\downarrow\uparrow\downarrow\rangle-
|\downarrow\uparrow\downarrow\uparrow\rangle+
\rho^{(a)}{\bigl\{}
|\downarrow\downarrow\uparrow\uparrow\rangle+
|\uparrow\uparrow\downarrow\downarrow\rangle-
|\downarrow\uparrow\uparrow\downarrow\rangle-
|\uparrow\downarrow\downarrow\uparrow\rangle{\bigr\}}
+O(\epsilon),\qquad a=1,2,
\label{vectors}
\eeq
with
\beq
\rho^{(1)}=\frac{i(1-b^2)}{1+b^2},\quad \rho^{(2)}=\frac{i(1+b^2)}{2(b^2-1)}\ .
\eeq
Here we used self-explanatory notations for the ``spin up'',
$|\uparrow\rangle$, and  ``spin down'', $|\downarrow\rangle$, states
of the edge spins.

Note, that the case $\eta=\pi/4$ corresponds $m=1$ and $L=2$ in
\eqref{rat-eta}. In terms of the variable $x$ the complete string
\eqref{compstr} for this case consists of two roots $x_{1},x_{2}$,
constrained by one relation $x_{2}=-x_{1}$.
Obviously,  at $\epsilon=0$ the zeroes of $\As_\pm(x)$, given by
\eqref{qpmlimit},  are the
complete 2-strings  at $x_1=b^{\pm1}$ or $x_1=\bar{b}^{\pm1}$.
Their positions are given in terms of the very same parameter $b$,
which defines the embedding of the eigenvectors \eqref{vectors}
into the eigenspace of the degenerate eigenvalue
\eqref{b0-def} in the sector with
two spins up.
Remind that the parameter $b$ can be arbitrary;
it determines the direction of the 2-variable limit \eqref{efexp}.

It is interesting to see what happens in some particular cases of
\eqref{efexp}. First, consider the zero-field case when $\varphi\equiv0$
from the very beginning and $\eta$ is approaching the value $\pi/4$.
In Eq.\eqref{efexp} this corresponds $b^2=\pm1$. As obvious from
\eqref{qpmlimit} and \eqref{b-bar} the limiting values of
$\As_\pm(x)$ then coincide and the complete 2-string
can take one of the two fixed positions with $x_1=1$ or $x_1=i$.

The second interesting case is when $\eta$ is set to its rational
value $\eta\equiv\pi/4$ while the $\varphi$ is
approaching zero from arbitrary values. In Eq.\eqref{efexp} this
corresponds to $b=b_0^{\pm1}$, where $b_0$ is defined in
\eqref{b0-def}. Note, also the corresponding value of $\bar{b}_0^2=b_0^{-2}$.
We see that the complete 2-strings in this case occupy precisely the
distinguished positions determined by Eq.\eqref{zftl1}.
Indeed, taking into account \eqref{b0-def} it is easy to see that
 Eq.\eqref{zftl1} is satisfied with $\Cs(0)=2$ and ${\Qo_\pm(x)\equiv
\As_\pm(x)}$ (it does not matter which solution in \eqref{qpmlimit} is taken,
because  $\As^{(1)}_\pm(x)=\As^{(2)}_\mp(x)$ when $b=b_0$).
We would like to stress that the
distinguished string positions arising from \eqref{zftl1}
can only be achieved by considering the limit from arbitrary values
of the field with fixed rational values of $\eta$.

Finally, note that degenerate eigenstates in the 6V- and 8V-models
were also considered in \cite{FM01,Bax02,Deg02}

\newpage
\nsection{Analytic continuation of the eigenvalues}\label{cont-sect}
The eigenvalues of the $\T$- and $\Q$- matrices of the 8V/SOS-model have
very interesting analytic properties with respect to the field parameter
$\varphi$. In this section we will study these properties by a combination
of analytic and numerical techniques. We consider
the disordered regime of the 8V-model with real $\q$ and $\eta$ in the range
\beq
0<\q<1,\qquad \pi/4<\eta<\pi/2.\label{dis-reg}
\eeq
Note that the parameter $\tau$ in this case is purely imaginary,
$\tau=i|\tau|$.

\subsection{Bethe Ansatz equation}
We will show that different eigenvalues
of the transfer matrix of the symmetric 8V-model
can be obtained from each other by the analytic
continuation in the variable $\varphi$.
In doing this we will, obviously, need to consider
$\varphi$ as an independent variable.
Therefore, for the eigenvalues $\Qs_\pm(u)$ we assume
the SOS-model periodicity \eqref{qper4} which allows arbitrary
values of $\varphi$. In practice it is more convenient to
work with a simply related variable, which interpolate the
values of $\varphi_+$ (remind that it is related to
$\varphi$ by \eqref{1rel}). We will denote this variable
by the ``upright'' symbol $\phi$.
The Weierstrass factorization for
the eigenvalues was discussed in Sect.\ref{wron-8v-sect}.
It is convenient
to parameterize the zeroes of $\Qs_\pm(u)$ as
\beq
u_k^\pm=\pi/2+i\alpha_k^\pm\ .\label{u-alpha}
\eeq
Note, that the product representations \eqref{wei} has an obvious  ambiguity.
For example, the translation of any single root $\a_k^+$ by
the period $2\pi |\tau|$ complemented by the $2\pi$-shift of $\varphi_+$,
\beq
\a_k^+\to\a_k^+-2\pi
i\tau,\qquad \varphi_+\to\varphi_++2\pi\ ,\label{phishift}
\eeq
leaves $\Qs_+(u)$ unchanged
(more precisely, $\Qs_+(u)$  acquires an
inessential scalar factor).
Using this freedom one can always bring all
zeroes of $\Qs_\pm(u)$ to the periodicity rectangular
\beq
-\pi|\tau|\leq\mbox{Re}\,\a_k^\pm\leq+\pi|\tau|,\qquad
-\pi/2\leq\mbox{Im}\,\a_k^\pm\leq\pi/2,\qquad k=1,\ldots,N.
\label{rect}
\eeq
With this convention the values of the exponents $\varphi_\pm$ in \eqref{wei}
cannot, in general, be restricted to any finite domain.

Introduce two functions
\beq
\Phi_1(\a)=\frac{1}{2\pi i} \log\frac{\vt_3(\eta+i\a|q)}{\vt_3(\eta-i\a|q)},
\qquad
\Phi_2(\a)=
\frac{1}{2\pi i}
\log\frac{\vt_1(2\eta+i\a|q^2)}{\vt_1(2\eta-i\a|q^2)}\ .\label{Phi-def}
\eeq
Consider the Bethe Ansatz equations of the following general form
\beq
{N}\Phi_1(\a_k)-\sum_{{j=1}}^N
\Phi_2(\a_k-\a_j)=
-n_k +\frac{2\eta\phi}{\pi^2},\label{lba}
\eeq
where the numbers $n_1,n_2,\ldots,n_N$ and $\phi$ are given,
while the (complex) variables $\a_1,\a_2,\ldots,\a_N$ are considered
as unknowns.
The numbers $n_k$
take half-odd integer values for even $N$ and
integer values for odd $N$. Taking into account \eqref{wei} and
\eqref{u-alpha} it is easy to see that the logarithmic form
of the Bethe Ansatz equations \eqref{BApm} for $\Qs_\pm(u)$ coincide
with \eqref{lba} provided $\a_k$ and $\phi$ therein are replaced
by $\a_k^\pm$ and $\varphi_\pm$.

Obviously, the numbers $\{n_k\}$ depend on the choice of branches of
logarithms in \eqref{Phi-def}.
In the considered case of the 8V-model, which involves elliptic
functions, an extra
care is required to define these branches.
Since the functions $\Phi_{1,2}$ are periodic
\beq
\Phi_i(\a+i\pi)=\Phi_i(\a),\qquad i=1,2
\eeq
one only needs to specify them in the
periodicity strip $0\le{\rm Im}\a<\pi$.
We choose the cuts
\beq
\Bigl\{\Bigl(m+\frac{1}{2}\Bigr)\pi|\tau|+i\Bigl(\frac{\pi}{2}+n\pi-\eta\Bigr),
\Bigl(m+\frac{1}{2}\Bigr)\pi|\tau|+i\Bigl(\frac{\pi}{2}+n\pi+\eta\Bigr)\Bigr\},
\quad m,n\in\mathbb{Z}
\eeq
for the function $\Phi_1(\a)$ and the cuts
\beq
\{m\pi|\tau|+i(n\pi-2\eta),m\pi|\tau|+i((n-1)\pi+2\eta)\},\quad m,n\in\mathbb{Z}
\eeq
for $\Phi_2(\a)$ (the latter are shown in Fig. \ref{Phicuts}).
\begin{figure}[ht]
\begin{picture}(360,215)\put(50,0){\begin{picture}(340,200)\put(0,100)
{\vector(1,0){360}}\put(179.8,192){\vector(0,1){15}}
\put(179.8,0){\line(0,1){8}}
\put(179.8,66){\line(0,1){68}}{\multiput(0,0)(0,126){2}{
\multiput(0,0)(138,0){3}{\put(42,12){\circle*{2}}\put(42,62){\circle*{2}}
\put(42,12){\arc{7.5}{-7.45}{-1.95}}\put(42,62){\arc{7.5}{-4.35}{1.2}}
\multiput(0,0)(3,0){2}{\multiput(0,0)(0.5,0){2}
{\put(40,15.5){\line(0,1){43}}}}}}}
\multiput(0,0)(0,126){2}{\multiput(0,12)(0,50){2}
{\dashline[-10]{5}[2](0,0)(360,0)}}
\put(170,103){$0$}\put(143,142){$\pi-2\eta$}\put(162,178){$2\eta$}
\put(186,200){$\mbox{Im}\,\a$}\put(184,51){$2\eta-\pi$}\put(184,16){$-2\eta$}
\put(335,105){$\mbox{Re}\,\a$}\put(180,100){\circle*{3}}
\multiput(-138,0)(138,0){3}{\qbezier(210,145)(200,159)(182,160)\put(184,159.8)
{\vector(-1,0){2}}\qbezier(150,55)(160,41)(178,40)
\put(176,40.2){\vector(1,0){2}}}
\multiput(42,68)(276,0){2}{\dashline[-10]{5}[2](0,0)(0,66)}
\put(12,105){$-\pi|\tau|$}\put(296,105){$\pi|\tau|$}\end{picture}}\end{picture}
\caption{ The principal Riemann
sheet of the function  $\Phi_2(\a)$.}\label{Phicuts}
\end{figure}
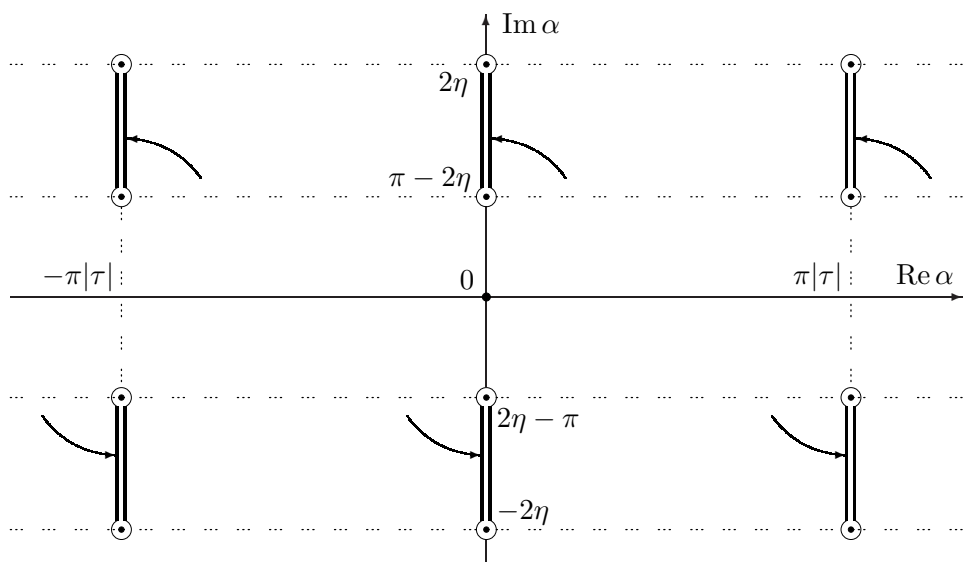
We will assume that the
values of \eqref{Phi-def} appearing in \eqref{lba} are always taken on
the principal sheet of the corresponding Riemann surface defined by the
condition
\beq
\Phi_i(-\a)=-\Phi_i(\a), \qquad i=1,2\ .
\eeq
The values on the cuts are taken from the right(left) side
for the upper(lower) cuts in Fig.\ref{Phicuts}, as shown by arrows.
Note that with these definitions the values of $\Phi_1(\a)$ and
$\Phi_2(\a)$ on the real axis of $\a$ are unbounded (this does not happen
in the trigonometric case). They monotonically decrease (or increase) with
the increase of $\a$. We should warn the reader that this property
is never achieved with {\em ad hoc} computer definitions of the
logarithms in \eqref{Phi-def}.

Every eigenvalue of the transfer matrix
corresponds to some set\footnote{%
The numerical examples considered below suggest that these
sets are the same for $\Qs_+(u)$ and
$\Qs_-(u)$, provided the phases of the logarithms in \eqref{lba}
are defined as
described above.} of the numbers $\{n_k\}$.
Even when the branches of the logarithms are
completely fixed, there is
always an ambiguity in $\{n_k\}$  related to the re-definition of
$\phi$. Indeed, the transformation
\beq
\a_k \to\a_k,\qquad n_k\to n_k+m,\qquad \phi\to
\phi+m\,\frac{\pi^2}{2\eta},
\qquad m\in\Zbbd\ ,\label{shift}
\eeq
leaves Eq.\eqref{lba} unchanged.

Further, the numbers $\{n_k\}$ do not uniquely determine
solutions of \eqref{lba}.
Different solutions corresponding to different
eigenvalues could be related with the same set
$\{n_k\}$
{}\footnote{%
With the exception of a few simple eigenvalues
(e.g. the band of
largest eigenvalues \cite{JKM73}) these sets are very poorly
understood and their
enumeration is a difficult (but, perhaps, not hopeless)
problem. To our knowledge the only case where this enumeration
problem was completely solved \cite{BLZ03} is the $c<1$ conformal field theory
(CFT) with $\Delta\to\infty$.
This CFT can be obtained in the scaling limit of the eight-vertex
(six-vertex) model.}.
In any case once the numbers $n_k$ are fixed
the roots $\a_k=\a_k(\phi)$ solving \eqref{lba}
become functions of the complex variable
$\phi$.  It turns out that the latter are multivalued functions with algebraic
branching points.
Below we will investigate their monodromy properties
for various eigenvalues of the transfer matrix.

\subsection{Differential form of the Bethe Ansatz equations}

The behavior of the roots $\a_k(\phi)$ under variations of $\phi$
can be effectively studied with a differential form of the Bethe Ansatz
equations.
Differentiating (\ref{lba}) with respect to $\phi$ one immediately
obtains a system of ordinary linear differential equations
\beq
\sum_{k=1}^N A({\a})_{jk}\,\frac{\partial \a_k(\phi)}{\partial \phi}
=\frac{2\,\eta}{\pi^2 },
\qquad j=1,\ldots,N\ ,\label{odeal}
\eeq
The $N$ by $N$ matrix $A(\a)_{jk}$ is given by
\beq
A({\a})_{jk}=\Phi_2'(\a_j-\a_k)+\delta_{jk}
\bigl\{N\Phi_1'(\a_j)-\sum_{l=1}^N\Phi_2'(\a_j-\a_l)
\bigr\}\label{odemat}
\eeq
where $\Phi'_{1,2}(\a)$ denote derivatives of the functions
\eqref{Phi-def} with respect to their argument $\a$.

Using the identity
\beq
\frac{\t_k'(x+y|\q)}{\t_k(x+y|\q)}+\frac{\t_k'(x-y|\q)}{\t_k(x-y|\q)}
{=\frac{\t_4'(2x|\q)}{\t_4(2x|\q)}+\sigma_k
\t_2(0|\q)\t_3(0|\q)\,\frac{\t_1(2x|\q)\t_{5-k}(x+y|\q)\t_{5-k}(x-y|\q)}
{\t_4(2x|\q)\t_k(x+y|\q)\t_k(x-y|\q)}}
\label{ident1}
\eeq
where $k=1,2,3,4$ and $\sigma_{1,4}=-\sigma_{2,3}=1$, one can show
that $\Phi_{1,2}'(\a)$ are meromorphic double-periodic function
of the variable $\a$,
\beq
\Phi_1'(\a)=\xi-\chi\,
\frac{\vt_3(0|\q)\vt_2(\eta+i\a|\q)\vt_2(\eta-i\a|\q)}
{\vt_2(0|\q)\vt_3(\eta+i\a|\q)\vt_3(\eta-i\a|\q)},\label{phi13a}
\eeq
\beq
\Phi_2'(\a)=\xi+\chi\,
\frac{\vt_2(2\eta|\q)\vt_4(i\a|\q)}
{2\vt_1(2\eta+i\a|\q^2)\vt_1(2\eta-i\a|\q^2)},
\label{phi13}
\eeq
where
two constants $\xi$, $\chi$ are given by
\beq
\xi=\frac{\t_4'(2\eta|\q)}{2\pi\,\t_4(2\eta|\q)},\qquad
\chi=
\frac{\t_2^2(0|\q)\t_1(2\eta|\q)}{2\pi\t_4(2\eta|\q)}.
\eeq

For our purposes we need to study the
differential equations \eqref{odeal} only with the initial conditions
corresponding to some eigenvalue of the symmetric 8V-model
(typically the ground state eigenvalue). In this case the
matrix $A(\a)$ is, in general, non-singular. Eq.\eqref{odeal}
can be then solved for the derivatives $\partial_{\phi}
{\a}_k(\phi)$ and defines
locally analytic functions $\a_k(\phi)$ of the variable $\phi$.
Exceptions occur at certain
root configurations, corresponding to special values of $\phi$, where
the matrix $A(\a)$ becomes singular.
These singular configurations correspond to the branching
points of the solutions in the complex
$\phi$-plane.

Let us rewrite (\ref{odeal}) in the form
\beq
\det\Vert A(\a)\Vert \>
\frac{\partial \a_j(\phi)}{\partial \phi}
=\frac{2\eta}{\pi^2}
\sum_{k=1}^NA_{jk}^{\mbox{\tiny adj}}(\a),\label{odeal1}
\eeq
where $A^{\mbox{\tiny adj}}$ stands for the adjoint matrix and
assume $\a(\phi_0)=(\a_1(\phi_0),\ldots,\a_N(\phi_0))$
is such that
\beq
\det\Vert A(\a(\phi_0))\Vert =0. \label{odedet}
\eeq
The type of the branching depends on the rank of $A(\a(\phi_0))$.
The most important case is when
\beq
\mathrm{rank}\,A(\a(\phi_0))=N-1\ .\label{oderank}
\eeq
Expanding the determinant $\det\Vert A(\a(\phi))\Vert $ near the
point $\phi=\phi_0$, one obtains
\beq
\det\Vert A(\a(\phi))\Vert =\sum_{i=1}^N c_i\,(\a_i(\phi)-\a_i(\phi_0))
+\mathrm{higher\>order\> terms},\label{detexp}
\eeq
where $c_i$ denote some constants.
Putting this back into \eqref{odeal1} and
keeping first order terms only, one
gets
\beq
\frac{\partial \a_j(\phi)}{\partial \phi}
\,
\sum_{i=1}^N c_i(\a_i(\phi)-\a_i(\phi_0))
=B_j \label{odeal2}
\eeq
where $B_j$ denotes the RHS of \eqref{odeal1} evaluated at
$\phi=\phi_0$. It is easy to see that all $B_j$ are non-zero,
otherwise the rank of $A(\phi_0)$ would have been less than $N-1$.
Summing \eqref{odeal2} over $j$ with the coefficients $c_j$ one obtains
\beq
\frac{\partial}{\partial \phi}\,
\Big(\sum_{i=1}^N c_i(\a_i(\phi)-\a_i(\phi_0)\Big)^2=2\sum_{i=1}^Nc_iB_i\ .
\eeq
Integrating over $\phi$ and substituting the result back to \eqref{detexp}
one gets
\beq
\det\Vert A(\a)\Vert =
\Big(2\sum_{i=1}^Nc_iB_i\Big)^{1/2}\,\sqrt{\phi-\phi_0}+O((\phi-\phi_0))
\eeq
Finally, from \eqref{odeal2} it follows that at $\phi\to\phi_0$
\beq
\a_j(\phi)-\a_j(\phi_0)= D_j\sqrt{\phi-\phi_0}+O((\phi-\phi_0)),\qquad
D_j=2B_j\,\Big(2\sum_{i=1}^Nc_iB_i\Big)^{-1/2},\label{alroot}
\eeq
Therefore, the condition \eqref{oderank}
implies the square root branching points for all $\a_j$.
In principle, higher-order branching points
are possible if the rank of the matrix
$A(\phi_0)$ drops below $N-1$.
 However, we found that the solutions
of the Bethe Ansatz equations for $N=2,3,4$ contain the second order
branching points only.

Note, that for even $N$ the eigenvalues $\Qs_\pm(u)$
for non-degenerate states
possess the more restrictive periodicity \eqref{Neven}.
Their imaginary (quasi-)period
is equal to $\pi\tau$ rather than $2\pi\tau$.
This is also true for the
SOS-model, see Eq.\eqref{qper5}.
Therefore, one can split the roots $\a_k(\phi)$ into the two
subsets $\{\a_1,\a_2,\ldots,\a_{N/2}\}$ and
$\{\a_{N/2},\a_{N/2+1},\ldots,\a_{N}\}$
obtained from each other by a uniform shift of all roots
by the period $\pi|\tau|$,
\beq
\a_{N/2+k}(\phi)-\a_k(\phi)=\pi|\tau|,\qquad k=1,2,\ldots,N/2.\label{halfalpha}
\eeq
It is easy to see that
under this substitution
the system (\ref{lba}) reduces to only $N/2$ equations,
provided the phases $\{n_{N/2+1},\ldots,n_N\}$
are properly fixed in terms of
$\{n_1,\ldots,n_{N/2}\}$. Further, the constraints \eqref{halfalpha}
are also compatible with
the differential equations \eqref{odeal}.
If these constraints hold for the initial data
they will continue to hold for an arbitrary
$\phi$.
Therefore for even $N$
one needs to consider only a half of the
Bethe roots, for instance, the first $N/2$ roots.
The matrix $A(\a)$ in \eqref{odeal} effectively reduces
to a $N/2\times N/2$ matrix.
This considerably simplifies the numeric analysis for
even values of $N$.

\subsection{Overview of the procedure}\label{sect-over}
Below we will give a complete classification
of the eigenvalues for the symmetric 8V-model with $N=2,3,4$. In
most cases
we present explicit analytic
expressions for the eigenvalues obtained by a direct diagonalization
of the transfer matrix. Whenever purely analytic results are not possible,
we provide required numeric values for a particular choice of the parameters
$\q$ and $\eta$. Depending on convenience we will use either of the
transfer matrices $\T(u)$ and $\T^B(v)$ related by \eqref{tq-redef},
assuming that the parameters $u$ and $v$ are always connected by
\eqref{uv-rel0}.
The corresponding eigenvalues,
denoted here by ${\mathsf \Lambda(u)}$ and ${\bf \Lambda}^B(v)$
respectively, are related as
\beq
{\mathsf \Lambda}(u)
=(-iq^{1/4})^N\,e^{iuN}{\bf\Lambda}^B(v),\qquad v=u+\frac{\pi\tau}{2}\ .
\label{num2}
\eeq
In the previous sections we also considered the eigenvalues
of the ``higher'' transfer matrices $\Ts_k(u)$, defined by \eqref{fus1}.
In this Section we will denote
them as ${\mathsf\Lambda}^{(k)}(u)$.
With these notations ${\mathsf\Lambda}(u)\equiv{\mathsf\Lambda}^{(2)}(u)$.

The transfer matrix of the 8V-model
commutes \cite{Sut70,Bax72b} with the XYZ-Hamiltonian
\beq
{\bf H}_{XYZ}=-\frac{1}{2}\, \sum_{j=1}^N
\big(J_x\,\sigma^{(j)}_1\sigma^{(j+1)}_1+
J_y\,\sigma^{(j)}_2\sigma^{(j+1)}_2+
J_z\,\sigma^{(j)}_3\sigma^{(j+1)}_3\big)\label{ham0}
\eeq
provided the constants $J_x,J_y,J_z$ satisfy \eqref{J-rat}. Below
we will assume the normalization
\beq
J_x=1\ .\label{xyz-norm}
\eeq
The eigenvalues of ${\bf H}_{XYZ}$
can be found from \eqref{ht} which relates this Hamiltonian
to the logarithmic derivative of the transfer matrix $\T^B(v)$.

For each $N$ there are $2^N$ eigenvalues
${\mathsf \Lambda}_i(u)$, $i=0,1,\ldots,2^N-1$.
Below  we will assume a definite ordering of ${\mathsf \Lambda}_i(u)$.
We will first split them according to the momenta ${\mathcal P}$ of the
corresponding eigenstates\footnote{%
The momenta ${\mathcal P}$ are eigenvalues of the {\em shift operator},
$\T^B(\eta)$ which cyclically shifts the edge
spins $\sigma^{j}\to\sigma^{j-1}$ to the left by one lattice step.}.
Then, in each sector with a fixed value
of $\mathcal{P}$ we will
arrange ${\mathsf \Lambda}_i(u)$ according
to eigenvalues of the XYZ-Hamiltonian \eqref{ham0}.

For every $i=0,\ldots,2^N-1$
let $\{\a^{(i)}\}$ denote the solution of \eqref{lba}, corresponding
to the eigenvalue ${\Ls}_i$.
We will say that the contour $C_{ij}\equiv C_{ij}(\zeta)$,
$\zeta\in[0,1]$ connects two eigenvalues ${\Ls}_i$ and
${\Ls}_j$ (corresponding to the values $\phi_i$ and
$\phi_j$)
if
\beq
C_{ij}(0)=\phi_i,\quad C_{ij}(1)=\phi_j,\quad
\{\a^{(i)}\}\stackrel{C_{ij}}\rightarrow\{\a^{(j)}\},\quad
{\Ls}_i\stackrel{C_{ij}}\rightarrow\,\pm{\Ls}_j,\label{Ccont}
\eeq
where we allowed for an ambiguity which can be
compensated by an overall shift of all the numbers $n_k$. Indeed,
the only effect of the transformation \eqref{shift} is that the
corresponding eigenvalue of the
transfer matrix ${\mathsf \Lambda}(u)$ acquires an extra sign factor
\beq
{\mathsf \Lambda}(u)\to (-1)^m {\mathsf \Lambda}(u)\ .\label{sign}
\eeq

As a simple illustration consider the ground state eigenvalue.
In the regime \eqref{dis-reg} all roots
$\{\a_k\}$ are real, so they can be arranged in an increasing sequence
$\a_1<\a_2<\cdots<\a_N$.  For the ground state the numbers $n_k$ take
consecutive integer or half-an-odd integer values
\beq
n_k=k-\frac{N+1}{2},\quad k=1,\ldots,N.\label{gphases}
\eeq
Further, for even $N$ the ground state eigenvalue $\La_0(u)$ of the 8V-model
is non-degenerate. It corresponds to
$\phi=0$. At this point the roots solving \eqref{lba}
are symmetrically distributed on the interval $(-\pi|\tau|,\pi|\tau|)$,
such that $\a_k(0)=-\a_{N-k+1}(0)$. Let us take this configuration as an
initial condition for \eqref{odeal} and trace its evolution for real
$\phi$.
When $\phi$ increases in the
positive direction all the
roots monotonously move in the negative direction. At $\phi=\pi$
the leftmost root reaches the boundary of the periodicity rectangular
\eqref{rect}, $\alpha_1(\pi)=-\pi|\tau|$, while one of the middle
roots hits the origin $\alpha_{N/2+1}(\pi)=0$. The remaining $N-2$
roots arrange symmetrically around the origin.
The value $\phi=\pi$ belongs to the set  of the exponents \eqref{fi8v} of
the symmetric 8V-model.
The resulting
solution of the Bethe Ansatz equation corresponds to the
next-to-leading eigenvalue of the transfer matrix, $\La_1(u)$.
Thus, the contour $C_{01}$ in this case is a straight line from $\phi=0$ to
$\phi=\pi$.
Further, at $\phi=2\pi$ the first root becomes
\beq
\a_1(2\pi)=\a_N(0)-2\pi|\tau|\ ,
\eeq
while the remaining roots take the original $\phi=0$
positions, but shifted one step left,
\beq
\a_k(2\pi)=\a_{k-1}(0),\qquad k=2,\ldots,N\ .
\eeq
The whole process is illustrated in Fig. \ref{Movroots}. With an account of
\eqref{phishift} the resulting root configuration is completely
equivalent to initial one at $\phi=0$. Thus the eigenvalue
$\La_0(u)$ is a periodic function of $\phi$ with the period
$2\pi$.
\begin{figure}[ht]
\begin{picture}(400,160)\put(50,15){\begin{picture}(360,120)\put(0,10)
{\vector(1,0){370}}\put(180,0){\vector(0,1){130}}\multiput(0,60)(0,50){2}
{\dashline[-10]{5}[2](0,0)(360,0)}\multiput(42,10)(276,0){2}
{\dashline[-10]{5}[2](0,0)(0,110)}\put(82,10){\circle*{4}}\put(103.7,10)
{\circle*{4}}\put(118.2,10){\circle*{4}}\put(140,10){\circle*{4}}
\put(220,10){\circle*{4}}\put(242,10){\circle*{4}}\put(256,10)
{\circle*{4}}\put(278,10){\circle*{4}}\put(42,60){\circle*{4}}
\put(95,60){\circle*{4}}\put(111,60){\circle*{4}}\put(127,60){\circle*{4}}
\put(180,60){\circle*{4}}\put(234,60){\circle*{4}}\put(249,60){\circle*{4}}
\put(265,60){\circle*{4}}\put(2.4,110){\circle*{4}}\put(82.2,110){\circle*{4}}
\put(103.7,110){\circle*{4}}\put(118.2,110){\circle*{4}}\put(140,110)
{\circle*{4}}\put(220,110){\circle*{4}}\put(242,110){\circle*{4}}\put(256,110)
{\circle*{4}}\qbezier(82,10)(42,60)(2.4,110)\qbezier(103.7,10)(97,60)(82.2,110)
\qbezier(118,10)(111,60)(104,110)\qbezier(140,10)(125,60)(118,110)
\qbezier(220,10)(180,60)(140,110)\qbezier(242,10)(236,60)(220,110)
\qbezier(256,10)(249,60)(242,110)\qbezier(278,10)(263,60)(256,110)
\dashline{5}(359,10)(277.4,110)\multiput(277.4,110)(40.8,-50){3}{\circle{4}}
\put(78,115){$\a_1$}\put(97,115){$\a_2$}\put(114,115){$\a_3$}
\put(136,115){$\a_4$}\put(214,115){$\a_5$}\put(235,115){$\a_6$}
\put(252,115){$\a_7$}\put(273,115){$\a_8$}
\put(78,0){$\a_1$}\put(100,0){$\a_2$}\put(116,0){$\a_3$}\put(138,0){$\a_4$}
\put(218,0){$\a_5$}\put(240,0){$\a_6$}\put(254,0){$\a_7$}\put(276,0){$\a_8$}
\put(25,-2){$-\pi|\tau|$}\put(308,-2){$\pi|\tau|$}\put(185,125){$\phi$}
\put(166,100){$2\pi$}\put(171,50){$\pi$}
\put(172,0){$0$}\end{picture}}\end{picture}
\caption{The $\phi$-dependence of the Bethe roots
for the ground state at $N=8$.}\label{Movroots}
\end{figure}
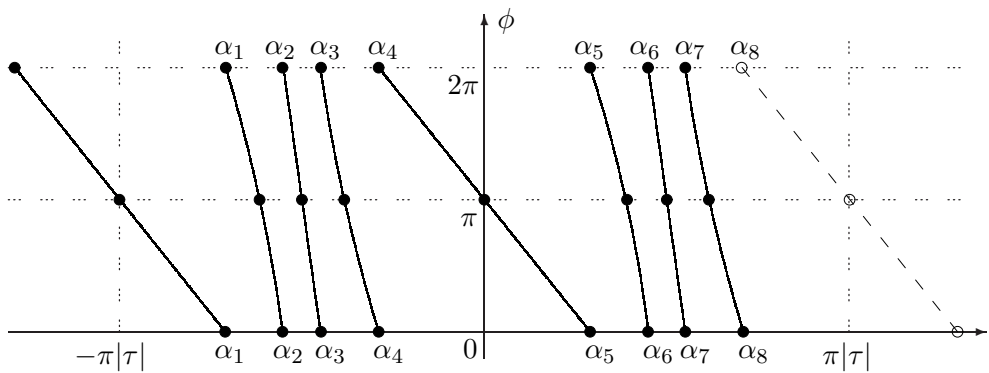

It is instructive to examine what happens to the exponent
$\psi$ in \eqref{qper4} for the same course variations of $\phi$.
It can be calculated from \eqref{2rel}. We plotted the result
in Fig.\ref{psiN8}. It is a $2\pi$-periodic function
which vanishes at the values $\phi=k\pi$, $k\in\Zbbd$, corresponding
the symmetric 8V-model, in accordance with \eqref{fi8v}.
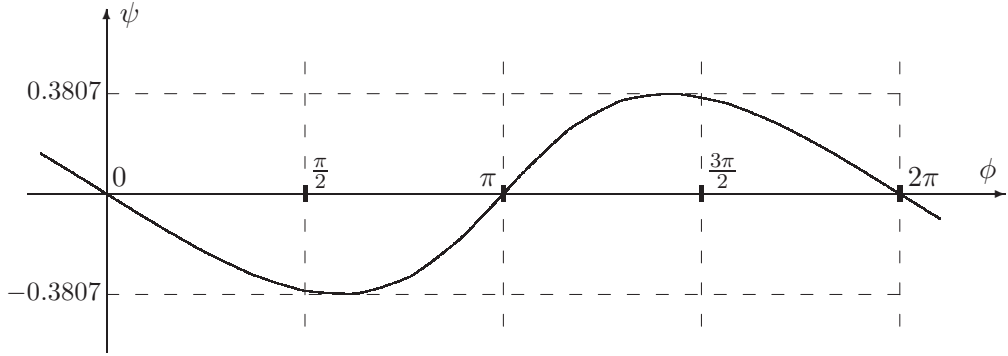
\begin{figure}[ht]
\begin{picture}(400,160)\put(50,15){\begin{picture}(360,120)\put(0,60)
{\vector(1,0){370}}\put(30,0){\vector(0,1){130}}
\qbezier(5.00,75.3)(15.0,69.3)(25.0,63.1)
\qbezier(25.0,63.1)(35.0,56.9)(45.0,50.7)
 \qbezier(45.0,50.7)(55.0,44.7)(65.0,39.1)
 \qbezier(65.0,39.1)(75.0,34.0)(85.0,29.5)
 \qbezier(85.0,29.5)(95.0,25.8)(105.,23.3)
 \qbezier(105.,23.3)(115.,22.0)(125.,22.3)
 \qbezier(125.,22.3)(135.,24.5)(145.,28.8)
 \qbezier(145.,28.8)(155.,35.4)(165.,44.2)
 \qbezier(165.,44.2)(175.,54.5)(185.,65.5)
 \qbezier(185.,65.5)(195.,75.8)(205.,84.6)
 \qbezier(205.,84.6)(215.,91.2)(225.,95.5)
 \qbezier(225.,95.5)(235.,97.7)(245.,98.0)
 \qbezier(245.,98.0)(255.,96.7)(265.,94.2)
 \qbezier(265.,94.2)(275.,90.5)(285.,86.0)
 \qbezier(285.,86.0)(295.,80.9)(305.,75.3)
\qbezier(305.,75.3)(315.,69.3)(325.,63.1)
 \qbezier(325.,63.1)(335.,56.9)(345.,50.7)\put(0,96){\small $0.3807$}
\multiput(105,57)(75,0){4}{\multiput(-0.5,0)(0.5,0){3}{\line(0,1){6}}}
\multiput(105,10)(75,0){4}{\dashline[-10]{5}(0,0)(0,100)}
\multiput(30,22)(0,76){2}{\dashline[-0]{5}(0,0)(300,0)}
\put(360,66){$\phi$}\put(35,125){$\psi$}\put(-8,20){\small $-0.3807$}
\put(107,66){$\frac{\pi}{2}$}\put(257,66){$\frac{3\pi}{2}$}
\put(32,63){$0$}\put(171,63){$\pi$}\put(333,63){$2\pi$}
\end{picture}}\end{picture}
\caption{The dependence between the exponents $\psi$ on $\phi$
for the ground state eigenvalue at $N=8$.}
\label{psiN8}
\end{figure}

In the remainder of this section we analyze
the solutions of the Bethe Ansatz equations
for the small size chains with $N\le4$. All numerical
calculations are performed for the following
case
\beq
q=\frac{1}{10},\quad \eta=\frac{7\pi}{20}. \label{numpar}
\eeq
This choice of the parameter $\eta$ is motivated by the following
considerations. First, it allows us to test
the behavior of Bethe roots near the special point $\eta=\pi/3$, since
the value \eqref{numpar} is quite close to it, $\eta-\pi/3\approx0.05$.
Second, it provides us with all advantages of the rational case
 without running into the problems
of degenerate states. Indeed, the value \eqref{numpar} corresponds to
$m=7$ and $L=10$ in \eqref{rat-eta}, while the
degenerate states can only occur for $N>L$.
Below we will apply the following procedure:
\begin{enumerate}[(i)]
\item Analytically diagonalize the transfer matrix of the symmetric
8V-model and thus determine the eigenvalues $\Ls_i(u)$,
$i=1,\ldots,2^N$.

\item Calculate the corresponding higher eigenvalues $\Ls_i^{(10)}(u)$ using
the formula \eqref{deter}.
\item Use the functional relations \eqref{tl} or \eqref{zftl} with $L=10$
to find zeroes of eigenvalues
$\Qs_\pm(u)$. For even $N$ this is straightforward since
these eigenvalues coincide, however, for odd $N$ one needs to correctly share
zeroes between $\Qs_+(u)$ and $\Qs_-(u)$.
\item Find the branching points of the eigenvalues with respect to
  the field variable $\phi$.
\item On the Riemann surface of the largest eigenvalue $\Ls_0(u)$ try
  to find paths connecting it to other eigenvalues $\Ls_i(u)$
  with the exponents from the discrete set \eqref{fi8v} of the
  symmetric 8V-model.
\end{enumerate}
The case $N=2$ is simple and does not really require the steps
(ii)-(iii).

The results are presented in tables of two types. In the first one
we list eigenvalues of operators $\Pcal$, $\Scal$, $\Rcal$ and ${\bf H}_{XYZ}$
(the latter are denoted as $\Es_i$) for each eigenvalue $\Ls_i(u)$.
Remind that we  use the normalization \eqref{xyz-norm}.
We also present there numerical values of $\Lambda_i^B=\Lambda_i^B(\pi/2)$
and $\Ls_i=\Ls_i(\pi/2)$ (remind that $\Lambda^B(v)$ and
$\Ls(u)$ are related by \eqref{num2}).
The second type tables contain the Bethe
roots and the values of the phases $n_k$ in \eqref{lba}.

Note, that here we did not attempt to diagonalize the infinite
dimensional transfer matrices of the SOS-model at arbitrary field
and compare the results against the corresponding exact
expressions for the eigenvalues. We hope to address this question in
the future.

\subsection{The case $N=2$}
This case is very
simple and all eigenvalues can be calculated analytically. However,
it is quite illustrative and reveals some general properties of the
Bethe Ansatz equations (\ref{lba}).

Four eigenvalues of the transfer-matrix ${\bf T}_B(v)$ for $N=2$
are given by
\beq
{\bf\Lambda}^B_0=2ab+c^2+d^2,\quad {\bf\Lambda}^B_1=a^2+b^2+2cd,\quad
{\bf\Lambda}^B_2=a^2+b^2-2cd,\quad {\bf\Lambda}^B_3=2ab-c^2-d^2.
\eeq
Corresponding eigenvalues of the Hamiltonian are given by
\beq
\Es_0=-J_x-J_y+J_z,\quad \Es_1=-J_x+J_y-J_z,\quad
\Es_2=J_x-J_y-J_z,\quad \Es_3=J_x+J_y+J_z.
\eeq
For the symmetric 8V-model with even $N$ the eigenvalues
${\Qs}_i^+(u)$ and ${\Qs}_i^-(u)$  coincide,  therefore there is no
need to distinguish them.
Their analytic expressions read
\bea
&&\Qs_0(u)=\t_1(u-\frac{\pi}{2}-\frac{\pi\tau}{2}|q^2)
\t_1(u-\frac{\pi}{2}+\frac{\pi\tau}{2}|q^2)=
\frac{\t_2(0|q)}{2}e^{-i\frac{\pi\tau}{4}}\t_3(u|q)\\
&&\Qs_1(u)=e^{iu}\,\t_1(u-\frac{\pi}{2}+\pi\tau|q^2)
\t_1(u-\frac{\pi}{2}|q^2)=
\frac{\t_2(0|q)}{2}e^{-i\frac{\pi\tau}{2}}\t_2(u|q)\\
&&\Qs_2(u)=e^{iu}\,\t_1(u+\pi\tau|q^2)\t_1(u|q^2)=i
\frac{\t_2(0|q)}{2}e^{-i\frac{\pi\tau}{2}}\t_1(u|q)\\
&&\Qs_3(u)=\t_1(u-\frac{\pi\tau}{2}|q^2)\t_1(u+\frac{\pi\tau}{2}|q^2)=
\frac{\t_2(0|q)}{2}e^{-i\frac{\pi\tau}{4}}\t_4(u|q).
\eea
The descriptive properties of the eigenvalues are given
Table \ref{tableEig2}.
\begin{table}[ht]
\begin{center}\begin{tabular}{|c|c|c|c|c|c|c|}\hline
$i$&$\Es_i$&${\bf\Lambda}_i^{B^{\phantom{i}}}$&${\Ls}_i$&$\Pcal$&$\Scal$&
$\Rcal$\\ \hline $0$ & $-2.02064769$ & $1.33767166$ & $2.06595956$ & $1$ &
$-1$&$1$\\ \hline $1$&$0.02064769$&$0.66200143$&$-0.04265730$&$1$&$1$&$1$
\\ \hline $2$&$0.84245604$&$0.38998228$&$-0.89156871$&$1$&$1$&$-1$\\ \hline
$3$&$1.15754396$&$-0.28568795$&$1.21704816$&$-1$&$-1$&$-1$\\ \hline
\end{tabular}\end{center}\caption{Descriptive properties of the
eigenvalues for $N=2$.}\label{tableEig2}
\end{table}
Table \ref{tableN2}
contains the Bethe roots, the values of the field parameter
$\phi=\varphi_+=\varphi_-$ and
the phases $n_k$ for all eigenvalues.
\begin{table}[ht]\begin{center}
\begin{tabular}{|c|c|c|c|c|}\hline
$i$  & $\a_1$ &$\a_2$&$\phi$ &$n_1,n_2$\\ \hline
$0$  &$i{\pi\tau}{/2}$ & $-i{\pi\tau}{/2}$ &$0$ &$\{-1/2,1/2\}$\\ \hline
$1$ &$i\pi\tau$ & $0$ &$\pi$&$\{-1/2,1/2\}$\\ \hline
$2$  &$i\pi\tau-i\pi/2$ & $-i\pi/2$ &$\pi$&$\{-1/2,1/2\}$ \\ \hline
$3$  &$i\pi\tau/2-i\pi/2$ & $-i\pi\tau/2-i\pi/2$ &$0$ &$\{-3/2,-1/2\}$\\ \hline
\end{tabular}\end{center}\caption{Roots and phases for $N=2$.}\label{tableN2}
\end{table}

Despite being entire functions
of the spectral parameter $u$,
the eigenvalues $\Ls_i(u)$ and $\Qs_i(u)$ are multivalued
functions of the field variable $\phi$. Actually, for $N=2$
their analytic properties are relatively simple since they are
determined by the properties of only one Bethe root $\a_1(\phi)$.
Indeed, with an account of \eqref{halfalpha}
the Bethe Ansatz equations \eqref{lba}
reduce to a single equation
\beq
2\Phi_1(\a_1)=\frac{2\eta}{\pi^2}(\phi+\pi)-(n_1+\hf)\label{n2ba},
\eeq
which defines the function $\a_1(\phi)$.  Its  branching points
are determined by the condition (\ref{odedet}), which in
this case reduces to
\beq
\Phi_1'(\a_1)=0.\label{n2bp}
\eeq
Solving this equation for $\a_1$ and substituting the result
into \eqref{n2ba} one gets potential locations of the branching points
$\phi^{(br)}$ in the $\phi$-plane.
\beq
\phi^{(br)}=\pm\phi_0+2\pi m +\frac{\pi^2}{2\eta} n,
\qquad m,n\in\Zbbd
\eeq
where $\Phi_1'(\a_1(\phi_0))=0$. No explicit analytic expressions for
$\phi_0$ and $\a_1(\phi_0)$ are available. Their asymptotic expansions
for small $\q$
read
\beq
\phi_0=\pi+\frac{2\pi i
  \sin2\eta}{\eta}\
\Big[\,\q+\q^3\Big(\,\frac{5}{3}+\frac{\cos 4\eta}{3}\,\Big)
+O(\q^5)\Big],\qquad \q\to0\label{phi-q0},
\eeq
\beq
\a_1(\phi_0)=i\pi\tau-\frac{i\pi}{4}
-i\q \cos 2\eta +i\q^3 \left(\frac{3}{2}\cos 2\eta-\frac{1}{6}\cos
6\eta\right)+O(\q^5),\qquad \q\to0\ .\label{al-q0}
\eeq
The numerical values
\beq
\phi_0=\pi+0.46954959i,\qquad \a_1(\phi_0)=-0.727625178i+i\pi\tau.
\eeq
for the case \eqref{numpar} are well approximated by the asymptotic
formulae \eqref{phi-q0} and \eqref{al-q0}.

Fig.\ref{fig3} shows two sheets of the Riemann surface
of the function
$\a_1(\phi)$ corresponding to the ground state eigenvalue $\Ls_0(u)$.
The principal
sheet is defined by the condition $\a_1(0)=i\pi\tau/2$.
This sheet
contains two cuts $(\phi_0,\pi+i\infty)$ and $(\phi_0^*,\pi-i\infty)$
in the strip $0<\mbox{Re}\,\phi<2\pi$ (they are shown by solid lines).
The same pattern is repeated with the period $2\pi$.
The second sheet can be reached from the first one by crossing
either of the solid line cuts in Fig.\ref{fig3} from left to right.
This sheet contains two new branch cuts (shown by dashed lines) in
addition to the same cuts as on the first sheet.

To demonstrate that all the solutions of Bethe Ansatz
equations for $N=2$ can be
obtained from each other by an analytic continuation in $\phi$ we
performed numerical calculations using the differential equations
(\ref{odeal}). For $N=2$ it is a single equation
\beq
\frac{d\a_1}{d\phi}=\frac{\eta}{\pi^2\Phi_1'(\a_1)}
\eeq
where $\Phi'_1(\a)$ is given by \eqref{phi13a}.

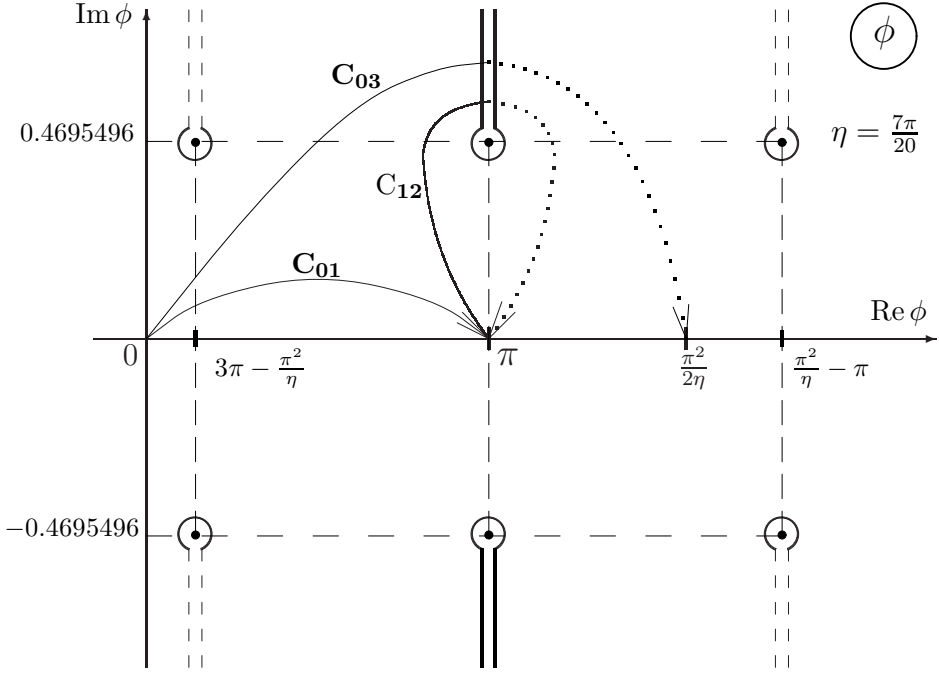
\begin{figure}[ht]
\setlength{\unitlength}{0.35mm}
\begin{picture}(370,265)\put(10,-20){
\begin{picture}(330,270)
\put(200,146){\allinethickness{0.3mm}\line(0,1){8}}\put(50,150)
{\vector(1,0){320}}\put(275,146){\allinethickness{0.3mm}\line(0,1){8}}
\put(70,25){\vector(0,1){250}}\put(61,140){\large$0$}\put(203,140)
{\Large$\pi$}\put(272,136){$\frac{\pi^2}{2\eta}$}
\multiput(197.5,230)(5,0){2}
{\Thicklines\line(0,1){45}}\multiput(0,0)(0,-205){2}
{\multiput(0,0)(222.8,0){2}{\multiput(86.1,230)(5,0){2}
{\dashline{5}(0,0)(0,45)}}}
\multiput(-111.4,0)(111.4,0){3}{
\put(199.8,224.3){\arc{13}{-1.2}{4.35}}
\put(199.8,224.24){\arc{12.5}{-1.2}{4.35}}
\put(199.8,224.18){\arc{12}{-1.2}{4.38}}\put(199.8,75.7){\arc{13}{-4.35}{1.2}}
\put(199.8,75.76){\arc{12.5}{-4.35}{1.2}}
\put(199.8,75.82){\arc{12}{-4.38}{1.2}}}
\spline(70,150)(90,165)(135,175)(180,165)(200,150)\path(200,150)(190,160)
\path(200,150)(188,156)\path(200,150)(210,160)\path(200,150)(205,164)
 \qbezier(200,240)(177,238)(175,220) \qbezier(175,220)(177,180)(200,150)
{\allinethickness{0.3mm} \qbezier[8](200,240)(223,238)(225,220)
 \qbezier[15](225,220)(223,180)(200,150)}
\multiput(-111.4,0)(111.4,0){3}{\put(200,224.6){\circle*{3}}}
\spline(70,150)(110,200)(150,240)(180,253)(200,255)\path(275,150)(276,163)
{\allinethickness{0.3mm} \qbezier[25](200,255)(260,250)(275,150)}
\multiput(0,0)(0,-150){2}{\drawline[-50](70,225)(311.4,225)}
\multiput(0,0)(111.4,0){3}{\drawline[-30](88.6,225)(88.6,75)}
\multiput(197.5,70)(5,0){2}{\Thicklines\line(0,-1){45}}
\path(275,150)(270,162)\put(16,75){\small $-0.4695496$}
\put(96,136){\small$3\pi-\frac{\pi^2}{\eta}$}
\put(315,136){\small$\frac{\pi^2}{\eta}-\pi$}
\multiput(88.6,147)(222.8,0){2}{\Thicklines\line(0,1){6}}
\multiput(-111.4,0)(111.4,0){3}{\put(200,75.4){\circle*{3}}}\put(125,175)
{$\bf C_{01}$}\put(157,205){$\bf \mbox{C}_{12}$}\put(140,247){$\bf C_{03}$}
\put(22,225){\small $0.4695496$}\put(350,265)
{\allinethickness{0.3mm}\circle{25}}
\put(346,263){\bf\Large$\phi$}\put(330,225){\bf\large$\eta=\frac{7\pi}{20}$}
\put(345,157){Re$\,\phi$}\put(43,270){Im$\,\phi$}\end{picture}}\end{picture}
\caption{The principal sheet of the Riemann surface which
contains the eigenvalue ${\Ls_0}$ for $N=2$.}\label{fig3}
\end{figure}
The results are illustrated in Fig.\ref{fig3} by three contours
$C_{01}$, $C_{12}$ and $C_{03}$, which connect corresponding
eigenvalues (we used the definition (\ref{Ccont})).
The contour $C_{01}$ goes between points on the first sheet while the
contours $C_{12}$ and $C_{03}$
start on the first and end on the second sheet.
Note that the ending point of the contour $C_{03}$ is $\pi^2/(2\eta)$
rather than $0$ as it should be according to the Table \ref{tableN2}.
As explained above, this does not affect the Bethe roots, but changes
the sign of the eigenvalue as in (\ref{sign}),
\beq
{\Ls}_0\stackrel{C_{03}}\longrightarrow\,-{\Ls}_3.
\eeq
Apparently it is possible to find a different contour $C_{03}$ ending
at the correct point $\phi=0$ on some other sheet of the same  Riemann
surface. We leave this as an exercise to the reader.

Note that in the trigonometric limit $\q=0$ the vertical distance
between upper and lower cuts vanishes
and neighboring cuts join together.
In the opposite low temperature limit
$q=1$ all cuts move to infinity and disappear.
In both cases the Riemann surface of the eigenvalues
splits into disjoint components, though in two different ways.
In particular, for $\q=0$ the connected parts of the
Riemann surface no longer possess
the $2\pi$-periodicity in $\phi$.

\subsection{\bf The case $N=3$.}

This case is quite important and will be analyzed in more details.
When $N$ is odd all eigenvalues of the transfer matrix of the 8V-model
are doubly degenerated. So altogether there are four different
eigenvalues. Two of them have the momentum $\mathcal{P}=1$, while for the
other two $\mathcal{P}=\omega^{\pm1}$, $\omega=e^{2\pi i/3}$.

A direct diagonalization of the transfer-matrix leads to the following
analytic expressions for the two eigenvalues of ${\bf T}_B(v)$
with $\mathcal{P}=1$,
\beq
{\bf\Lambda}^B_{0,1}(v)=\t_1^3(2\eta|\q)\frac{\t_1(v|\q)
\t_1(v+v_{0,1}|\q)\t_1(v-v_{0,1}|\q)}
{\t_1(\eta|\q)\t_1(\eta+v_{0,1}|\q)\t_1(\eta-v_{0,1}|\q)}
\eeq
where zeros $v_{0,1}$ satisfy
the following transcendental equations
\beq
\frac{\vt_4^2(0|\q)}{\vt_3^2(0|\q)}
\frac{\vt_3^2(v_{0,1}|\q)}{\vt_4^2(v_{0,1}|\q)}
=\frac{1+7\zeta^2(\g-1)+3\g-2\zeta\Bigl[1+
\g\pm4\sqrt{\g^2-\zeta\g(\g+1)+\zeta^2(1-\g+\g^2)}\Bigr]}
{(\g\zeta-1)[\zeta(3\g+5)-3\g-1]},\label{N3trans}
\eeq
with the constants $\zeta$ and $\g$ defined in (\ref{zetagam}).
For the choice (\ref{numpar}) its solutions read\footnote{%
Note that Eq.(\ref{N3trans}) for $\eta=\pi/3$ has a
solution $v_0=0$, corresponding to the simple ground state eigenvalue
${\bf\Lambda}^B_{0}(v)=\t_1(v|\q)^N$, considered in \cite{BM05,BM06a}.}
\beq
v_0=0.3749001333,\quad
v_1=\frac{\pi}{2}+0.29695326\,i\label{zerosL1}
\eeq

\begin{table}[ht]
\begin{center}
\begin{tabular}{|c|c|c|c|c|c|}\hline
$i$ & $\Es_i$ & ${\bf\Lambda}_i^{B^{\phantom{i}}}$ &
${\Ls}_i$ & $\Pcal$ & $\Scal$ \\ \hline
$0$ & $-1.85546785$ & $1.11943180$ & $1.40606193$ & $1$ &
$\pm1$ \\ \hline
$1$ & $0.69792389$ & $0.25256796$ & $-1.45228936$ &
$1$ & $\pm1$ \\ \hline
$2$ & $0.57877198$ & $-0.14650972$ & $0.65945338$ &
$\omega$ & $\pm1$ \\ \hline
$3$ & $0.57877198$ & $-0.14650972$ & $0.65945338$ &
$\omega^{{-1}^{\phantom{|}}}$ & $\pm1$ \\ \hline
\end{tabular}
\end{center}
\caption{Descriptive properties of the eigenvalues for $N=3$.}
\label{table3}
\end{table}

The remaining two eigenvalues
\beq
{\bf\Lambda}^B_{2,3}(v)=
\frac{\t_1(2\eta|\q)\t_1(v-2\eta|\q)\t_1(v+\eta|\q)^2}{\t_1(\eta|\q)}-
\omega^{\mp
  1}\frac{\t_1(2\eta|\q)^2\t_1(\eta|\q)\t_1(2v|\q)}{\t_1(v|\q)}.
\label{N3eig3}
\eeq
correspond to
$\mathcal{P}=\omega^{\pm1}$.
Their numerical values in Table \ref{table3}
are real and coincide with each other
only due to the choice of the symmetric point $v=\pi/2$ where the second
term in (\ref{N3eig3}) vanishes. For a generic real $v$
these eigenvalues are complex and non-degenerate.

\subsubsection{The eigenvalue ${\Ls}_0$}

Consider the ground state eigenvalue ${\Ls}_0$.
The numerical calculation with Eq.\eqref{tl} referred to in the step
(iii) in Sect.\ref{sect-over}
above require a separation
of zeroes between $\Qs^+(u)$ and $\Qs^-(u)$. Since the total number
of zeroes, $2N=6$, is small we used a simple trial and error method.
Taking an arbitrary subset of 3 zeroes for $\Qs^+(u)$ and substituting
the product \eqref{wei} into the TQ-equation \eqref{TQp} one
gets a relation which must be fulfilled
identically in the variable $u$ at some
(yet undermined) value of $\varphi_+$. If the subset of zeroes is chosen
incorrectly then no such value exist and the process should be repeated.

The numerical results (with the choice (\ref{numpar}))
are collected in Table~\ref{table4} and plotted in Fig.\ref{fig4}.
\begin{table}[bt]
\begin{center}
\begin{tabular}{|c|c|c|c|c|c|}\hline
 & \multicolumn{3}{c|}{\mbox{Zeros in} $u$} & {$\phi_\pm$}
 &$n_1,n_2,n_3$ \\
\hline
 $\Qs^{+^{\phantom{I}}}_0(u)$ &
 ${\ds\frac{\pi}{2}-\frac{\pi\tau}{2}}^{\phantom{I}}_{\phantom{I_I}}$ &
 ${\ds\frac{\pi}{2}+\frac{\pi\tau}{2}-0.42746i}$&
 ${\ds\frac{\pi}{2}+\frac{\pi\tau}{2}+0.42746i}$&
 ${\ds -\frac{\pi}{2}}$&$\{-1,0,1\}$\\ \hline
 $\Qs^{-^{\phantom{I}}}_0(u)$ &
 ${\ds\frac{\pi}{2}-\frac{\pi\tau}{2}-0.42746i}$&
 ${\ds\frac{\pi}{2}-\frac{\pi\tau}{2}+0.42746i}$&
 ${\ds\frac{\pi}{2}+\frac{\pi\tau}{2}}^{\phantom{I}}_{\phantom{I_I}}$ &
 ${\ds +\frac{\pi}{2}}$&$\{-1,0,1\}$\\ \hline
 ${\Ls}_0(u)$ &
 ${\ds\pm\frac{\pi\tau}{2}}^{\phantom{I}}_{\phantom{I_I}}$ &
 ${\ds\pm\frac{\pi\tau}{2}-0.374900}$ &
 ${\ds\pm\frac{\pi\tau}{2}+0.374900}$ &
 \multicolumn{2}{}{}
  \\ \cline{1-4}
 ${\Ls}_0^{(3)}(u)$ &
 ${\ds\pm\frac{\pi\tau}{2}}^{\phantom{I}}_{\phantom{I_I}}$ &
 ${\ds\pm\frac{\pi\tau}{2}-1.068576}$ &
 ${\ds\pm\frac{\pi\tau}{2}+1.068576}$ &
 \multicolumn{2}{}{}
  \\ \cline{1-4}
\end{tabular}
\end{center}
\begin{center}
\caption{Zeros of the ground state eigenvalues for $N=3$.}\label{table4}
\end{center}
\end{table}
Each of the eigenvalues $\Qs^\pm(u)$ has exactly $N$ zeros in
the periodicity rectangular
\beq
0\leq\mbox{Re}(u)\leq\pi,\quad
-\pi|\tau|\leq\mbox{Im}(u)\leq\pi|\tau|.\label{fundr}
\eeq
Any of the eigenvalues ${\Ls}^{(k)}(u)$
has $2N$ zeros in the same domain (their imaginary period is $\pi\tau$).
The corresponding Bethe roots are
\beq
\a_1^+=-\frac{1}{2}{\pi|\tau|},\quad\a_2^+=\frac{1}{2}{\pi|\tau|}-q_0,
\quad\a_3^+=\frac{1}{2}{\pi|\tau|}+q_0,
\quad q_0=0.427465646.\label{n3roots}
\eeq
and the field parameter $\phi_+=-\pi/2$.
The roots $\a_k^-$ can be obtained from $\a_k^+$ by either
a reflection $\a_k^-=-\a_{N-k}^+$ or a shift
\beq
\a_k^-=\a_{k+1}^+ -\pi|\tau|,\qquad k=1,\ldots,N-1,\qquad
\a_N^-=\a_1^++\pi|\tau|\ .\label{qpmshift}
\eeq
Note also that
the sets $\{\a_k^+\}$ and $\{\a_k^-\}$
precisely interlace each other (see, e.g., Fig.\ref{fig9}).
The above remarks relating zeroes of the ground state eigenvalues
$\Qs_0^+(u)$ and $\Qs_0^-(u)$  apply for arbitrary odd $N$.

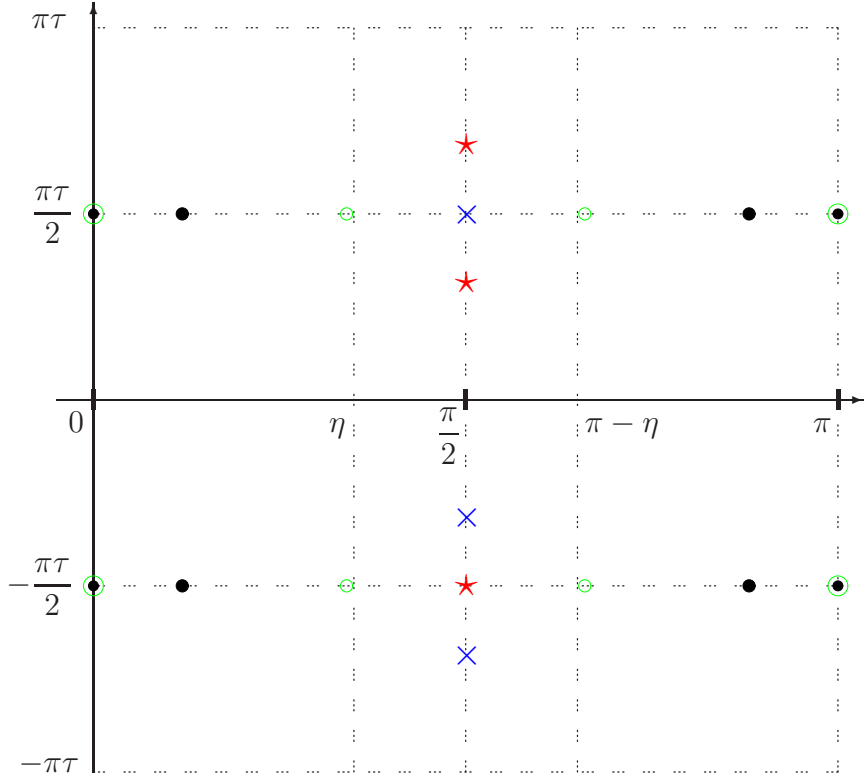
\begin{figure}[ht!]\setlength{\unitlength}{0.33mm}\begin{picture}(340,320)
\put(50,10){\begin{picture}(340,270)
\put(10,150){\vector(1,0){325}}\put(25,0){\vector(0,1){310}}
\multiput(25,146)(150,0){3}{{\allinethickness{0.5mm}\line(0,1){8}}}
\multiput(25,0)(0,75){5}{\dashline[-10]{5}[2](0,0)(300,0)}
\multiput(175,0)(150,0){2}{\dashline[-10]{5}[2](0,0)(0,300)}
\multiput(130,0)(90,0){2}{\dashline[-10]{5}[2](0,0)(0,300)}
\put(163,133){\large$\ds\frac{\pi}{2}$}
\put(15,137){\large $0$}\put(315,137){\large $\pi$}
\put(0,222){\large$\ds\frac{\pi\tau}{2}$}
\put(-10,72){\large$-\ds\frac{\pi\tau}{2}$}
\put(-5,0){\large$-\ds{\pi\tau}$}\put(0,300){\large$\ds{\pi\tau}$}
\put(120,137){\large$\ds\eta$}\put(223,137){\large$\ds\pi-\eta$}
\put(169.6,221){\color{blue}\Large$\times$}
\multiput(169.6,43.2)(0,55.6){2}{\color{blue}\Large$\times$}
\put(170.8,71){\color{red}\LARGE$\star$}
\multiput(170.8,193.2)(0,55.6){2}{\color{red}\LARGE$\star$}
\multiput(0,0)(0,-150){2}{\multiput(25,225)(300,0){2}{\circle*{4}}
\multiput(60.8,225)(228.4,0){2}{\circle*{5}}}\multiput(0,0)(0,-150){2}
{\multiput(25,225)(300,0){2}{\color{green}\circle{8}}
\multiput(127,225)(96,0){2}{\color{green}\circle{5}}}
\end{picture}}\end{picture}
\caption{Zeros of $\Qs^+_0(u)$, $\Qs^-_0(u)$,
${\Ls}_0(u)$, ${\Ls}_0^{(3)}(u)$
for $N=3$, marked by {\color{red}\Large$\bf\star\>$},
{\color{blue}\large$\times\>$},
{\large$\bullet\>$},
{\color{green}\large$\circ\>$}, accordingly.}\label{fig4}\end{figure}

\subsubsection{The eigenvalue ${\Ls}_1$.}

This is the smallest eigenvalue for $N=3$ (it has the largest
eigenvalue of the Hamiltonian).
The numerical zeroes (for the case \eqref{numpar}) are presented in
Table \ref{table5} and plotted in Fig.\ref{fig5}. As clearly seen
from the picture the zeroes of $\Qs^+_1(u)$ and
$\Qs^-_1(u)$ form solitary 3-strings.
\begin{table}[ht]\begin{center}\begin{tabular}{|c|c|c|c|c|c|}\hline
&\multicolumn{3}{c|}{\mbox{Zeros in $u$}}&{$\phi_\pm$}&$n_1,n_2,n_3$\\\hline
$\Qs^{+^{\phantom{I}}}_1(u)$&${\ds\frac{\pi}{2}+\frac{\pi\tau}{2}+0.90837673}$&
${\ds\frac{\pi}{2}+\frac{\pi\tau}{2}}^{\phantom{I^I}}_{\phantom{I_I}}$&
${\ds \frac{\pi}{2}+\frac{\pi\tau}{2}-0.90837673}$&${\ds -\frac{3\pi}{2}}$&
$\{-1,0,1\}$\\ \hline$\Qs^{-^{\phantom{I}}}_1(u)$&
${\ds\frac{\pi}{2}-\frac{\pi\tau}{2}+0.90837673}$&
${\ds\frac{\pi}{2}-\frac{\pi\tau}{2}}^{\phantom{I^I}}_{\phantom{I_I}}$&
${\ds \frac{\pi}{2}-\frac{\pi\tau}{2}-0.90837673}$&${\ds +\frac{3\pi}{2}}$&
$\{-1,0,1\}$\\ \hline ${\Ls}_1(u)$ &
${\ds\frac{\pi}{2}\pm\frac{\pi\tau}{2}+0.29655326i}$ &
${\ds\pm\frac{\pi\tau}{2}}^{\phantom{I}}_{\phantom{I}}$ &
${\ds\frac{\pi}{2}\pm\frac{\pi\tau}{2}-0.29655326i}$& \multicolumn{2}{}{}
\\ \cline{1-4} ${\Ls}_1^{(3)}(u)$ &${\ds\pm\frac{\pi\tau}{2}+1.10116874}$ &
${\ds\pm\frac{\pi\tau}{2}}^{\phantom{I}}_{\phantom{I}}$ &
${\ds\pm\frac{\pi\tau}{2}-1.10116874}$ & \multicolumn{2}{}{}\\ \cline{1-4}
\end{tabular}\end{center}
\caption{Zeros for the smallest eigenvalue at $N=3$.}\label{table5}\end{table}

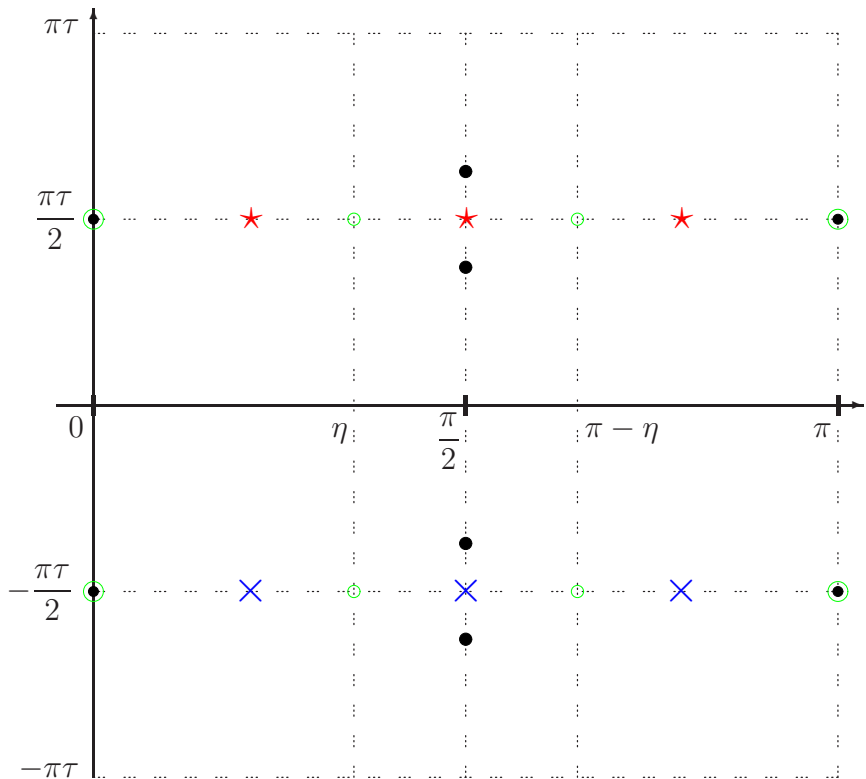
\begin{figure}[ht]
\setlength{\unitlength}{0.33mm}\begin{picture}(340,320)
\put(50,10){\begin{picture}(340,310)\put(10,150){\vector(1,0){325}}
\multiput(25,146)(150,0){3}{{\allinethickness{0.5mm}\line(0,1){8}}}
\multiput(25,0)(0,75){5}{\dashline[-10]{5}[2](0,0)(300,0)}
\multiput(175,0)(150,0){2}
{\dashline[-10]{5}[2](0,0)(0,300)}
\multiput(130,0)(90,0){2}{\dashline[-10]{5}[2](0,0)(0,300)}
\put(163,133){\large$\ds\frac{\pi}{2}$}\put(25,0)
{\vector(0,1){310}}\put(15,137){\large $0$}
\put(315,137){\large $\pi$}\put(,222)
{\large$\ds\frac{\pi\tau}{2}$}\put(-10,72)
{\large$-\ds\frac{\pi\tau}{2}$}\put(-5,0){\large$-\ds{\pi\tau}$}\put(5,300)
{\large$\ds{\pi\tau}$}\put(121,137){\large$\ds\eta$}\put(223,137)
{\large$\ds\pi-\eta$}
\multiput(84.2,221.2)(86.7,0){3}{\color{red}\LARGE$\star$}
\multiput(81.2,70.6)(86.7,0){3}{\color{blue}\LARGE$\times$}
\multiput(0,0)(0,-150){2}{\multiput(25,225)(300,0){2}{\circle*{4}}
\multiput(175,205.7)(0,38.6){2}{\circle*{5}}}\multiput(0,0)(0,-150){2}{
\multiput(25,225)(300,0){2}{\color{green}\circle{8}}
\multiput(130,225)(90,0){2}{\color{green}\circle{5}}}
\end{picture}}\end{picture}
\caption{Zeros of $\Qs^+_1(u)$, $\Qs^-_1(u)$, ${\Ls}_1(u)$, ${\Ls}_1^{(3)}(u)$
for $N=3$, marked by {\color{red}\Large$\bf\star\>$}, {\color{blue}\large$\times\>$},
{\large$\bullet\>$},
{\color{green}\large$\circ\>$}, accordingly.}\label{fig5}
\end{figure}

It is not difficult to check that these solutions of the Bethe
Ansatz equations are
connected with each other by an analytic
continuation in $\phi$. Consider the path
from $\phi=3\pi/2$ to $\phi=-3\pi/2$ along the real axis
of $\phi$ (with a small imaginary part for a better convergence).
It turns out that the solution $\{\a_1^-,\a_2^-,\a_3^-\}$ continued
along this path transforms into another solution
\beq
\{\widetilde\a^+_k\}=\{\a_1^+-4\pi|\tau|,\a_2^++2\pi|\tau|,\a_3^++2\pi|\tau|\},
\label{modsec2}
\eeq
where $\a^+_i$, $i=1,2,3$ correspond to zeros of $\Qs^+_1(u)$ from
the Table \ref{table5}. Remind that $\a_k$ and $u_k$ are related by
\eqref{u-alpha}. It is easy to check that shifts by $2\pi k|\tau|$
in \eqref{modsec2}  can be compensated by a proper choice of $n_k$ without
changing the field $\phi_+=-3\pi/2$.
Table \ref{table5} contains the resulting values of $n_k$
corresponding to the roots
$\{\a_1^+,\a_2^+,\a_3^+\}$.

The problem of the analytic continuation between different
eigenvalues of the transfer matrix is much more complicated.
The most difficult part of this procedure
is a numerical study of the Riemann surface of the eigenvalues.
For $N\ge3$ the difficulty lies in the fact that the branching
condition \eqref{oderank}
just defines an $(N-1)$-dimensional hyper-surface in the space of Bethe
roots which, by itself,
does not correspond to any particular value of the field $\phi$.
To obtain the structure of the branching points in the $\phi$-plane
one needs to determine
how this hyper-surface intersects with required solutions of the
Bethe Ansatz equations.
In practice we used the differential equations \eqref{odeal} and
numerically studied monodromy properties of the
solutions when $\phi$ was varied around
random loops in the complex plane.

Remarkably, we found that the largest and smallest eigenvalues,
$\Ls_0(u)$ and $\Ls_1(u)$, do indeed correspond to different branches
of the same (multivalued) function of $\phi$.
The contour $C_{01}$ in the complex $\phi$-plane
connecting these eigenvalues is shown in
Fig.\ref{fig6}. As evident from the figure,
the structure of branching cuts is extremely complicated.
This figure shows the (upper half-plane) cuts
on just two sheets of the corresponding
Riemann surface. The first sheet (solid line cuts) contains the eigenvalue
$\Qs_0^-(u)$ at $\phi=\pi/2$, while the second one (dashed line cuts)
contains the eigenvalue $\Qs_1^-(u)$ at $\phi=3\pi/2$.
Note that the contour $C_{01}$ leaves the first sheet
when it goes under the cut at ${\rm Re}\,\phi=0$ and arrives
to the second of these sheets from under the (dashed line) cut at
${\rm Re}\,\phi=\pi-\a$. In between these two points the contour crosses
additional cuts on other sheets of the Riemann surface, which are not
shown in the picture.

\begin{figure}[ht!]\centering
\setlength{\unitlength}{0.45mm}
\begin{picture}(350,180)
\put(0,-15){\begin{picture}(350,200)
\put(50,50){\vector(1,0){270}}\multiput(0,0)(105,0){3}{\put(70,45)
{\line(0,1){17}}}\multiput(0,0)(105,0){3}{\multiput(68,65.6)(4,0){2}
{\dashline[10]{3}(0,0)(0,25)}\put(70,63){\arc{6}{-1.0}{4.1}}\put(70,63)
{\circle*{2}}}\multiput(0,30)(105,0){3}{\multiput(68,65.6)(4,0){2}
{\line(0,0){75}}\put(70,63){\arc{6}{-1.0}{4.1}}\put(70,63){\circle*{2}}}
\multiput(10,40)(105,0){2}{\multiput(68,65.6)(4,0){2}
{\dashline[10]{3}(0,0)(0,65)}
\put(70,63){\arc{6}{-1.0}{4.1}}\put(70,63){\circle*{2}}
\dashline[10]{3}(70,10)(70,62)}
\multiput(95,40)(105,0){2}{\multiput(68,65.6)(4,0){2}
{\dashline[10]{3}(0,0)(0,65)}
\put(70,63){\arc{6}{-1.0}{4.1}}\put(70,63){\circle*{2}}
\dashline[10]{3}(70,10)(70,62)}
\multiput(52.5,55)(105,0){2}{\multiput(68,65.6)(4,0){2}{\line(0,0){50}}
\put(70,63){\arc{6}{-1.0}{4.1}}\put(70,63){\circle*{2}}
\dashline[10]{3}(70,0)(70,63)}
\dashline[10]{3}(55,63)(280,63)\dashline[10]{3}(55,93)(280,93)
\put(300,55){\mbox{Re}$\,\phi$}
\dashline[10]{3}(55,103)(270,103)\dashline[10]{3}(55,118)(225,118)
\multiput(70,47)
(52.5,0){5}{{\allinethickness{0.4mm}\line(0,1){6}}}
\put(300,170){\allinethickness{0.3mm}
\circle{22}}{\thicklines \qbezier(85,63)(100,55))(122,50)
 \qbezier(85,63)(55,80)(68,115)
 \qbezier[55](72,124)(125,210)(165,125) \qbezier[20](165,125)(173,110)(168,75)
 \qbezier[10](168,75)(168.5,57.5)(175,57) \qbezier[5](175,57)(182,57.5)(185,61)
 \qbezier[20](185,61)(200,78)(227,50)}
\put(203,67.5){\arr{-7.72741}{-2.07055}{-7.72741}{2.07055}}
\put(65,177){\mbox{Im}$\,\phi$}
\put(145,156){\arr{-6.99696}{3.87848}{-4.1203}{6.85734}}
\dashline[10]{3}(55,167)(123,167)
\put(27,65){\small $0.0855202$}\put(30,120){\small $0.858157$}
\put(50,169){\small $2.0$}
\put(25,87){\small $0.5238261$}\put(25,105){\small $0.6874681$}
\put(65,38){\large$0$}
\put(120,35){$\ds\frac{\pi}{2}$}\put(172,38){\large$\pi$}
\put(222,35){$\ds\frac{3\pi}{2}$}
\put(275,38){\large$2\pi$}\put(70,30){\small$\a=0.070359$}
\put(80,36){\vector(0,1){10}}
\put(150,42){\small$\pi-\a$}\put(183,42){\small$\pi+\a$}
\put(250,42){\small$2\pi-\a$}
\put(300,168){\bf\Large$\phi$}\put(195,72){$\bf \mbox{C}_{01}$}
\end{picture}}\end{picture}
\caption{Structure of the cuts on the principal sheet Riemann surface
containing ${\Ls}_0$ (solid lines) and on the sheet containing
${\Ls}_1$ (dashed lines). Only the upper half-plane is shown as the
arrangement of the cuts is reflection-symmetric with respect to the
real axis of $\phi$}\label{fig6}
\end{figure}
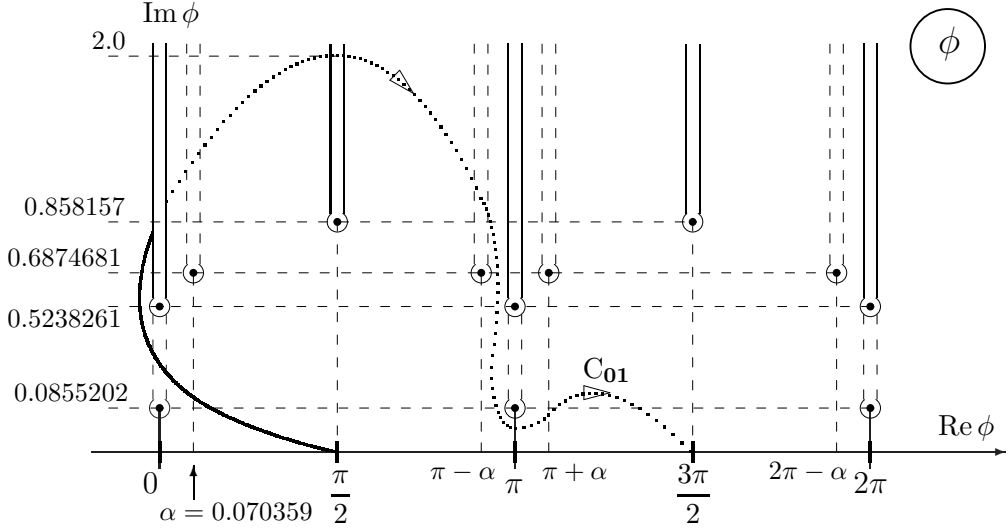

Note that in the limit $\eta\to\pi/3$ the zeroes of $\Qs^\pm_1(u)$ form
complete 3-strings which cancel out from
the TQ-relation. As a result one gets
\beq
{\Ls}_1(u)=-\t_4^3(u+\frac{\pi}{3}|\q)-\t_4^3(u-\frac{\pi}{3}|\q),
\qquad \eta=\frac{\pi}{3}.
\eeq
The minus signs arise from
the phase factors $\exp(\pm2i\eta\phi/\pi)=-1$ at $\phi=3\pi/2$.
When $\eta\to\pi/3$ the parameter $\a$ tends to zero
(this parameter determines positions of several
cuts in Fig.\ref{fig6}). In addition to that
the dashed line cut with $\mbox{Re}\,\phi=\pi$ and
its symmetric reflection in the lower half plane
both approach the real axes.
Therefore, the contour $C_{01}$ gets pinched twice
in both vertical and horizontal directions.
As a result
the complete string solution corresponding to
${\Ls}_1(u)$ for $\eta=\pi/3$ is no longer analytically connected
to the ground state eigenvalue ${\Ls}_0(u)$.

\subsubsection{The eigenvalues ${\Ls}_{2,3}$}

It is enough to consider
only one eigenvalue ${\Ls}_2$ with $\mathcal{P}=\omega$,
since all results for ${\Lambda}_3$ can be obtained by a complex conjugation.
The analytic formula for ${\bf\Lambda}_2^B(v)$ is given in (\ref{N3eig3}).
The Bethe roots and phases for $\Qs^-_2(u)$ read
\beq
\{\a_1^-,\a_2^-,\a_3^-\}=\{-0.24023993,-0.45552631\pm0.75825341\,i\}
,\quad \{n_1,n_2,n_3\}=\{0,0,1\}\label{N3lam3}
\eeq
and the exponent $\varphi_-=\pi/2$.
Similarly, for $\Qs^+_2(u)$ one has
\beq
\a_k^+=\a_k^++\pi|\tau|,
\quad \varphi_+=-\frac{5\pi}{2},\quad \{n_1,n_2,n_3\}=\{0,0,1\}.
\eeq
For the eigenvalue ${\Ls}_3$
all $\a_i^\pm$, $\varphi_\pm$ and $n_k$ change their sign with respect
to those for $\Ls_2$.

The problem of the analytic continuation for these eigenvalues and, in
particular, the question about their connection to the ground state
eigenvalue $\Ls_0(u)$ have not been considered.

\subsection{The case $N=4$.}

This is the last case where we systematically study
all eigenvalues of the transfer matrix.
As before the eigenvalues will be first grouped according to their
momenta, which in this case takes four possible values $\Pcal=1,-1,\pm i$.
In total there are $16$ eigenvalues, all non-degenerate.
Below we will show that all the eigenvalues with $\mathcal{P}=\pm1$
(there are ten such eigenvalues) correspond to
different branches of the same multivalued function of $\phi$ and
explicitly present all paths on the Riemann surface, which
connect these eigenvalues to each other.

\subsubsection{The sector $\mathcal{P}=1$.}
This sector contains $6$ eigenvalues listed in Table~\ref{table6}.
Three of them are given by the formula
\beq
{\bf\Lambda}_i^B=
a^4+b^4+2ab(a^2+b^2)\Bigl[\frac{x_i+(\g+1)(\zeta-1)}{1-\g}\Bigr]-
2a^2b^2\Bigl[\frac{(\g+1)[\zeta^2(\g-1)-(\zeta+1)x_i]-x_i^2}
{(\g-1)^2}\Bigr],
\label{cubic}
\eeq
where $i=0,3,5$ and the constants $x_0,x_3$ and $x_5$  satisfy the same
cubic equation
\beq
 x^3+x[2\zeta(\g+1)^2
-(\zeta^2+1)(3-2\g+3\g^2)]+2(\zeta-1)^2(\zeta+1)(\g-1)^2(\g+1)=0.
\eeq
with $\zeta$, $\gamma$ defined in (\ref{zetagam}).
With the choice \eqref{numpar} this equation has three real roots
\beq
x_0=12.98056849,\quad x_3=-3.41421744,\quad x_5=-9.56635105.
\eeq
The remaining three eigenvalues are polynomially expressed
through the Boltzmann weights \eqref{weights}
\beq
{\bf\Lambda}_1^B=(a^2+b^2)(c^2+d^2)+(a^2+b^2+c^2+d^2)(ab+cd)+4abcd,
\eeq
\beq
{\bf\Lambda}_2^B=(a^2+b^2)(c^2+d^2)+(a^2+b^2+c^2+d^2)(ab-cd)-4abcd,
\eeq
\beq
{\bf\Lambda}_4^B=a^4+b^4-2c^2d^2.
\eeq
The corresponding eigenvalues of the XYZ-Hamiltonian \eqref{ham0} read
\beq
\Es_1=-2J_1,\quad \Es_2=-2J_2,\quad \Es_4=-2J_3.
\eeq
\begin{table}[ht]\begin{center}
\begin{tabular}{|c|c|c|c|c|c|c|}\hline$i$&$\Es_i$&
${\bf\Lambda}_i^{B^{\phantom{I}}}$&$\Ls_i$&$\Pcal$&$\Scal$&$\Rcal$\\ \hline
$0$ &$-2.75822456$ & $1.30674491$& $2.36255547$ &$1$ &$1$ &$1$\\ \hline
$1$ &$-2.00000000$&$0.94124717$ &$0.45441276$ &$1$ &$-1$ &$1$\\ \hline
$2$ &$-1.17819166$&$0.61623116$ &$-0.92097245$ &$1$ &$-1$ &$-1$\\ \hline
$3$ &$0.72548274$&$0.14757534$ & $-1.33882675$&$1$ &$1$ &$1$\\ \hline
$4$ &$0.86310373$&$0.12908441$ &$-1.21915423$&$1$ &$1$ &$-1$\\ \hline
$5$ &$2.03274182$&$0.05569128$ & $0.61370849$&$1$ &$1$ &$1$\\ \hline
\end{tabular}\end{center}
\caption{Properities of the eigenvalues
with $\mathcal{P}=1$ for $N=4$.}\label{table6}\end{table}

Table~\ref{table7} contains positions of the Bethe roots $\a_{1,2}$,
the values of the field $\phi$ and phases $n_k$ appearing in \eqref{lba}.
Note that, since $N$ is even the roots satisfy (\ref{halfalpha})
and it is enough to present only two roots $\a_{1,2}$.

\begin{table}[ht]
\begin{center}
\begin{tabular}{|c|c|c|c|c|}\hline
$i$  & $\a_1$ & $\a_2$ &$\phi$ & $n_1,n_2,n_3,n_4$\\ \hline
$0$ & $-1.41965608$ & $-0.88292901$ & $0$ & $\{-3/2,-1/2,1/2,3/2\}$  \\ \hline
$1$  & $i\pi\tau$ & ${\ds i{\pi\tau}/{2}}$ & $\pi$ &
  $\{-3/2,-1/2,1/2,3/2\}$ \\
 \hline
$2$   & ${\ds i\pi\tau-i{\pi}/{2}}$ &
${\ds i{\pi\tau}/{2}}$ & $\pi$  & $\{-3/2,-1/2,1/2,3/2\}$  \\ \hline
$3$  & $i\pi\tau-0.59503973\,i$ & $0.59503973\,i$ & 0 &
 $\{-3/2,-1/2,1/2,3/2\}$  \\ \hline
$4$   & ${\ds i\pi\tau-i{\pi}/{2}}$ & $0$ & $0$ &
 $\{-3/2,-1/2,1/2,3/2\}$  \\ \hline
$5$  & ${\ds-1.65004854-i{\pi}/{2}}$  & ${\ds-0.65253655-i{\pi}/{2}}$ & $0$ &
 $\{-5/2,1/2,-1/2,5/2\}$  \\ \hline
\end{tabular}
\end{center}
\caption{Bethe roots, values of the field
 and phases for $\mathcal{P}=1$ and $N=4$.}\label{table7}\end{table}

\subsubsection{The sector $\mathcal{P}=-1$.}

This sector contains 4 eigenvalues. All of them are polynomial in
Boltzmann weights and the corresponding Bethe roots can be found explicitly.
\beq
{\bf\Lambda}^B_6=2a^2b^2-c^4-d^4
\eeq
\beq
{\bf\Lambda}^B_7=2(a^2b^2+c^2d^2)-(a^2+b^2)(c^2+d^2)
\eeq\beq
{\bf\Lambda}^B_8=-(a^2+b^2)(c^2+d^2)+(a^2+b^2+c^2+d^2)(ab-cd)+4abcd
\eeq
\beq
{\bf\Lambda}^B_9=-(a^2+b^2)(c^2+d^2)+(a^2+b^2+c^2+d^2)(ab+cd)-4abcd
\eeq
The associated eigenvalues of the XYZ-Hamiltonian are
\beq
\Es_6=2J_3,\quad \Es_7=0,\quad \Es_8=2J_2,\quad \Es_9=2J_1.
\eeq

\begin{table}[ht]\begin{center}\begin{tabular}{|c|c|c|c|c|c|c|}\hline
$i$ & $\Es_i$ & ${\bf\Lambda}_i^{B^{\phantom{I}}}$ &
$\Ls_i$ &$\Pcal$& $\Scal$ & $\Rcal$ \\ \hline
$6$ & $-0.86310373$& $-0.51124109$ & $1.25718614$ &$-1$ &$1$ &$-1$\\ \hline
$7$ & $0$& $-0.27935395$  &$1.63561756$ &$-1$ &$1$ &$1$\\ \hline
$8$ & $1.17819166$& $-0.09456292$ & $0.86905645$&$-1$ &$-1$ &$-1$\\ \hline
$9$ & $2.00000000$& $-0.05570662$ & $-0.54254103$&$-1$ &$-1$ &$1$\\ \hline
\end{tabular}\end{center}
\caption{Properties of eigenvalues for $\Pcal=-1$ and
 $N=4$.}\label{table8}\end{table}

\begin{table}[ht]
\begin{center}
\begin{tabular}{|c|c|c|c|c|}\hline
$i$  & $\a_1$ & $\a_2$ &$\phi$ & $n_1,n_2,n_3,n_4$\\ \hline
$6$&$i\pi\tau/2-i\pi/2$&$i\pi\tau/2$&$0$&$\{-5/2,-3/2,-1/2,1/2\}$\\ \hline
$7$&$i\pi\tau/2-i(\pi/2-\eta)$&$i\pi\tau/2+i(\pi/2-\eta)$&$0$&
$\{-3/2,-1/2,1/2,3/2\}$\\ \hline $8$&$i\pi\tau/2-i\pi/2$&$0$&$-\pi$&
$\{-7/2,-1/2,-3/2,3/2\}$\\ \hline $9$&$i\pi\tau/2-i\pi/2$&$-i\pi/2$&$-\pi$&
$\{-5/2,1/2,-1/2,5/2\}$  \\ \hline\end{tabular}\end{center}
\caption{Bethe roots, values of the field and phases with $\Pcal=-1$ at $N=4$.}
\label{table9}\end{table}

Using the differential equations \eqref{odeal}
we have found that all solutions of the Bethe Ansatz equations \eqref{lba}
corresponding to $\Pcal=\pm1$ can be obtained from the
ground state solution by the analytical
continuation in $\phi$. For the numerical calculations we always assumed
the values \eqref{numpar}.
Consider the principal sheet of the Riemann surface containing
the ground state solution at $\phi=0$. As explained
before the cut structure on this sheet is always symmetric
with respect to the real axis and periodic with the period $2\pi$.
Therefore we consider only the upper half of the periodicity strip.
There is only one cut $(\pi+0.67535\, i,\pi+i\,\infty )$ on this sheet
shown by solid lines in Fig.\ref{fig7}. Encircling the branching point
(as in the contour  $C_{12}$) brings one to the second sheet
containing four additional cuts (shown by dashed lines in
Fig.\ref{fig7}). In this figure
we also shown the contours which connect ${\Ls}_{0}$ with five other
eigenvalues ${\Ls}_{1}$, ${\Ls}_{2}$, ${\Ls}_{3}$, ${\Ls}_{4}$, $-{\Ls}_{6}$.
The change of sign for ${\Ls}_{6}$ is related with the transformation
\eqref{shift} and \eqref{sign} with $m=1$.
The horizontal coordinates of the cuts shown on the
figures have been numerically fitted by considering a few different
values of $\eta$ in the vicinity $\eta=7\pi/20$.

\begin{figure}[ht]
\setlength{\unitlength}{0.45mm}
\begin{picture}(350,165)(20,20)
\put(175,46){\allinethickness{0.3mm}\line(0,1){8}}
\put(50,50){\vector(1,0){270}}
\put(280,46){\allinethickness{0.3mm}\line(0,1){8}}
\put(70,25){\vector(0,1){150}}
\multiput(132.75,115)(4,0){2}{\dashline[10]{3}(0,0)(0,35)}
\put(280,50){\line(0,1){125}}
\multiput(82.75,120)(4,0){2}{\dashline[10]{3}(0,0)(0,35)}
\put(134.8,111){\arc{7}{-1.0}{4.1}}
\multiput(262.75,120)(4,0){2}{\dashline[10]{3}(0,0)(0,35)}
\put(84.8,116){\arc{7}{-1}{4.10}}
\multiput(212.75,115)(4,0){2}{\dashline[10]{3}(0,0)(0,35)}
\put(264.8,116){\arc{7}{-1}{4.10}}
\multiput(172.75,120)(4,0){2}{\thicklines\line(0,1){55}}
\put(214.8,111){\arc{7}{-1}{4.10}}
\put(300,170){\allinethickness{0.3mm}\circle{20}}
\put(57,180){\mbox{Im}$\,\phi$}
\multiput(84.8,116)(90,0){3}{\circle*{2}}
\multiput(134.8,111)(80,0){2}{\circle*{2}}
\multiput(84.8,50)(90,0){3}{\dashline[30]{2}(0,0)(0,66)}
\put(174.8,116.5){\arc{8}{-1}{4.10}}
\multiput(134.8,50)(80,0){2}{\dashline[30]{2}(0,0)(0,60)}
\put(174.8,116.44){\arc{7.5}{-1}{4.10}}
\put(80,35){{\small${ 3}\pi$}$-\frac{\pi^2}{\eta}$}
\put(210,35){$\frac{\pi^2}{2\eta}$}
\put(125,35){{\small${ 2}\pi$}$-\frac{\pi^2}{2\eta}$}
\put(250,35){$\frac{\pi^2}{\eta}-\pi$}
\dashline[-20]{2}(70,116)(254,116)\dashline[-20]{2}(70,111)(214,111)
\put(296,168){\bf\Large$\phi$}\put(60,35){\large$0$}\put(170,35)
{\Large$\bf\pi$}
\put(280,35){2\Large$\bf\pi$}\put(290,137){
\bf\large$\eta=\frac{7\pi}{20}$}
\put(41,115){\small $0.67535$}\put(41,107){\small $0.63561$}
\put(300,55){\mbox{Re}$\,\phi$}
{\thicklines\spline(70,50)(80,53)(122.5,60)(165,53)(175,50)
\spline(175,50)(160,105)(165,120)(167.5,122)(170,124.5)(173,125)
 \qbezier[35](177,125)(200,122)(175,50) \qbezier[15](175,50)(195,65)(215,50)
 \qbezier(70,50)(190.6,80)(191,115) \qbezier(191,115)(190.8,140)(177,140)
 \qbezier[65](172,140)(130,139)(70,50) \qbezier(70,50)(155,126)(172,128)
 \qbezier[25](177,130)(224,137.5)(225,110)
 \qbezier[45](225,110)(224,80)(70,50)}
\put(145,57){$\bf C_{01}$}\put(187,60){$\bf C_{26}$}\put(146,95){$\bf C_{12}$}
\put(222,95){$\bf C_{04}$}\put(88,96){$\bf C_{03}$}
\put(109,100.5){\arr{3.63192}{7.12805}{6.70936}{4.35711}}
\put(131,57.9){\arr{-7.72741}{-2.07055}{-7.72741}{2.07055}}
\put(211,52.9){\arr{-7.79496}{1.79961}{-5.85083}{5.45599}}
\put(206,88.9){\arr{5.03456}{6.21717}{7.46864}{2.86694}}
\put(162,108.9){\arr{2.73616}{-7.51754}{-1.38919}{-7.87846}}
\end{picture}\caption{The principal sheet of the Riemann surface
containing the eigenvalue ${\Ls}_0$ for $N=4$.}\label{fig7}\end{figure}
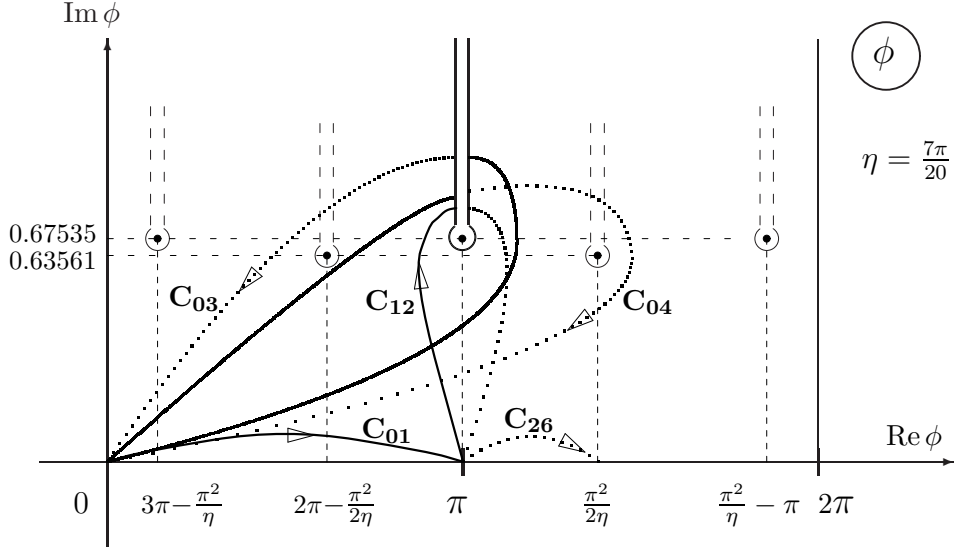

\begin{figure}[ht!]\centering
\setlength{\unitlength}{0.45mm}
\begin{picture}(350,160)(20,20)
\put(50,50){\vector(1,0){270}}\put(70,25){\line(0,1){55}}
\put(70,170){\vector(0,1){5}}
\put(130,49){\allinethickness{0.3mm}\line(0,1){5}}
\put(280,49){\allinethickness{0.3mm}\line(0,1){5}}\put(280,50){\line(0,1){125}}
\multiput(115,49)(45,0){3}{\allinethickness{0.3mm}\line(0,1){5}}
\multiput(0,0)(90,0){2}{\multiput(113,140)(4,0){2}{\dashline[10]{3}(0,0)(0,30)}
\put(115,136){\arc{7}{-1.0}{4.1}}\put(115,136){\circle*{2}}
\put(115,50){\dashline[30]{2}(0,0)(0,85)}}
\multiput(0,0)(90,0){3}{\multiput(68,85)(4,0){2}{\dashline[10]{3}(0,0)(0,85)}
\put(70,81){\arc{7}{-1.0}{4.1}}\put(70,81){\circle*{2}}
\put(70,50){\dashline[30]{2}(0,0)(0,60)}}\put(130,33){\vector(0,1){8}}
\multiput(0,0)(0,55){2}{\dashline[-20]{2}(70,81)(280,81)}
\put(300,170){\allinethickness{0.3mm}\circle{20}}
\put(97,40){\small$\frac{\pi^2}{2\eta}-\pi$}
\put(118,27){\small$2\pi-\frac{\pi^2}{2\eta}$}
\put(152,40){\small$\frac{\pi^2}{\eta}-2\pi$}
\put(195,40){\small$\frac{3\pi^2}{2\eta}-3\pi$}
\put(235,40){\small$\frac{2\pi^2}{\eta}-4\pi$}
\put(296,168){\bf\Large$\phi$}\put(60,35){\large$0$}
\put(280,35){2\Large$\bf\pi$}
\put(285,145){\bf\large$\eta=\frac{7\pi}{20}$}\put(41,135){\small $1.15202$}
\put(31,80){\small $0.14175$}\put(300,60)
{\mbox{Re}$\,\phi$}\put(57,180){\mbox{Im}$\,\phi$}
{\thicklines  \qbezier(70,50)(92.5,60)(115,50) \qbezier(70,50)(53,89)(68,90)
 \qbezier[15](70,50)(87,89)(72,90) \qbezier(115,50)(98,142)(113,143)
 \qbezier[30](115,50)(132,142)(117,143) \qbezier[20](70,50)(92.5,90)(115,50)
 \qbezier(70,50)(43,109)(68,110) \qbezier[40](130,50)(95,109)(72,110)}
\put(99,55){\arr{-7.72741}{-2.07055}{-7.72741}{2.07055}}
\put(88,70){\arr{7.72741}{2.07055}{7.72741}{-2.07055}}
\put(79,75){\arr{-2.07055}{7.72741}{2.07055}{7.72741}}
\put(95,97.5){\arr{-6.99696}{3.87848}{-4.1203}{6.85734}}
\put(124,115){\arr{-2.07055}{7.72741}{2.07055}{7.72741}}
\put(77,110){$\bf C_{47}$}\put(75,90){$\bf C_{43}$}\put(91,73){$\bf C_{95}$}
\put(89,58){$\bf C_{48}$}\put(127,115){$\bf C_{89}$}
\end{picture}
\caption{The sheet of the Riemann surface containing ${\Ls}_4$ for $N=4$.}
\label{fig8}
\end{figure}
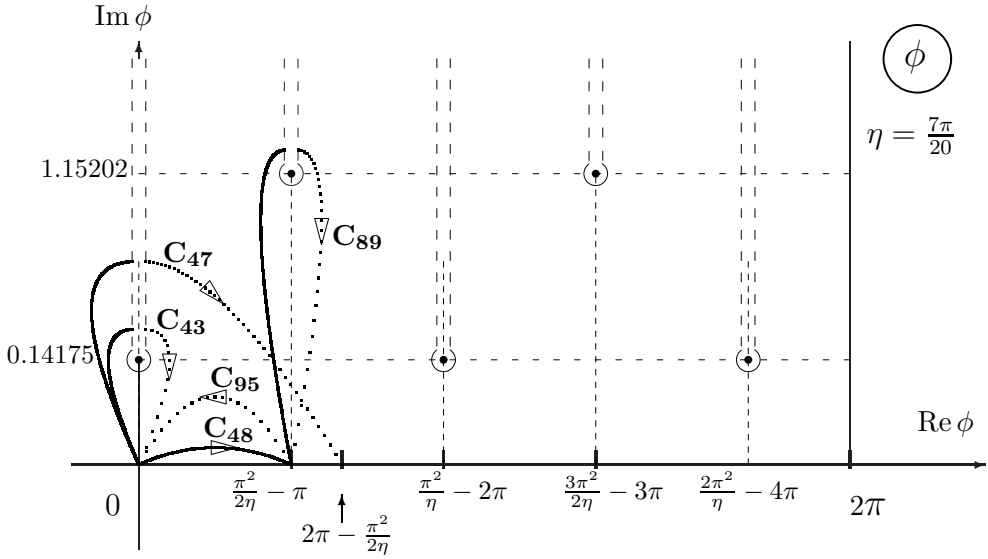

Paths to the other four eigenvalues $\Ls_5$, $\Ls_7$, $\Ls_8$ and
$\Ls_9$ are more conveniently described on a different sheet of the
Riemann surface which contains the eigenvalue $\Ls_4$ at $\phi=0$.
The structure of cuts on this sheet and connecting contours are shown
in Fig.\ref{fig8}. Again at the points $\pi^2/(2\eta)-\pi$
and $2\pi-\pi^2/(2\eta)$ we have to take into account
\eqref{sign} with $m=\pm1$. For $\Ls_7$
we used the shift \eqref{phishift}
to move the Bethe roots to the region \eqref{rect}.

\subsubsection{The sector $P=i$.}

There are three complex eigenvalues
with $\Pcal=i$ and three complex conjugated eigenvalues
with $\Pcal=-i$. Again it is sufficient to consider only $\Pcal=i$.
\beq
{\bf\Lambda}^B_{10}=2(a^2b^2-c^2d^2)+i(a^2-b^2)(c^2-d^2)
\eeq
\beq
{\bf\Lambda}^B_{11}=(ab+cd)(a^2+b^2-c^2-d^2)+i(a^2-b^2)(c^2-d^2)
\eeq
\beq
{\bf\Lambda}^B_{12}=(ab-cd)(a^2+b^2-c^2-d^2)+i(a^2-b^2)(c^2-d^2).
\eeq
Note that for all of them the corresponding energies $\Es_i$ of the
XYZ-Hamiltonian are zero.
In general, these eigenvalues are complex. However, at the symmetric
point $v=\pi/2$ where $a=b$, all eigenvalues are real.
All information about these eigenvalues is collected
in Tables \ref{table10}, \ref{table11}. The parameters of the
eigenvalues with the momentum $\Pcal=-i$ are
obtained by negating of all $\a_i$'s, $\phi$'s and  $n_k$'s.

\begin{table}[ht]\begin{center}\begin{tabular}{|c|c|c|c|c|c|c|}\hline
$i$ & $\Es_i$ & ${\bf\Lambda}_i^{B^{\phantom{I}}}$ &
$\Ls_i$ &$\Pcal$& $\Scal$ & $\Rcal$ \\ \hline
$10$ & $0$& $0.12908441$  &$1.25718614$ &$i$ &$1$ &$-1$\\ \hline
$11$ & $0$& $-0.09456292$ &$-0.92097245$ &$i$ &$-1$ &$-1$\\ \hline
$12$ & $0$& $-0.05570662$ &$-0.54254103$ &$i$ &$-1$ &$1$\\ \hline
\end{tabular}\end{center}
\caption{Eigenvalues with $\Pcal=i$ at $N=4$.}\label{table10}\end{table}

\begin{table}[ht]
\begin{center}
\begin{tabular}{|c|c|c|c|c|}\hline
$i$  & $\a_1$ & $\a_2$ &$\phi$ & $n_1,n_2,n_3,n_4$\\ \hline
$10$&$ -1.50555656-{i\pi}/{2}$&
  $-0.79702853$&$0$&$\{-5/2,-1/2,-1/2,3/2\}$\\
\hline
$11$&$-1.89851386-{i\pi}/{2}$&$-1.55536378$&$\pi$&
$\{-3/2,-3/2,1/2,1/2\}$\\ \hline
$12$&${3i\pi\tau}/{4}-0.40566532\,i$&${3i\pi\tau}/{4}+0.40566532\,i$&$\pi$&
$\{-3/2,-3/2,1/2,1/2\}$\\ \hline
\end{tabular}\end{center}
\caption{Bethe roots, values of the field and phases for $\Pcal=i$ and $N=4$.}
\label{table11}\end{table}

The problem of analytic continuation for these eigenvalues has not
been considered.

\subsection{The case  $N=13$}

The purpose of this section is to analyze a
distribution of zeroes for the ground state
eigenvalues for sufficiently large value of $N$.
We deliberately choose an odd value of $N$ when
$\Qs_+(u)$ and $\Qs_-(u)$  are
linearly independent and their quantum Wronskian is nonzero. In this
section it is more convenient to work with the original Baxter's
normalization of
eigenvalues $Q_\pm^B(v)$ and the variable $v$
\beq
\Qs_\pm(u)=e^{ivN/2} \,Q_\pm^B(v),\qquad v=u+\pi\tau/2.\label{q2qb}
\eeq
Introduce the linear combinations $Q_{1,2}(v)$,
\beq
Q_\pm^B(v)=(Q_1(v)\pm Q_2(v))/2,\label{q2q12}
\eeq
such that
\beq
Q_{1,2}(v+\pi)=(-1)^{(N-1)/2}Q_{2,1}(v)\label{q12per}
\eeq

Numerical zeroes of eigenvalues can be easily calculated from the
Bethe Ansatz equations \eqref{lba} with the set of phases \eqref{gphases}.
Here we demonstrate an alternative numerical method which works well
for the considered case.
For any odd $N$ the ground state eigenvalue ${\bf\Lambda}_0^B(v)$
an odd function of $v$, having $N$ zeroes
in its periodicity rectangular \eqref{tper3}.  Using the identity
\beq
2\,\vt_1(x+y|\q)\,\vt_1(x-y|\q)=
\ov\vt_4(2x)\,\ov\vt_3(2y)-\ov\vt_3(2x)\,\ov\vt_4(2y), \qquad
\ov \vt_i(x)\equiv\vt_i(\frac{x}{2}|\q^{1/2})\ ,
\eeq
it can be represented as
\beq
{\bf\Lambda}_0^B(v)=\vt_1(v\,|\,\q) \,\sum_{k=0}^{(N-1)/2} t_k
\>\ov \vt_3(2v)^{k}\,\ov\vt_4(2v)^{(N-1-2k)/2}\ ,\qquad
\label{Tpolyn}
\eeq
where $t_k$ are constants.
Similarly, $Q_1(v)$  can be written as
\beq
Q_1(v)=2^{-N}\>\ov
\vt_3(v)\sum_{k=0}^{(N-1)/2}c_k\>\ov
\vt_3(v)^{2k}\,\ov\vt_4(v)^{N-1-2k}\ ,
\label{Qpolyn}
\eeq
with some unknown constants $c_k$.
In \cite{BM05} we used the representation \eqref{Qpolyn} for $\eta=\pi/3$,
however, it is valid for the ground state eigenvalues
with arbitrary values of $\eta$ (the case $\eta=\pi/3$
is special because all coefficients $c_k$ can be calculated explicitly
\cite{BM05}).
Substituting  \eqref{Tpolyn} and \eqref{Qpolyn}
into the TQ-equation and evaluating
it numerically for several values of the spectral parameter one gets
a bi-linear system of equations for the unknown coefficients
$t_k$ and $c_k$. For the choice (\ref{numpar})
its numerical solution is given in Table~\ref{table12}.
\begin{table}[ht]
\begin{center}
\begin{tabular}{|c|c|c|c|}\hline
$k$ & $c_k$ & $k$ & $c_k$ \\ \hline
$0$ & $\phantom{-}1.00000000$ & $3$ & $-26.42365231$ \\ \hline
$1$ & $-5.63412047$ & $4$ & $\phantom{-}28.59503534$ \\ \hline
$2$ & $\phantom{-}15.52969161$ & $5$ &$-18.21779332$ \\ \hline
\multicolumn{2}{c|}{} & $6$ &$\phantom{-}5.231349086$\\ \cline{3-4}
\end{tabular}
\end{center}
\begin{center}
\caption{Numerical values of the coefficients $c_k$
in \eqref{Qpolyn} for
$N=13$ in the case \eqref{numpar}.}\label{table12}
\end{center}
\end{table}
\begin{table}[ht]
\begin{center}
\begin{tabular}{|c|c|c|c|c|c|}\hline
$i$ & $\a_i^{-\phantom{J_k}}$ & $i$ & $\a_i^-$ & $i$ &$\a_i^-$  \\ \hline
$1$ & $-1.97251218$ & $5$ & $-1.00040409$ & $9$ & 0.91284796\\ \hline
$2$ & $-1.50058366 $ & $6$ & $-0.80200143$ & $10$ & 1.07830969\\ \hline
$3$ & $-1.30218100$ & $7$ & $-0.33007292$ & $11$ & 1.22427541\\ \hline
$4$ & $-1.15129255$ & $8$ & $\phantom{-}0.64848502$ & $12$ &
  1.38973713\\ \hline
\multicolumn{4}{c|}{} & 13 & 1.65410007\\ \cline{5-6}
\end{tabular}
\end{center}
\begin{center}
\caption{Bethe roots for the ground state at $N=13$.}\label{table13}
\end{center}
\end{table}

Using these results and (\ref{q2qb}--\ref{q12per}) one can find the
 Bethe roots $\a_k^\pm$,\  $k=1,\ldots,13$. The results for $\a_k^-$
are given in Table~\ref{table13}, moreover,
$\a_i^+=-\a_{N-i+1}^-$ and
$\varphi_\pm=\mp{\pi}/{2}$.

Further, zeros for $\Ls_0(u)$ and $\Ls_0^{(3)}(u)$ are
as follows.
The $2N=26$ zeros of $\Ls_0(u)$ read
\beq
u_k=\pm\frac{\pi\tau}{2},\quad
\pm\frac{\pi\tau}{2}\pm\{r_1,r_2,r_3,r_1^*,r_2^*,r_3^*\}\ ,
\eeq
where the signs can be chosen independently and $r^*_i$ stand for
complex conjugates of $r_i$.
Similarly, for $\Ls_0^{(3)}(u)$,
\beq
u_k=\pm\frac{\pi\tau}{2},\quad
\pm\frac{\pi\tau}{2}+\frac{\pi}{2}\pm\{s_1,s_2\},\quad
\pm\frac{\pi\tau}{2}+\frac{\pi}{2}\pm\{s_3,s_4,s_3^*,s_4^*\}.
\eeq
The numerical constants $r_i$, $s_i$
are given in Table~\ref{table14}.
The zeroes of eigenvalues are plotted in Fig.~\ref{fig9}.

\begin{table}[ht]
\begin{center}
\begin{tabular}{|c|c|c|}\hline
$k$ & $s_k$ & $r_k$  \\ \hline
$1$ & $0.27442683$ & $0.40999133+0.42208534\>i$\\ \hline
$2$ & $0.47123771 $ & $0.49344215+0.21629447\>i $\\ \hline
$3$ & $0.28395326+0.15112367\, i$ & $0.51703097+0.067457998\>i$\\ \hline
$4$ & $0.32747706+0.34260021\,i$ \\ \cline{1-2}
\end{tabular}
\end{center}
\begin{center}
\caption{Parameters $s_k$, $r_k$ which determine positions of zeros
$\Ls_0(u)$ and $\Ls_0^{(3)}(u)$.}\label{table14}
\end{center}
\end{table}

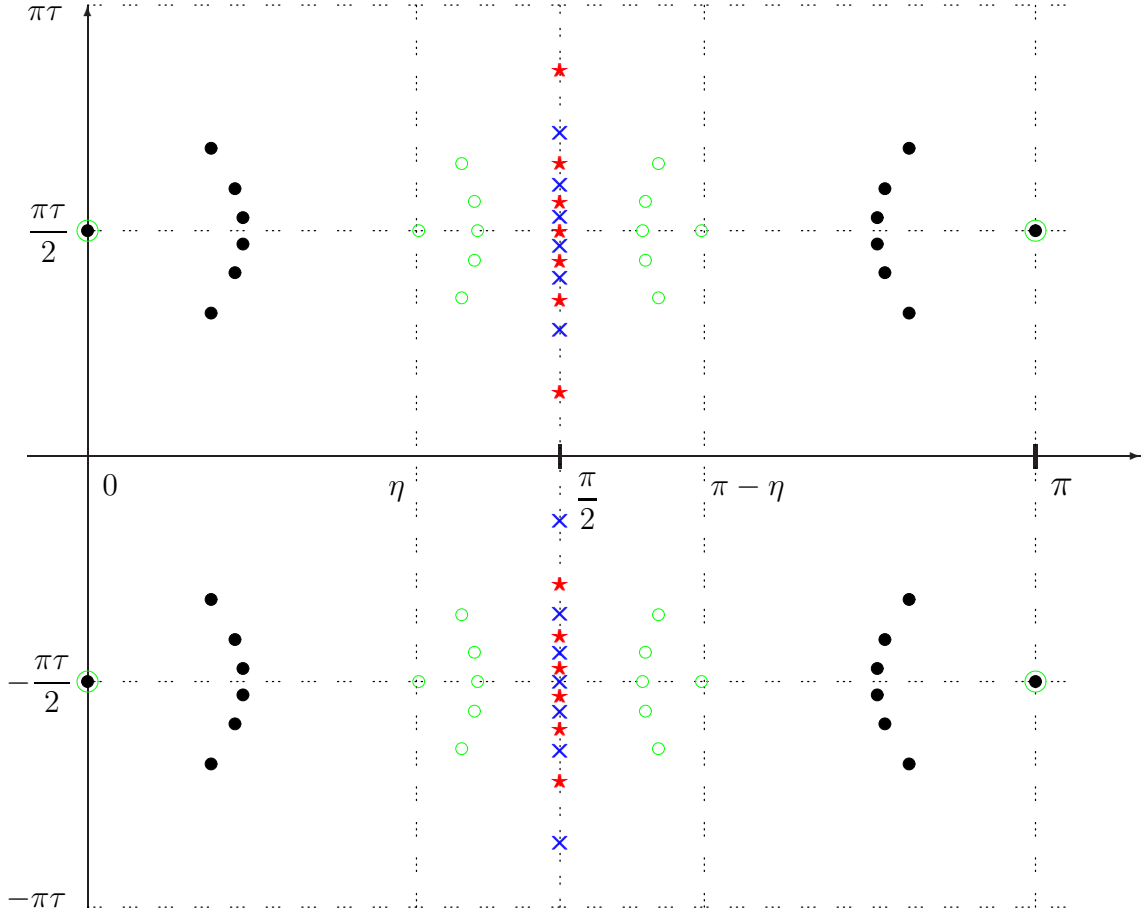
\begin{figure}[ht!]
\setlength{\unitlength}{0.4mm}
\begin{picture}(400,290)(40,0)
\put(20,5){\begin{picture}(400,300)\put(40,150){\vector(1,0){370}}
\put(60,0){\vector(0,1){300}}\put(217,146)
{\allinethickness{0.5mm}\line(0,1){8}}
\put(375,146){\allinethickness{0.5mm}\line(0,1){8}}\put(-157,0){\circle*{3}}
\dashline[-10]{5}[2](60,225)(385,225)\dashline[-10]{5}[2](60,300)(385,300)
\dashline[-10]{5}[2](60,75)(385,75)\dashline[-10]{5}[2](60,0)(385,0)
\dashline[-10]{3}[2](217,0)(217,300)\dashline[-10]{5}[2](375,0)(375,300)
\dashline[-10]{5}[2](169.3,0)(169.3,300)\dashline[-10]{5}[2](265,0)(265,300)
\put(65,137){\large $0$}\put(222,133){\large $\ds\frac{\pi}{2}$}
\put(380,137){\Large$\ds\pi$}\put(40,222){\large$\ds\frac{\pi\tau}{2}$}
\put(33,72){\large$-\ds\frac{\pi\tau}{2}$}\put(33,0){\large$-\pi\tau$}
\put(40,295){\large$\pi\tau$}\put(160,137){\large$\ds\eta$}
\put(267,137){\large$\ds\pi-\eta$}
\multiput(0,0)(315,0){2}{\multiput(0,0)(0,-150){2}{\put(60,225){\circle*{4}}}}
\multiput(217,225)(0,-150){2}{
\multiput(0,0)(232,0){2}{\multiput(0,0)(0,54.8){2}
{\put(-116,-27.4){\circle*{4}}}}
\multiput(0,0)(216,0){2}{\multiput(0,0)(0,28){2}
{\put(-108,-14){\circle*{4}}}}
\multiput(0,0)(210.8,0){2}{\multiput(0,0)(0,8.8){2}
{\put(-105.4,-4.4){\circle*{4}}}}}
\multiput(217,225)(0,-150){2}{\multiput(-157,0)(315,0){2}
{\color{green}\circle{7}}\put(47,0){\color{green}\circle{4}}
\put(-47,0){\color{green}\circle{4}}\put(27.4,0)
{\color{green}\circle{4}}\put(-27.4,0){\color{green}\circle{4}}
\multiput(0,0)(65.4,0){2}{\multiput(0,0)(0,44.6){2}
{\put(-32.7,-22.3){\color{green}\circle{4}}}}
\multiput(0,0)(56.8,0){2}{\multiput(0,0)(0,19.6){2}
{\put(-28.4,-9.8){\color{green}\circle{4}}}}}
\multiput(213,19)(0.4,0){2}{\color{blue}$\times$}
\multiput(213,49.7)(0.4,0){2}{\color{blue}$\times$}
\multiput(213,62.7)(0.4,0){2}{\color{blue}$\times$}
\multiput(213,72.5)(0.4,0){2}{\color{blue}$\times$}
\multiput(213,82.3)(0.4,0){2}{\color{blue}$\times$}
\multiput(213,95.3)(0.4,0){2}{\color{blue}$\times$}
\multiput(213,126)(0.4,0){2}{\color{blue}$\times$}
\multiput(213,189.7)(0.4,0){2}{\color{blue}$\times$}
\multiput(213,207)(0.4,0){2}{\color{blue}$\times$}
\multiput(213,217.7)(0.4,0){2}{\color{blue}$\times$}
\multiput(213,227.2)(0.4,0){2}{\color{blue}$\times$}
\multiput(213,238)(0.4,0){2}{\color{blue}$\times$}
\multiput(213,255.3)(0.4,0){2}{\color{blue}$\times$}
\multiput(214.2,276)(0.2,0){4}{\color{red}$\star$}
\multiput(214.2,245.25)(0.2,0){4}{\color{red}$\star$}
\multiput(214.2,232.3)(0.2,0){4}{\color{red}$\star$}
\multiput(214.2,222.5)(0.2,0){4}{\color{red}$\star$}
\multiput(214.2,212.7)(0.2,0){4}{\color{red}$\star$}
\multiput(214.2,199.7)(0.2,0){4}{\color{red}$\star$}
\multiput(214.2,169)(0.2,0){4}{\color{red}$\star$}
\multiput(214.2,105.3)(0.2,0){4}{\color{red}$\star$}
\multiput(214.2,88)(0.2,0){4}{\color{red}$\star$}
\multiput(214.2,77.3)(0.2,0){4}{\color{red}$\star$}
\multiput(214.2,67.7)(0.2,0){4}{\color{red}$\star$}
\multiput(214.2,57)(0.2,0){4}{\color{red}$\star$}
\multiput(214.2,39.7)(0.2,0){4}{\color{red}$\star$}
\end{picture}}\end{picture}\caption{
Zeros of $\Qs_0^+(u)$, $\Qs_0^-(u)$,
$\Ls_0(u)$, $\Ls_0^{(3)}(u)$ for $N=13$, marked by
{\color{red}$\bf\star\>$}, {\color{blue}$\times\>$},
$\bullet\>$, {\color{green}$\circ\>$}, accordingly.}\label{fig9}
\end{figure}

Note that
the sets $\{\a_k^+\}$ and $\{\a_k^-\}$
precisely interlace each other. This fact looks quite remarkable
given these sets are obtained from each other by the uniform translation
\eqref{qpmshift} over the distance $\pi|\tau|$ which is much larger
than the spacing between roots. For a numerical study of the Bethe
Ansatz equations on a large chain of even length (up to $N=512$)
we refer the reader to \cite{BBP87}.


\nsection{Conclusion}

In this paper we developed some new ideas in the classical
subject of Baxter's celebrated eight-vertex and solid-on-solid models.
Our primary observation concerns a (previously unnoticed) arbitrary
field parameter in the solvable solid-on-solid model. This
parameter is analogous to the horizontal field in the six-vertex
model. This fact might not be so surprising to experts, since all the
hard work has been done before and one just needs to lay side-by-side
the papers \cite{Bax73c,TF79,FV96} to realize that an arbitrary
field parameter is, in fact, required to describe
the continuous spectrum  of the
unrestricted solid-on-solid model.

The introduction of an arbitrary field allowed us
to develop a completely analytic
theory of the functional relations in the 8V/SOS-model.
The solutions of the Bethe Ansatz equations
are multivalued functions of the field variable, having algebraic
branching points.
It is plausible that many (if not all) eigenvalues of the transfer
matrix can be obtained from each other via analytic continuation in
this variable. To demonstrate this we performed a comprehensive
study of all eigenvalues for the 8V-model
for small chains of the length $N\le4$
with a combination of analytic and numeric techniques.
In particular, we saw in these cases
that the largest and smallest eigenvalues of the
transfer matrix are always connected by the analytic continuation.
This study was partially motivated by our attempts to understand
properties of the eigenvalues in the ferromagnetic regime
\cite{Sut95}, analytically connecting it with the disordered regime.
Note that the ferromagnetic regime eigenvalues are important in
connection with the AdS/CFT correspondence (see \cite{BSMZ03} and
references therein); we hope to study them elsewhere.

The field parameter is also important in our future
considerations \cite{BM06b} of
the quantum field theory limit of the 8V/SOS-model, where it becomes
the massive sine-Gordon model.
Note, that the connection between the largest and next-to-largest
eigenvalues in this model was previously
studied \cite{DT96} via the thermodynamic Bethe
Ansatz. The authors of \cite{DT96}
also considered the analytic continuation but in a different variable,
namely, the scaling variable, which
has no a direct analogue in the lattice theory.

It appears that the analytic structure of eigenvalues in the
eight-vertex/SOS model certainly deserves further studies.
Somewhat simpler (but still very interesting) structure
arises in the six-vertex model and, especially, in the
$c<1$ conformal field theory \cite{BLZ03}. In the latter case the Riemann
surface of the eigenvalues closes within each level subspace of the Virasoro
module.

\section*{Acknowledgments}

The authors thank R.J.Baxter, M.T.Batchelor, B.M.McCoy, K.Fabricius,
P.A.Pearce, S.M.Sergeev, F.A.Smirnov
and M.Bortz for useful remarks. One of us (VVB) thanks S.M.Lukyanov and
A.B.Zamolodchikov for numerous discussions of the analytic structure of
eigenvalues in solvable models. This work was supported by the Australian
Research Council.

\newpage


\app{The eight-vertex model} \label{appA}
In this Appendix we briefly summarize basic
properties of the symmetric eight-vertex (8V) model used in this paper.
For a more detailed information the reader should
consult with Baxter's original publications
\cite{Bax72,Bax73a,Bax73b,Bax73c,Bax82}.

Consider a square lattice of $N$ columns and $M$ rows, with the
toroidal boundary conditions. Each edge of the lattice carries a spin
variable $\a$ taking two values $\a=+$ and $\a=-$, corresponding to
the ``spin-up'' and  ``spin-down'' states of the edge.
Each vertex is assigned with a Boltzmann weight $R(\a,\a'|\b,\b')$
depending on the spin states $\a,\a',\b,\b'$ of the surrounding edges
arranged as in Fig.\ref{R-picture}.
\begin{figure}[ht]
\setlength{\unitlength}{0.4mm}
\begin{picture}(300,150)(0,0)
\multiput(180,69.5)(0,0.5){3}{\line(1,0){80}}
\multiput(219.5,29)(0.5,0){3}{\line(0,1){80}}
\put(73,67){\Large$ R(\a,\a'|\b,\b')\>\>=$}
\put(185,58){\Large$\a$}\put(245,58){\Large$\a'$}
\put(225,34){\Large$\b$}\put(225,99){\Large$\b'$}
\end{picture}
\caption{The arrangement of four spins around the vertex.}
\label{R-picture}
\end{figure}
There are only eight ``allowed''
vertex configurations, shown in Fig.\ref{8-vertices},
which have non-vanishing Boltzmann weights. These weights are not
arbitrary; they parameterized by only four
arbitrary constants $a,b,c,d$,
\beq
\o_1=\o_2=a,\qquad \o_3=\o_4=b,\qquad
\o_5=\o_6=c,\qquad \o_7=\o_8=d\ .
\eeq
The remaining eight
configurations are forbidden; their Boltzmann weight is zero.
\begin{figure}[hbt]
\begin{picture}(300,200)(-40,0)
\put(20,150){\line(1,0){50}}\put(45,125){\line(0,1){50}}
\multiput(110,149.25)(0,0.5){4}{\line(1,0){50}}
\multiput(134.25,125)(0.5,0){4}{\line(0,1){50}}
\put(200,149.25){\line(1,0){50}}
\multiput(225,125)(0.5,0){4}{\line(0,1){50}}
\multiput(290,150)(0,0.5){4}{\line(1,0){50}}
\put(314.25,125){\line(0,1){50}}
\put(20,70){\line(1,0){50}}\put(45,45){\line(0,1){50}}
\multiput(45,69.25)(0,0.5){4}{\line(1,0){25}}
\multiput(45,44.25)(0.5,0){4}{\line(0,1){25}}
\put(110,70){\line(1,0){50}}\put(135,45){\line(0,1){50}}
\multiput(110,69.25)(0,0.5){4}{\line(1,0){25}}
\multiput(134.25,70)(0.5,0){4}{\line(0,1){25}}
\put(200,70){\line(1,0){50}}\put(225,45){\line(0,1){50}}
\multiput(200,69.25)(0,0.5){4}{\line(1,0){25}}
\multiput(224.25,45)(0.5,0){4}{\line(0,1){25}}
\put(290,70){\line(1,0){50}}\put(315,45){\line(0,1){50}}
\multiput(315,69.25)(0,0.5){4}{\line(1,0){25}}
\multiput(314.25,70)(0.5,0){4}{\line(0,1){25}}
\put(50,120){\large$\o_1$}
\put(140,120){\large$\o_2$}
\put(230,120){\large$\o_3$}
\put(320,120){\large$\o_4$}
\put(50,40){\large$\o_5$}
\put(140,40){\large$\o_6$}
\put(230,40){\large$\o_7$}
\put(320,40){\large$\o_8$}
\end{picture}
\caption{Eight allowed vertex configuration and their Boltzmann weights.
Thin lines represent the ``spin-up'' states and the bold lines
represent the ``spin-down'' states of the edge spins}
\label{8-vertices}
\end{figure}
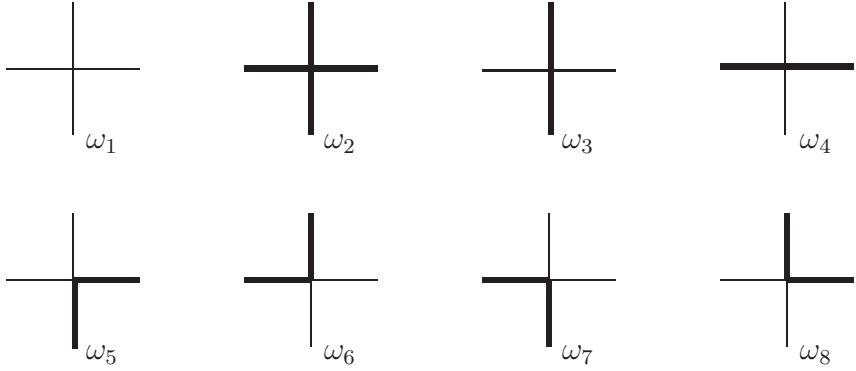

The partition function
\beq\label{pf}
Z=\sum_{\rm (spins)}\  \prod_{\rm (vertices)}  R(\a,\a'|\b,\b')
\eeq
is defined as the sum over all spin configurations of the
whole lattice, where each configuration  is counted with the weight
equal to the product of vertex weights over all lattice vertices.

The vertex weight matrix $R(\a,\a'|\b,\b')$ can be thought  as
a matrix acting in the direct product of the two two-dimensional
vector spaces $\Cbbd^2\otimes \Cbbd^2$, where the indices $\a,\a'$
refer to the first space and the indices $\b,\b'$ to the second.
It has an elegant representation \cite{Bax72}
\beq
\R=\sum_{k=1}^4\, w_k \ (\s_k\otimes\s_k)\ ,\label{Rw}
\eeq
where $\s_k$, $k=1,2,3,4$, are the Pauli matrices
\beq
\s_1=\left( \begin{array}{cc}
&1\\
1&\end{array} \right),\quad
\s_2=\left( \begin{array}{cc}
&-i\\
i&\end{array} \right),\quad
\s_3=\left( \begin{array}{cc}
1&\\
&-1\end{array} \right),\quad
\s_4=\left( \begin{array}{cc}
1&\\
&1\end{array} \right)\ ,
\eeq
and
\beq
w_1=\frac{c+d}{2},\qquad w_2=\frac{c-d}{2},\qquad
w_3=\frac{a-b}{2},\qquad w_4=\frac{a+b}{2}\ .\label{w-def}
\eeq
The matrix $\R$ can conveniently be
presented as a two-by-two block matrix acting in the first space with
the two-dimensional matrix blocks acting in the second space,
\beq
\R=\left( \begin{array}{cc}
\R_{++}&\R_{+-}\\
\R_{-+}&\R_{--}
\end{array} \right)\label{R-def}
\eeq
where
\beq
\R_{++}=\left( \begin{array}{cc}
a&\\
&b
\end{array} \right),\quad
\R_{+-}=\left( \begin{array}{cc}
&d\\
c&
\end{array} \right),\quad
\R_{-+}=\left( \begin{array}{cc}
&c\\
d&
\end{array} \right),\quad
\R_{--}=\left( \begin{array}{cc}
b&\\
&a
\end{array} \right)\ .\label{Rpm}
\eeq
The matrix $\R$ possesses simple symmetry relations, obvious from
\eqref{Rw},
\beq\label{simplesym}
\R=\left(\s_k\otimes\s_k\right)\,\R
\left(\s_k\otimes\s_k\right),\qquad k=1,2,3\ .
\eeq

The row-to-row transfer matrix $\T$ acts in space of states
of $N$ spins located on a horizontal row of vertical edges.
It is defined as the trace of the product of the two-by-two matrices
\eqref{R-def}
\beq
\T={\rm Tr}_{\Cbbd^{2}}\, \left(\R^{(1)}\,\R^{(2)}\cdots
\R^{(N)}\right),
\label{T-def}
\eeq
where the operator entries \eqref{Rpm} of the matrix $\R^{(j)}$ act on
the states of the $j$-th spin in the row. The partition function \eqref{pf} can
written as
\beq
Z={\rm Tr}\left( \T^M\right) \ ,
\eeq
where $M$ is the number of rows of the lattice. It follows from
\eqref{simplesym} that
\beq
[\T,{\mathcal S}]=[\T,{\mathcal R}]=0
\eeq
where the operators
\beq
{\mathcal S}=
\sigma^{(1)}_3\otimes\sigma^{(2)}_3\otimes\cdots\otimes\sigma^{(N)}_3,\qquad
{\mathcal R}=
\sigma^{(1)}_1\otimes\sigma^{(2)}_1\otimes\cdots\otimes\sigma^{(N)}_1
\label{rs-def}
\eeq
defined as the product of the local spin operators
$\sigma^{(j)}_3$ and $\sigma^{(j)}_1$ over the all sites,
$j=1,\ldots,N$,\  in a row.  Note, that for an odd $N$
these operators do not commute among themselves
(indeed, $\Rcal\Scal=(-1)^N\Scal\Rcal$)
and only one of them can be diagonalized
simultaneously with $\T(v)$. Below we will always assume the
basis where the operator $\Scal$ is diagonal (then for even $N$ the operator
$\Rcal$ will be diagonal as well).

Following \cite{Bax72} we
parameterize the Boltzmann weights $a$, $b$, $c$, $d$  as
\bea
&a=\rho
  \ \vt_4(2\eta\,|\,\q^2)\ \vt_4(v-\eta\,|\,\q^2)\ \vt_1(v+\eta\,|\,\q^2),&
\nonumber\\
&b=\rho
  \ \vt_4(2\eta\,|\,\q^2)\ \vt_1(v-\eta\,|\,\q^2)\ \vt_4(v+\eta\,|\,\q^2),
&\nonumber\\
&c=\rho
  \ \vt_1(2\eta\,|\,\q^2)\ \vt_4(v-\eta\,|\,\q^2)\ \vt_4(v+\eta\,|\,\q^2),
&\label{weights}\\
&d=\rho
  \ \vt_1(2\eta\,|\,\q^2)\ \vt_1(v-\eta\,|\,\q^2)\ \vt_1(v+\eta\,|\,\q^2),
&\nonumber
\eea
where we use the standard theta-functions \cite{WW}
\beq
\vt_i(v\,|\,\q),\>\> i=1,\ldots,4,\quad  \q=e^{{\rm i} \pi \tau},
\qquad \rm{Im}\,\tau>0,\label{theta-def}
\eeq
with the periods $\pi$ and $\pi \tau$.
Apart from the simple change
in the notation for the theta-functions
the above parametrization
is the same as that given by Eq.(8) of \cite{Bax72}.
The theta-functions $\HJ(z)$ and $\TJ(z)$ of the nome $q_B$
therein are given by
\beq
q_B=\q^2,\quad \HJ(z)=\vt_1(\frac{\pi z}{2\K_B}\,|\,\,\q^2),\quad \TJ(z)=
\vt_4(\frac{\pi z}{2\K_B}\,|\,\,\q^2),\label{tjacobi}
\eeq
where  $\K_B$ is the complete elliptic integral of the first kind of
the modulus
\beq
k=\frac{\vt_2(0\,|\,\q_B)}{\vt_3(0\,|\,\q_B)}\ .
\eeq
The variables $v$, $\eta$ and $\rho$
used in \cite{Bax72}
(hereafter denoted as $v_B$, $\eta_B$ and $\rho_B$) are related to those in
\eqref{weights} by
\beq
v=\frac{\pi v_B}{2\K_B}, \qquad \eta=\frac{\pi \eta_B}{2\K_B},\qquad
\rho=\rho_B\ .\label{baxvar}
\eeq
With these definitions the transfer matrix
\eqref{T-def} is the same
as that given by Eqs.(6)-(8) of ref \cite{Bax72}. We will denote it
as $\T^B(v)$, remembering that our variable $v$ is related to
Baxter's variable $v_B$ by \eqref{baxvar}.

Throughout this paper we fix the normalization factor $\rho$ in
\eqref{weights} as
\beq
\rho=2\,\ \vt_2(0\,|\,\q)^{-1}\ \vt_4(0\,|\,\q^2)^{-1}.\label{norm}
\eeq
Note, that the variables \eqref{w-def} can be then written
as
\beq
w_i=\frac{1}{2}\,\vt_1(2\eta\,|\,\q)\,
\frac{\vt_{5-i}(u\,|\,\q)}{\vt_{5-i}(\eta\,|\,\q)},\quad i=1,\ldots,4.
\label{wt}
\eeq
We shall also make use of two invariants
\beq
\zeta=\frac{cd}{ab}=
\biggl[\frac{\vt_1(2\eta|\q^2)}{\vt_4(2\eta|\q^2)}\biggr]^2,\quad
\g=\frac{w_3^2-w_2^2}{w_4^2-w_1^2}=
-\biggl[\frac{\t_1(\eta|\q)\t_4(\eta|\q)}{\t_2(\eta|\q)\t_3(\eta|\q)}\biggr]^2.
\label{zetagam}
\eeq

The transfer matrix $\T^B(v)$ commutes \cite{Sut70,Bax72b}
with the Hamiltonian of the $XYZ$-model
(remind that we are assuming the periodic boundary condition),
\beq
{\bf H}_{XYZ}=-\frac{1}{2}\, \sum_{j=1}^N
\big(J_x\,\sigma^{(j)}_1\sigma^{(j+1)}_1+
J_y\,\sigma^{(j)}_2\sigma^{(j+1)}_2+
J_z\,\sigma^{(j)}_3\sigma^{(j+1)}_3\big)\label{ham}
\eeq
provided
\beq
J_x:J_y:J_z= \frac{\vartheta_4(2\eta\,|\,\q)}{\vartheta_4(0\,|\,\q)}:
\frac{\vartheta_3(2\eta\,|\,\q)}{\vartheta_3(0\,|\,\q)}:
\frac{\vartheta_2(2\eta\,|\,\q)}{\vartheta_2(0\,|\,\q)}\ .\label{J-rat}
\eeq
.

Following \cite{Bax72b} let us show that this Hamiltonian is simply
related to the logarithmic derivative of the transfer matrix at $v=\eta$.
It follows from \eqref{Rw} and \eqref{wt} that
\beq
\vt_1(2\eta\,|\,\q)\,\Big[\frac{d}{dv} \log \T^B(v)\Big]_{v=\eta}=
N\,p_4'\,{\bf I}+\sum_{j=1}^N
\big(p_1'\,\sigma^{(j)}_1\sigma^{(j+1)}_1+
p_2'\,\sigma^{(j)}_2\sigma^{(j+1)}_2+
p_3'\,\sigma^{(j)}_3\sigma^{(j+1)}_3\big)\ ,\label{T-diff}
\eeq
where ${\bf I}$ denotes the unit operator. Here we used new variables
\beq
p_1=\frac{1}{2}(w_1-w_2-w_3+w_4),\quad p_2=\frac{1}{2}(-w_1+w_2-w_3+w_4)\ ,
\eeq
\beq
p_3=\frac{1}{2}(-w_1-w_2+w_3+w_4),\quad p_4=\frac{1}{2}(w_1+w_2+w_3+w_4)\ .
\eeq
Their $v$-derivatives evaluated at $v=\eta$ denoted as
\beq
p_i'=\frac{dp_i}{dv}\Big\vert_{v=\eta}\ .
\eeq
Note that
\beq
p_1+p_2+p_3+p_4=2w_4\ .
\eeq
Using \eqref{wt} on can readily show that
\beq
2p_i'=\vt_1'(0\,|\,\q)
\frac{\vt_{5-i}(2\eta\,|\,\q)}{\vt_{5-i}(0\,|\,\q)},\qquad i=1,2,3\ .
\eeq
Combining the last relation with \eqref{T-diff} one can express the Hamiltonian
\eqref{ham} as
\beq
{\bf H}_{XYZ}=-\frac{J_x \,\vt_4(0\,|\,\q)}{\vt_4(2\eta\,|\,\q)\,
\vt_1'(0\,|\,\q)}\,\Big\{
{\vt_1(2\eta\,|\,\q)}\,
\frac{d}{dv}\log \T^B(v)-{N p_4'}\,{\bf I}
\Big\}\Big|_{v=\eta}\label{ht}
\eeq

In the main text of the paper we mostly use the spectral parameter $u$
which is related to variable $v$ in \eqref{weights} as
\beq\label{uv-rel}
u=v-\pi\tau/2 \ .
\eeq
In \eqref{tq-redef} we
also define the renormalized transfer-matrix $\T(u)$
differing from $\T^B(u+\pi\tau/2)$ by a simple $u$-dependent factor.
Of course, this factor could have been included in the normalization
of the Boltzmann weights \eqref{weights}, but we prefer not to do so
and keep a clear distinction between two alternative
normalizations and, in fact, to use both of them.
In particular, the Baxter's normalization and
the variable $v$ are more
suitable in writing explicit
expressions for the eigenvalues of the transfer matrix
presented in Sect.~\ref{cont-sect}.

There are two related, but different, constructions of the
$\Q$-matrix for the 8V-model, given in Baxter's 1972 and 1973
papers \cite{Bax72} and
\cite{Bax73a}. The results of \cite{Bax72}
apply for rational $\eta$'s and an
arbitrary number of sites $N$, while the those of \cite{Bax73a} apply
to arbitrary $\eta$'s and even $N$. We will denote the corresponding
$\Q$-matrices as $\Q^{(72)}(v)$ and $\Q^{(73)}(v)$.
They obey the same TQ-equation (Eq.(4.2) of \cite{Bax72} and Eq.(87)
of \cite{Bax73a}),
\beq
\T^B(v)\,\Q^B(v)=\phi(v-\eta)\,\Q^B(v+2\eta)+\phi(v+\eta)\,\Q^B(v-2\eta),
\label{TQop}
\eeq
where
\beq
\phi^B(v)=\big(\vartheta_1(v\,|\,q)\big)^N\ .\label{phib-def}
\eeq
and possess  the same
periodicity properties
\beq
\Q^B(v+\pi)=\Scal\,\Q^B(v),\qquad
\Q^B(v+2\pi\tau)=\q^{-2N}\,e^{-2iv N}\,\Q^B(v).\label{qper}
\eeq
where $\Scal$  is defined in \eqref{rs-def} and
$\Q^B(v)$ stands for either of  $\Q^{(72)}(v)$ or $\Q^{(73)}(v)$.
Nevertheless, as noticed in \cite{McCoy1}, these two $\Q$-matrices
do not coincide. Of course, they can
only be compared for rational $\eta$ and even $N$
when both of them can be constructed.
As shown in \cite{McCoy1} the
matrices  $\Q^{(72)}(v)$ and $\Q^{(73)}(v)$
have different eigenvalues for degenerate
eigenstates of the transfer matrix. This phenomenon is obviously
related with the non-uniqueness of the eigenvectors for the
degenerate eigenstates \cite{Bax02} and demonstrates the impossibility
of an universal definition of the $\Q$-matrix in the zero-field
8V-model at rational $\eta$ (cf. \cite{BS90}).

The difference between  $\Q^{(72)}(v)$ and $\Q^{(73)}(v)$ can be
traced back \cite{McCoy1} to their commutativity properties with the operator
$\Rcal$. The matrix $\Q^{(73)}(v)$ (defined for even $N$ only)
always commutes with $\Rcal$.  The matrix $\Q^{(72)}(v)$ (defined for
all values of $N$)
does not commute with $\Rcal$, irrespectively to whether $N$ is odd or even.
On the other hand both these matrices commute with $\Scal$, so that
the first equation immediately translates into the (real period) periodicity
for the eigenvalues (the first relation in \eqref{qper3}).
The second equation in \eqref{qper} implies only the
$2\pi\tau$-periodicity relation (the second relation in
\eqref{qper3}).

To complete our discussion of various constructions for the
$\Q$-matrix, note that in \cite{Bax82} Baxter considered a modified
version of
his $\Q^{(73)}$ matrix. There he used yet another set of parameters, namely
the variables $v$ and $\lambda$, which we denote here as
$v^{(82)}$ and $\lambda^{(82)}$. They
are related to $v$
and $\eta$ in \eqref{weights} as
\beq
v^{(82)}=\frac{2i\K_B}{\pi}(2v-\pi),\qquad \lambda^{(82)}=
\frac{2i\K_B}{\pi}(\pi-2\eta)
\eeq
where $\K_B$ is the same as in \eqref{baxvar}.
Writing the $\Q$-matrix of \cite{Bax82} as
$\Q^{(82)}(v^{(82)},\lambda^{(82)})$ and that of \cite{Bax73a} as
$\Q^{(73)}(v,\eta)$, and using the explanations in Sect.7 of
\cite{Bax02} it is not difficult to verify that
\beq
\Q^{(73)}(v,\eta)=\Rcal\,\Dcal\,\Q^{(82)}(v^{(82)},\lambda^{(82)})
\eeq
where the operator
\beq
\Dcal=e^{+\frac{i\pi}{4}\sigma_3}\otimes e^{-\frac{i\pi}{4}\sigma_3}
\otimes e^{+\frac{i\pi}{4}\sigma_3}\otimes\cdots
\otimes e^{-\frac{i\pi}{4}\sigma_3}
\eeq
is defined similarly to those in \eqref{rs-def}. Using the
relations
\beq
\Rcal\,\Dcal=(-1)^{N/2}\Dcal\,\Rcal\,\Scal,\qquad
\Q^{(73)}(v+\pi\tau)=\q^{-N/2}\,e^{-iv N}\,\Rcal\,\Q^{(73)}(v),\label{qpera}
\eeq
one can show that
\beq
\Q^{(82)}(v^{(82)}+2\K'_B)=(q_B)^{-N/4}\,\exp(\pi v^{(82)} N/4\K_B)\,
\Rcal\Scal\,\Q^{(82)}(v^{(82)})
\eeq
which is exactly Eq.(10.5.43a) of \cite{Bax82}.

The partition function \eqref{pf} can be regarded as a function of
four parameters $w_1,w_2,w_3,w_4$, defined in \eqref{w-def}.  So
one can write it as $Z(w_1,w_2,w_3,w_4)$.
When both $M$ and $N$ are even, it possesses the symmetry
\beq
Z(w_1,w_2,w_3,w_4)=Z(\pm w_i,\pm w_j,\pm w_k,\pm w_l),\label{symm}
\eeq
where $\{i,j,k,l\}$ is an arbitrary permutation of $\{1,2,3,4\}$ and
all signs can be chosen independently.
Using this symmetry one can always rearrange $w_i$ so that
\beq
w_1>w_2>w_3>|w_4| \ . \label{region}
\eeq
The partition function-per-site, $\kappa(v)$, is defined as
\beq
\log \kappa(v)=\lim_{M,N\to\infty}\frac{1}{MN}\log Z=
\lim_{N\to\infty}\frac{1}{N}\log\,\Lambda_0^B(v)\label{kappa-def}
\eeq
where $\Lambda_0^B(v)$ is the maximum eigenvalue of the transfer-matrix
$\T^B(v)$. In the (unphysical) regime (\ref{region})
it reads \cite{Bax72}
\beq
\log\kappa(v)
=\log c+
2\sum_{n=1}^\infty\frac{\sinh^2((\tilde\tau_B-\l_B)n)
\ (\cosh(n\l_B)-\cosh(n\a_B))}
{n\sinh(2n\tilde\tau_B)\cosh(n\l_B)},\label{partition}
\eeq
where
\beq
\a_B=-2i v,\quad \l_B=-2i\eta,\quad \tilde\tau_B=-\pi i\,\tau.\label{Baxchoice}
\eeq

In this paper we consider the disordered regime\footnote{%
See \cite{Bax82} for the classification of the regimes of the 8V-model.
Beware that the variables $w_i$ therein are numbered differently as compared
to this work. Here we follow the original papers
\cite{Bax72,Bax73a,Bax73b,Bax73c}.}
\beq
0<\eta<\pi/2, \qquad \eta<v<\pi-\eta,
\eeq
which corresponds to a different ordering of the variables $w_i$, namely,
\beq
w_4>w_1>w_2>w_3>0.\label{disorder}
\eeq
This regime can be mapped into \eqref{region} by the following
transformation of the weights
\beq
a'=\frac{a-b+c-d}{2},\quad
b'=\frac{a-b-c+d}{2},\quad
c'=\frac{a+b+c+d}{2},\quad
d'=\frac{a+b-c-d}{2}\ ,\label{trans1}
\eeq
which is equivalent to
\beq
w_1'=w_4,\quad w_2'=w_1,\quad w_3'=w_2,\quad w_4'=w_3\ ,\label{trans2}
\eeq
where $w_1',w_2',w_3',w_4'$ are defined by \eqref{w-def} with
$a,b,c,d$ replaced by $a',b',c',d'$. It is easy to check that if
$w_1,w_2,w_3,w_4$ belong to the regime (\ref{disorder}) then
$w_1',w_2',w_3',w_4'$ belong to (\ref{region}).
Putting \eqref{weights} into \eqref{trans1} it is not very difficult to
show that the ``transformed'' weights $a',b',c',d'$ can be parameterized by
same formulae \eqref{weights} provided one makes the following substitution
\beq
v\to \frac{v-\pi/2}{\tau},\qquad
\eta\to\frac{\eta-\pi/2}{\tau},\qquad
\tau\to-\frac{1}{\tau},\qquad \rho\to\rho_1\rho\ ,
\eeq
where
\beq
\rho_1=-i(-i\tau)^{-1/2}
\exp\Bigl\{-\frac{i}{\pi\tau}\bigl[v(v-\pi)+3\eta(\eta-\pi)+\pi^2\bigr]\Bigr\}\ .
\eeq
Thus, to find the partition function-per-site
in the regime \eqref{disorder} one needs to substitute parameters $\alpha_B,
\l_B$ and $\tau_B$ in (\ref{partition}) by
\beq
\a_B=\frac{i(\pi-2v)}{\tau},\qquad
\l_B=\frac{i(\pi-2\eta)}{\tau},\qquad
\tilde\tau_B=\frac{i\pi}{\tau}\ ,
\eeq
and replace $c$ by $c'$. After straightforward calculations one obtains
\beq
\log\kappa(v)= \log\vt_1(v+\eta\,|\,\q)+\frac{2i\eta(v-\eta)}{\pi\tau}
+\sum_{n=1}^\infty\frac{\sinh(2\eta\, x_n)\,\sinh(2(v-\eta)\,x_n)}
{n \,\sinh(\pi \,x_n)\,\cosh((\pi-2\eta)\,x_n)} \ ,
\label{ourpart}
\eeq
where $ x_n=in/\tau$.
The above formula can be equivalently rewritten as
\beq
\log\kappa(v)=\log\vt_1(v\,|\,\q)
+\frac{i\eta(\pi-3\eta)}{\pi\tau}
+2\sum_{n=1}^\infty\frac{
\sinh(\eta\,x_n)
\sinh((\pi-3\eta)\,x_n)
\cosh((\pi-2v)\,x_n)}
{n\, \sinh(\pi\,x_n)\,
\cosh((\pi-2\eta)\,x_n)}\ .\label{ourpart1}
\eeq

Note that for $\eta=\pi/3$ the second and third terms in the last expression
vanish, so it simplifies to
\beq
\log\kappa(v)=\log\vt_1(v\,|\,\q), \qquad \eta=\pi/3 .\label{ourpart2}
\eeq
Recall that the  above derivation was based on the symmetry \eqref{symm}
valid for even values of $N$ only. However, since in the disordered regime
\eqref{disorder} the
Boltzmann weights \eqref{weights} are strictly  positive,
the result \eqref{ourpart2},
obviously, does not depend on the way
the limit $N\to\infty$ is taken. Eqs. \eqref{kappa-def} and \eqref{ourpart2}
imply that the asymptotics on the largest
eigenvalue of $\T^B(v)$ in this special
case has an extremely simple form $\Lambda_0^B(v)\simeq (\vt_1(v\,|\,\q))^N$.
Remarkably, as it was argued in \cite{Bax89,Str01c,BM05},
for odd values of $N$ this
asymptotics is {\em exact} even for a finite lattice,
\beq
\Lambda^B_0(v)=(a+b)^N=\vt_1^N(v\,|\,q),\quad N=2n+1,\quad \eta=\pi/3\ .
\label{largest}
\eeq
We have also checked this fact analytically up to $N=15$.

\def\cprime{$'$} \def\cprime{$'$}

\end{document}